\documentclass{article}

\usepackage{microtype}
\usepackage{graphicx}
\usepackage{subcaption}
\usepackage{booktabs} %
\usepackage{tabularx}
\usepackage{booktabs}
\usepackage{hyperref}

\makeatletter
\let\real@addcontentsline\addcontentsline
\makeatother

\usepackage[accepted]{icml2026}

\usepackage{amsmath}
\usepackage{amssymb}
\usepackage{mathtools}
\usepackage{amsthm}

\usepackage[capitalize,noabbrev]{cleveref}

\theoremstyle{plain}

\theoremstyle{definition}

\theoremstyle{remark}

\usepackage[textsize=tiny]{todonotes}
\usepackage{booktabs,multirow,adjustbox}
\usepackage{threeparttable}
\makeatletter
\IfFileExists{aas_macros.sty}{%
  \usepackage{aas_macros}%
}{%

}
\makeatother

\newcommand{\blue}[1]{#1}
\newcommand{\newblue}[1]{\textcolor{black}{#1}}

\usepackage[toc,page]{appendix}

\usepackage{titletoc}

\usepackage{stfloats}

\icmltitlerunning{\texttt{StarEmbed}: Benchmarking Time Series Foundation Models on
Astronomical Observations of Variable Stars}

\begin{document}

\twocolumn[
  \icmltitle{\texttt{StarEmbed}: Benchmarking Time Series Foundation Models on
  \\
  Astronomical Observations of Variable Stars}

  \icmlsetsymbol{equal}{*}
    \begin{icmlauthorlist}
      \icmlauthor{Weijian Li}{equal,cfmg,skai,cs}
      \icmlauthor{Hong-Yu Chen}{equal,cfmg,skai,cs}
      \icmlauthor{Nabeel Rehemtulla}{equal,skai,ciera,astro}
      \icmlauthor{Ved G. Shah}{skai,ciera,astro}
      \icmlauthor{Dongho Kim}{stats}
      \icmlauthor{Dennis Wu}{cfmg,skai,cs}
      \icmlauthor{Qinjie Lin}{cfmg,skai,cs}
      \icmlauthor{Adam A. Miller}{skai,ciera,astro}
      \icmlauthor{Han Liu}{cfmg,skai,cs,stats}
    \end{icmlauthorlist}

  \icmlaffiliation{cfmg}{Center for Foundation Models and Generative AI (CFMG), Northwestern University, Evanston, USA}
  \icmlaffiliation{ciera}{Center for Interdisciplinary Exploration and Research in Astrophysics (CIERA), Northwestern University, Evanston, USA}
  \icmlaffiliation{skai}{NSF – Simons AI Institute for the Sky (SkAI), Chicago, USA}
  \icmlaffiliation{cs}{Department of Computer Science, Northwestern University, Evanston, USA}
  \icmlaffiliation{astro}{Department of Physics and Astronomy, Northwestern University, Evanston, USA}
  \icmlaffiliation{stats}{Department of Statistics and Data Science, Northwestern University, Evanston, USA}

  \icmlcorrespondingauthor{Weijian Li}{weijianli@u.northwestern.edu}
  \icmlcorrespondingauthor{Hong-Yu Chen}{charlie.chen@u.northwestern.edu}
  \icmlcorrespondingauthor{Nabeel Rehemtulla}{nabeelr@u.northwestern.edu}
  \icmlcorrespondingauthor{Adam A. Miller}{amiller@northwestern.edu}
  \icmlcorrespondingauthor{Han Liu}{hanliu@northwestern.edu}

  \icmlkeywords{Machine Learning, ICML}

  \vskip 0.3in
]

\printAffiliationsAndNotice{\icmlEqualContribution}  %

\begin{abstract}
  
Current time series foundation model (TSFM) training corpora largely omit data with certain complexities like irregular temporal sampling. Astronomical time series of stellar fluxes (``light curves'') are available in immense quantities and exhibit irregular sampling, multiple variates, and heteroskedasticity. We introduce \texttt{StarEmbed}, the first public benchmark for light curves comprised of real observations of $\sim$40,000 stars expert-labeled across seven classes and evaluations in clustering, classification, and out-of-distribution (OOD) source detection. We benchmark TSFMs with differing architecture and training strategies as well as domain-specific transformers. Our results demonstrate that the \texttt{Chronos} family, despite being pre-trained on regularly sampled non-astronomical data, yields state-of-the-art (SOTA) performance in light curve clustering and OOD detection. While no TSFM strictly surpasses the classification performance of the long-established domain baseline, they do demonstrate excellent generalization abilities. \texttt{StarEmbed} marks a step toward universal light curve embeddings and improved TSFM performance on challenging data.

\end{abstract}

\section{Introduction}
\label{sec:intro}

\looseness=-1
\vspace{-0.5em}

\begin{figure*}
    \centering
    \includegraphics[width=0.60\textwidth]{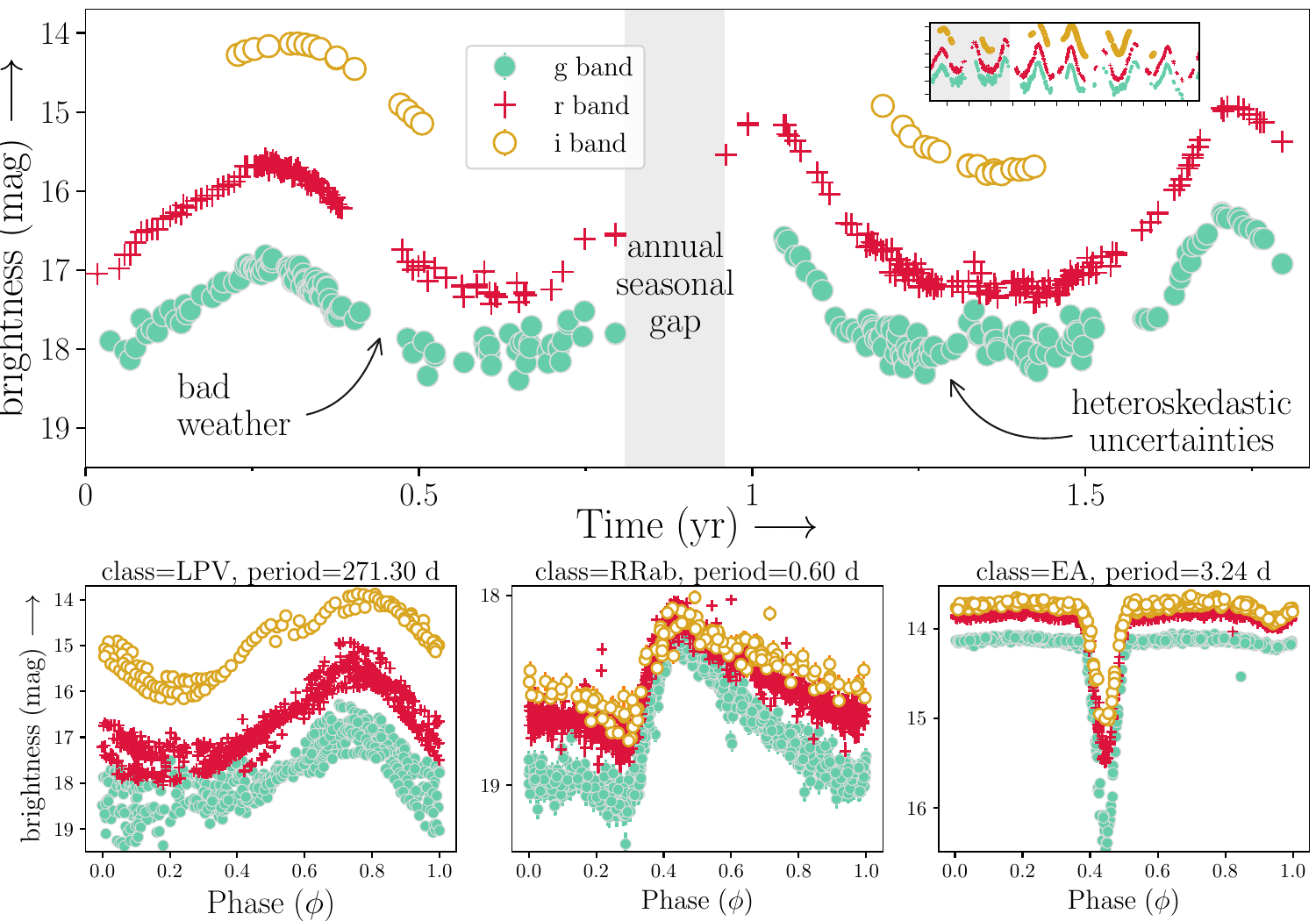}
    \includegraphics[width=0.34\linewidth]{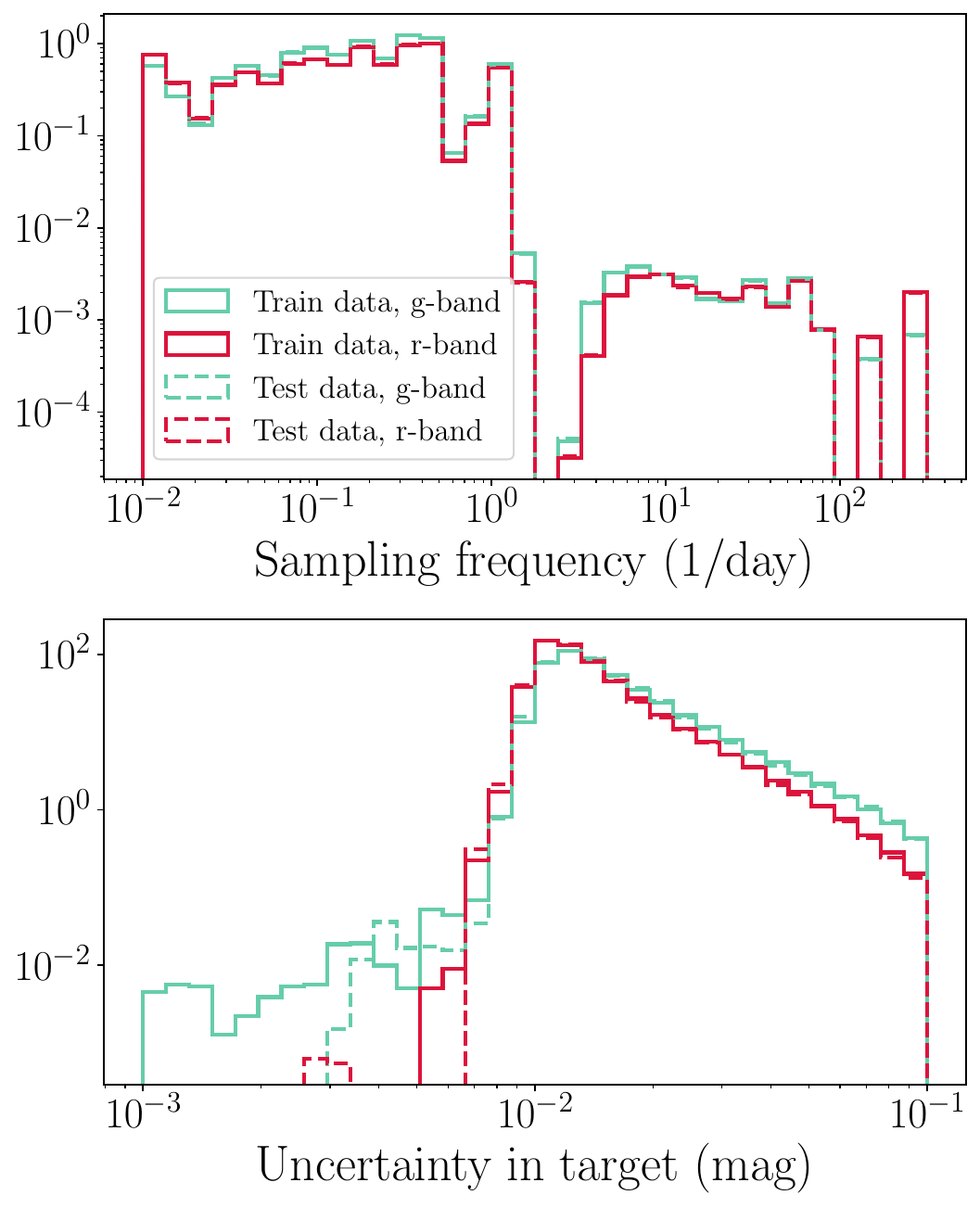}
    \caption{
    Data challenges represented in our data which are characteristic of astronomical time series.
    \textit{Upper left panel}: Observed light curve of a periodic variable star exhibiting multiple variates, heteroskedastic uncertainties, and large regular and irregular gaps. The inset shows the full $\sim$6.5\,years of observations by ZTF.
    \textit{Lower left panels}: Phase-folded light curves highlighting the periodic patterns in three different classes. Note that most stars have few $i$-band observations (orange points) so we exclude these data from our analysis.
    \textit{Upper right panel}: Histogram of sampling intervals (days between consecutive measurements).
    \textit{Lower right panel}: Histogram of measurement uncertainties. Sampling frequency and uncertainty values vary across multiple orders of magnitude, reflecting the considerable influence of these characteristics on the signals in the data.
    \vspace{-1.5 em}
    }    
    \label{fig:lc}
\end{figure*}

The adoption of time series foundation models (TSFMs), with pre-training corpora that span commerce, finance, electricity, and traffic data, is proliferating due to their highly capable, general-purpose representation learning capabilities \citep{zhou2021informer,nie2022time,yang2024research,woo2024unified}. The pre-training corpora, however, do not include the large quantities of publicly available astronomical time series which contain signatures that are rare in standard benchmarks: multiple variates, irregular temporal sampling, regular and irregular gaps, and heteroscedasticity (see Figure~\ref{fig:lc}). These challenges are unavoidable for the ground-based survey observatories which are generating petabyte-scale time series data sets, such as the Zwicky Transient Facility \citep[ZTF;][]{Bellm+2019a} and the Vera C.~Rubin Observatory \citep[Rubin;][]{Ivezic+2019}. Over 7\,yr of operations, ZTF alone has produced multi-variate time series for $\sim$10$^9$ stars, each containing $\sim$10$^3$ observations; during its first decade of full survey operations starting in 2026, Rubin will scale this by a factor of $10-100\times$. The different variates in a light curve represent different colors of starlight captured by placing a filter (or ``passband" / ``band") in the telescope's focal path, and the bands' relative values (as a function of time) hold nuanced astrophysical insights into the star. Challenges similar to those seen in light curves are also present in time series originating from other scientific domains, so there is both a pressing need and a unique opportunity to evaluate TSFMs on these real scientific data.

We focus this benchmark on \textit{periodic variable stars}: stars which exhibit brightness variations over regular, periodic intervals. These stars are uniquely valuable probes of astrophysics from stellar interiors to galactic structure and more \citep[e.g.,][]{Feast+1987, Clementini+2003, Genovali+2014, Catelan+2015, Ripepi+2017}. Despite the abundance of light curves, there is no standardized benchmark for assessing their embeddings. The absence of common datasets, class sets, and train-test splits has hindered fair, reproducible comparisons and obscured whether domain-specific pipelines outperform generic representations from foundation models \citep[cf.,][]{pan2024astromlab}.

We introduce \texttt{StarEmbed}, the first public benchmark for rigorous, standardized evaluation of SOTA TSFMs on astronomical observations. \texttt{StarEmbed} integrates multi-band ZTF light curves for $\sim$40k stars with expert-vetted labels over seven astrophysical classes. Embeddings are evaluated on unsupervised clustering, supervised classification, and OOD source detection in zero-shot and fine-tuned settings. Here, we present results for three TSFM families (\texttt{Moirai}, \texttt{Chronos}, and \texttt{Time-MoE}), a domain-specific transformer (\texttt{Astromer}), and the long-standing domain baseline of hand-crafted feature extraction.

We describe related works and models (\Cref{sec:related_works}), our dataset (\Cref{sec:datasets}), methodologies (\Cref{sec:eval}), results (\Cref{sec:benchmark_results}), and interpretations of our results (\Cref{sec:discussion_conclusions}).

\vspace{-1em}

\section{Related Works and Models}
\label{sec:related_works}

\vspace{-0.5em}

The classification of periodic variable stars has long been a task of significant interest. As such, both pre- \citep[e.g.,][]{debosscher07} and post-deep learning models \citep[e.g.,][]{Moreno-Cartagena+2025} have been thoroughly explored. Here, we briefly summarize the evolution of the models used in the astrophysics literature, the TSFMs we benchmark, and the baselines we compare them against.

\vspace{-0.5em}
\subsection{Supervised Classifiers}
\vspace{-0.5em}

The first AI models to classify variable stars used manually engineered features combined with support vector machines \citep{debosscher07} or gradient boosted decision trees \citep{Richards+2011, Sesar+2017, 2019AJ....158..257B}. The set of extracted features varies across studies but is consistently dozens large and includes Fourier coefficients, variability amplitude and skewness, and goodness-of-fit statistics for matches to model templates \citep{Nun+2015, Kim+2016, Malanchev+2021}. Deep learning methods have also been used to eliminate feature engineering by adopting recurrent neural networks \citep[RNNs;][]{2019PASP..131k8002M, 2020MNRAS.493.2981B, shah2025oracle} and transformers \citep{cabrera2024atat, Moreno-Cartagena+2025}. Deep learning efforts, however, do not perform meaningfully better than hand-crafted feature extraction: accuracies of RNN models are $\pm1-3\%$ of hand-crafted features across multiple variable star datasets \citep[see, e.g.,][]{naul2018recurrent}. As a result, we choose to use hand-crafted features to establish our baseline performance.

\textbf{Baseline Model: } We extract features from the light curves with the \texttt{FATS} \citep{Nun+2015} and \texttt{light\_curve} \citep{Malanchev+2021} software packages. Example features include: the best-fit Lomb-Scargle \citep{Lomb+1976, Scargle+1982} period, the scatter, skewness, kurtosis, and other variability statistics. In total, we define 69 features per band, yielding a total embedding size of 138 for the two-band ZTF data (see Appendix~\ref{app:handcrafted} for a full feature list with explanations). In Appendix~\ref{app:HCF_feat_importance} we present relative feature importance statistics which find that the period of variability contributes the most to the embedding vector's classification capabilities. Before passing the 138-size embedding to task heads, we normalize each feature to have zero mean and unit variance. While very effective, hand-crafted features rely heavily on domain knowledge, can be brittle to data quality issues, and are expensive to compute.

\vspace{-0.5em}
\subsection{Astrophysics Embedding Models}
\vspace{-0.5em}

Semi-supervised and self-supervised techniques have also recently been applied to light curves to learn general representations of the data and to later perform downstream tasks. Many of these have been limited to single categories of astrophysical phenomena, for instance, variational autoencoders \citep{2020ApJ...905...94V}, sparse autoencoders \citep{2025MNRAS.537..931D}, and contrastive learning \citep{zhang2024maven} have been applied to explosive transients. A few foundation models have been developed for variable star light curves, such as \texttt{FALCO} \citep{FALCO} and \texttt{Astromer} \citep{donoso2023astromer, astromer2}. Unlike \texttt{FALCO}, \texttt{Astromer-1} and \texttt{Astromer-2} are designed to generalize across observatories, so we adopt them as domain-specific foundation models for discerning what advantages in-domain pre-training yields. 

\textbf{\texttt{Astromer}} \citep{donoso2023astromer, astromer2} \texttt{Astromer-1} was pre-trained using self-supervised learning on 1.5 million single-band light curves from the MACHO survey \citep{Alcock+2000}. The model’s output is a fixed-length embedding of 256-dimensions taken from the final attention layer. \texttt{Astromer-2} functions very similarly but increases the number of model parameters from $0.66$M to $5.4$M and adopts an uncertainty-weighted loss function for pre-training.

\vspace{-0.5em}
\subsection{Time Series Foundation Models}
\vspace{-0.5em}

TSFMs have been shown to consistently outperform the traditional \emph{one-dataset-per-model} schema in multiple fields, including finance, climate science, and commerce \citep[e.g.,][]{yue2022ts2vec, woo2024unified, ansari2024chronos}. With strong performance that scales with model and dataset size, they are a promising tool for driving the future of AI for time series \cite{Edwards+2024, Pan+2024}. Astronomy, despite having an enormous collection of light curves, has yet to explore the potential of TSFMs, which may prove transformative in the ability to accomplish multiple downstream tasks. Furthermore, these light curve datasets provide a unique opportunity to evaluate TSFMs with minimal risk of data leakage because they are not included in the widely used pre-training corpora. Each of the TSFMs we consider here have multiple versions with varying parameter counts; in the main text we consider the ``small" and ``tiny" variants of \texttt{Moirai} and \texttt{Chronos}, respectively, and the smallest available \texttt{Time-MoE} variant, \texttt{Time-MoE-base}. Larger variants are evaluated in \cref{app:model_size}, and we detail computational cost in \cref{tab:flops_L200}.

\textbf{\texttt{Moirai}} \citep{woo2024unified, Liu+2025_moirai2} is a pre-trained TSFM designed for universal forecasting across diverse sampling frequencies, dimensionalities, and data distributions. \texttt{Moirai-1} uses a multi-patch-size projection scheme, an any-variate attention mechanism that scales to arbitrary numbers of variables, and a flexible mixture-distribution output head. It is trained on LOTSA, an open archive of 27 billion observations spanning nine domains. \texttt{Moirai-2} simplifies this framework with a decoder-only architecture and quantile forecasting. 
By removing masked-token objectives, multi-patch inputs, and mixture-distribution outputs, \texttt{Moirai-2} emphasizes data efficiency and scalability.

\textbf{\texttt{Chronos}} \citep{ansari2024chronos} is also a pre-trained TSFM trained for universal forecasting. It treats forecasting as a language-modeling problem by quantizing each time-series point into a fixed vocabulary and training off-the-shelf transformer models (T5-style models with 20M to 710M parameters) with an cross-entropy loss. Augmented by TSMixup and Gaussian-process–generated synthetic data, \texttt{Chronos} is pre-trained on a large collection of public datasets and evaluated on 42 benchmarks. The models deliver strong probabilistic forecasts which are ahead of classical and deep-learning baselines on in-domain data and in zero-shot settings. 
\texttt{Chronos-bolt} further improves the inference speed by changing from autoregressive decoding to generating multiple forecasting steps in one run. 
It also changes from point-wise quantization into patching mechanism. 

\textbf{\texttt{Time-MoE}} \citep{Shi+2024_TimeMoE} utilizes a sparse mixture-of-experts architecture to achieve efficient and scalable universal forecasting. It dynamically routes inputs to a small collection of specialized experts rather than relying on dense parameter sharing. This enables scaling up to greater parameter counts with more favorable scaling in computational cost. It is also trained on a massive and heterogeneous time series corpora and demonstrates excellent performance across domains and tasks. Its mixture-of-experts architecture offers an alternative philosophy to a TSFM relative to the \texttt{Moirai} and \texttt{Chronos} families. 

\vspace{-0.5em}
\subsection{Random Embeddings As a Sanity Check Baseline}
\vspace{-0.5em}

We generate values for a 256-dimensional vector from a $\mathcal{U}[0,1)$ distribution as a proxy for a light curve embedding to serve as a marker for the performance floor. These vectors carry no information on the data, so any improvement relative to them suggests the alternative models captured some useful information in their embeddings.
\vspace{-0.5em}

\section{Dataset}
\label{sec:datasets}
\vspace{-0.5 em}

The benchmark dataset includes multi-variate time series observations of periodic variable stars. 
The data are obtained over 6.5\,yr of ZTF operations of imaging the entire Northern sky every few nights and are accessed through the 23rd ZTF data release (DR).\footnote{\url{https://irsa.ipac.caltech.edu/Missions/ztf.html}} ZTF observes in three different bands, $g$, $r$, and $i$\footnote{Most ZTF sources have very few or no observations in $i$ band and we therefore exclude it from the benchmark.} (see left side of \Cref{fig:lc}) with effective wavelengths of $\sim$4746\AA, 6366\AA, and 7829\AA, respectively \citep{Dekany+2020}. The flux is presented in magnitudes (an astronomy-specific unit), while the time is recorded as the modified Julian date. 

The right side of Fig.~\ref{fig:lc} highlights the broadness of sampling frequencies and the significant irregularity in the temporal sampling of this data. The frequencies of two observations per day mean that observing from the ground can only occur at nighttime and it is always daytime 0.5 days after one observation. The sampling is consistent across bands and splits. Many factors can affect the uncertainties of the flux measurements, which is reflected in their large spread.\footnote{We observe a slight over-density of $g$ band measurements with very small uncertainties in the train split. Noting the logarithmic scale on the y-axis, this is not a concern as these represent only a very small number of observations.} These features are in stark contrast to those in time series datasets commonly used for TSFM pretraining. LOTSA \cite{woo2024unified}, e.g., only includes time series without spread in their sampling frequencies, i.e. only regularly sampled time series, and nearly all time series in LOTSA do not include uncertainties on the target or multiple variates. Therefore, these astronomical time series data represents a special out-of-domain test for the TSFMs we explore here.

While many catalogs of periodic variable stars exist in the literature, most are produced by low-capacity AI models. Although benchmarks built off of these catalogs can be useful \citep[e.g.,][]{Fei+2024}, we avoid these to prioritize label quality. We instead take the Catalina Surveys Periodic Variable Star Catalog \citep[CSPVS;][]{Drake+2014}, a human-labeled catalog created by the Catalina Real-Time Transient Survey \citep{Drake+2009}. We extract ZTF DR23 light curves for each CSPVS star. Stars that lack a ZTF light curve are omitted; we also remove (i) observations flagged as ``bad'' in ZTF DR23; (ii) light curves with $<32$ total observations; (iii) light curves that lack both $g$ and $r$ observations; and (iv) light curves from classes with fewer than 350 total examples. The resulting dataset contains $\sim$40,000 light curves of expert-labeled stars across seven classes. See App.~\ref{app:classes_intro} for a detailed astrophysical description of each class.

Nature produces an imbalance between the number of examples in each class, which is further exacerbated by each class having a different detection efficiency. To ensure that each split gets a representative number of examples from each class, we sample each into the train, validation, and test splits independently in a 7:1:2 ratio. Table~\ref{tab:data} shows the counts of examples per class in each split. We release our train/validation/test splits on HuggingFace.\footnote{\url{https://huggingface.co/datasets/StarEmbed/ZTF_40k}}

\subsection{ZTF Dataset in Context: Relevance to Other Observatories}

Variable star science has an expansive scope that extends beyond ZTF and periodic variables; the \texttt{StarEmbed} benchmark is designed to allow for the addition of new datasets and metrics in future expansions. This flexibility is crucial to the long-term health of this benchmark as numerous new time-domain facilities, like Rubin, will begin releasing floods of data in the coming years. Each new survey has unique observational capabilities and priorities that will affect the resulting embeddings and downstream task performance. As the largest astronomical time-domain experiment to date, ZTF is an apt choice for building a preparatory benchmark dataset as ZTF data has already been used to explore emerging AI areas like multi-modality, pre-training, and transformers \citep[e.g.,][]{Duev+2021, Carrasco-Davis+2021, Gagliano+2023, Allam+2023, zhang2024maven, Rehemtulla+2024, Rehemtulla+2026}.

All current and future datasets associated with the \texttt{StarEmbed} benchmark are and will be made public with a license for reuse. We host a leader board tracking model performance on each task on our website\footnote{\url{https://www.starembed.com/}} as well as an interactive data visualization tool called the StarEmbed Explorer where a user can peruse the data comprising the benchmark.\footnote{\url{https://see.starembed.com/}} No personal or sensitive information is present in these datasets because they consist of astronomical observations.
\vspace{-0.5em}

\begin{table}[t]
\captionsetup{skip=1pt}
    \centering
    \caption{Number of periodic variable stars in our data.}
    \label{tab:data}
    \setlength{\tabcolsep}{4pt}
    \renewcommand{\arraystretch}{0.6}
    \small
    \begin{tabularx}{\columnwidth}{l *{7}{>{\centering\arraybackslash}X}}
        \toprule
        Class & EW & EA & RRab & RRc & RRd & RS~CVn & LPV \\
        \midrule
        Train & 18998 & 2889 & 1386 & 3233 & 298 & 942 & 255 \\
        Val. & 2690 & 410 & 194 & 463 & 42 & 134 & 35 \\
        Test & 5387 & 818 & 397 & 926 & 83 & 276 & 70 \\
        \midrule
        Total & 27075 & 4117 & 1977 & 4622 & 423 & 1352 & 360 \\
        \bottomrule
    \end{tabularx}
\end{table}

\section{Evaluation Methodology}
\label{sec:eval}
\vspace{-0.5em}
We assess the quality of embeddings for unsupervised clustering, supervised classification, and OOD source detection. Together, these give a comprehensive view of the intrinsic structure captured by the embeddings and their usefulness for downstream tasks. Below we detail the experimental settings including training procedures and metrics used to evaluate the embeddings. We release our code to reproduce our benchmark experiments.\footnote{\url{https://github.com/skai-institute/StarEmbed}} To maintain consistency throughout the benchmark, we use identical embedding sizes across different models whenever possible (i.e., when there exists such a pre-trained version of the model). \texttt{Astromer-1}, \texttt{Astromer-2}, \texttt{Chronos-Bolt}, \texttt{Chronos} and the Random Embeddings all have embedding size of 256. We adopt the smallest available pre-trained \texttt{Moirai} model, \texttt{Moirai}, which uses an embedding size of 384.

\vspace{-0.5em}
\subsection{Unsupervised Clustering}
\label{sec:methods_clustering}
\vspace{-0.5em}

Our unsupervised clustering evaluation assesses the alignment of structure in the embedding's feature space with known periodic variable star classes. Specifically, we use the K-means and Ward's method clustering algorithms, both configured to produce $N = 7$ clusters corresponding to the seven astrophysical classes in the dataset. Before executing the clustering algorithms, we normalize all embeddings to the standard normal because the clustering methods compute Euclidean distances which are sensitive to the scale of entries. The resulting cluster memberships are assigned to class predictions using the Hungarian matching algorithm \citep{kuhn1955hungarian}.

We produce uncertainties on performance metrics by repeating the K-means algorithm with 10 different initializations, choosing the clustering with the lowest within-cluster variance. Ward’s clustering is deterministic, so no uncertainties are provided. We compute three literature-standard metrics to comprehensively evaluate the clustering performance \citep{huang2020partially, monnier2020deep, sun2024lsenet, li2024image}. \textbf{Normalized Mutual Information} (NMI) measures the mutual information between the cluster assignments and the true class labels and is normalized to the range [0,1] with higher values indicating better alignment between the clusters and the true classes. \textbf{Adjusted Rand Index} (ARI) evaluates pairwise clustering agreements. It considers how often pairs of light curves are in the same cluster versus the same true class. An ARI of unity indicates perfect clustering, ARI$\sim$0 indicates random clustering, and ARI$<$0 indicates clustering worse than random. \textbf{Macro-averaged F1 score} (F1) is the average of the harmonic means of precision and recall computed independently for each class. It treats minority and majority classes equally, and is thus sensitive to the class imbalance inherent to our dataset.

This evaluation setting allows us to probe the separable class structure in the embeddings without any supervised training. We also visualize the embedding spaces with the uniform manifold approximation and projection (UMAP) algorithm, yielding an intuitive, qualitative view of clustering performance in Appendix~\ref{app:umap}.

\vspace{-0.5em}
\subsection{Supervised Classification}
\label{sec: 4.2}
\vspace{-0.5em}

Complementary to unsupervised clustering, we also evaluation the embeddings in a supervised classification scenario where the variable star class labels are directly predicted from the embeddings. We comprehensively evaluate the relevant information content of the embeddings by running this evaluation with four different classification heads: $k$-nearest neighbors ($k$-NN), a logistic linear probe, a random forest (RF), and a multi-layer perceptron (MLP). This simulates a scenario where one uses a pre-trained model to produce embeddings used as feature inputs for a classification task but does not fine-tune the embedding model (hence ``zero-shot'' in terms of the embedding model). 

Both $k$-NN and logistic probes are standard in the embedding evaluation literature \citep{caron2021emerging, zhou2021ibot, neelakantan2022text}, as they are simple methods that directly reflect separability in the embedding space.
RF is included, in part, because it is widely used in the astronomical literature for periodic variable star classification \citep{naul2018recurrent,2021AJ....161..141S,pimentel2022deep} and it achieves SOTA performance across many datasets \citep{naul2018recurrent}. An MLP is used as a modern, higher capacity, non-linear, deep-learning option. Detailed information on the classifiers, the hyperparameter optimization, and the final hyperparameters for each embedding model can be found in \blue{Appendix~\ref{app:hyperparameters}}.

We report standard multi-class classification metrics to summarize the performance of each embedding model and classification head pairing. These include accuracy, the macro-averaged F1 (see Sec.~\ref{sec:methods_clustering}), precision (the false positive rate subtracted from one), and the recall (the true positive rate).

\vspace{-0.5em}
\subsection{\blue{Out-of-Distribution (OOD) Source Detection}}
\vspace{-0.5em}

Identifying variable stars physically unlike those in labeled training sets is of great astrophysical interest. To test the effectiveness of the embeddings for detecting such OOD sources, we first create embeddings for the ZTF light curves of CSPVS stars with too few examples to be included in the train/validation/test splits ($\beta$-Lyrae, Blazhko, Anomalous Cepheids, Cepheid-II, HADS, LADS, ELL, Hump, PCEB, and EAup; see Sec.~\ref{sec:datasets}). We define these as OOD sources; they are represented in the \texttt{StarEmbed} dataset as an additional ``anom" data split. The embeddings of the OOD sources are mixed with the test set and run through a ``multi-class isolation forest" \citep[MCIF;][]{Gupta+2025_MCIF} where a separate isolation forest is fit to the embeddings of each of the seven inlier classes in the training set. The minimum of the seven isolation forest scores assigned to a given star's embedding is its OOD score. Isolation forest is a popular method for finding astrophysical outliers \citep{Malanchev+2021}, and \cite{Gupta+2025_MCIF} show that following this multi-class prescription yields better macro-averaged performance than a single isolation forest in many settings, including for periodic variables. Here, the performance is benchmarked with the fraction of sources in the top~$N^{\textrm{th}}$ percentile of OOD scores which are genuine OOD sources: the OOD purity.

\vspace{-0.5em}

\section{Benchmark Results}
\label{sec:benchmark_results}

We highlight the benchmark results below for each of the three downstream tasks: unsupervised clustering, supervised classification, and OOD detection. Besides, we also include extensive analyses and TSFM training experiments.
\vspace{-0.5em}

\subsection{Unsupervised Clustering}
\label{sec:results_unsupervised_clustering}

\begin{table*}[t!]
\captionsetup{skip=1pt}
  \centering
    \setlength{\tabcolsep}{5pt}
    \renewcommand{\arraystretch}{0.95}
    \small
  \caption{Results of unsupervised clustering with K-means and Ward. The best results are highlighted in \textbf{bold}, and the second-best results are \underline{underlined}. The \texttt{Chronos} models perform very well on this unseen data, placing first or second in all metrics and universally better than \texttt{Moirai} and the \texttt{Astromer} models. However, hand-crafted features perform the best overall.
  Model FLOPs are in \cref{tab:flops_L200}.
  }
  \resizebox{0.8\textwidth}{!}{
  \begin{tabular}{@{}lcccccc@{}}
    \toprule

      Methods & NMI (K-Means) & ARI (K-Means) & F1 (K-Means) & NMI (Ward) & ARI (Ward) & F1 (Ward) \\
    \midrule

\texttt{Astromer-1}        & 0.0041(0.0001) & 0.0017(0.0011) & 0.1660(0.0014) & 0.0041 & 0.0001 & 0.1652 \\
\texttt{Astromer-2}        & 0.0082(0.0010) & 0.0192(0.0078) & 0.1590(0.0042) & 0.0091 & 0.0310 & 0.1600 \\
\texttt{Moirai}      & 0.1749(0.0017) & 0.0981(0.0028) & 0.2831(0.0034) & 0.1476 & 0.0828 & 0.2612 \\
\texttt{\newblue{Moirai-2}}          & 0.2017(0.0019) & 0.0967(0.0060) & 0.3007(0.0095) & 0.1850 & 0.0603 & 0.2655 \\
\texttt{Chronos}      
& \underline{0.2374(0.0082)} 
& \textbf{0.1596(0.0029)} 
& 0.3110(0.0362) 
& 0.1890 
& 0.1217 
& \textbf{0.3671} \\
\texttt{Chronos-Bolt} 
& 0.2120(0.0033) 
& \underline{0.1306(0.0125)} 
& \underline{0.3128(0.0027)}
& \underline{0.2273}
& \textbf{0.1553} 
& \underline{0.3662} \\

\texttt{\newblue{Time-MoE}}    
& 0.2069(0.0108) 
& 0.0913(0.0090) 
& 0.3101(0.0227) 
& 0.1946 
& 0.0944 
& 0.3385 \\
Hand-crafted Features      
& \textbf{0.2700(0.0058)} 
& 0.1197(0.0092) 
& \textbf{0.3960(0.0271)} 
& \textbf{0.2508} 
& \underline{0.1319}
& 0.3323 \\

    \bottomrule
  \end{tabular}
  }
  \label{table:clustering-results-combined}
  \vspace{-0.8em}
\end{table*}

\begin{table*}[htbp]
\large
\captionsetup{skip=1.5pt}
  \caption{\blue{Results of benchmark supervised classification. The best results are highlighted in \textbf{bold}, and the second-best are \underline{underlined}. The $k$-NN and logistic classifiers are deterministic, so we report the 1-run performance; We report the mean and standard deviation of runs with $10$ seeds for RF and MLP. The hand-crafted features are SOTA with \texttt{Chronos} a clear second-best. See \cref{app:flops} for computation analysis of each model.}}
  \resizebox{\textwidth}{!}{
  \begin{tabular}{@{}llccccccccc@{}}
    \toprule
    \textbf{Classifier} & \textbf{Metric} & \textbf{\texttt{Astromer-1}} & \textbf{\texttt{Astromer-2}} & \textbf{\texttt{Moirai}} & \textbf{\texttt{\newblue{Moirai-2}}} & \textbf{\texttt{Chronos}} & \textbf{\texttt{Chronos-Bolt}} & 
    \textbf{\texttt{\newblue{Time-MoE}}} & \textbf{Random} & \textbf{HC features} \\
    \midrule
    \multirow{4}{*}{$k$-NN}
       & Accuracy  & 0.644 & 0.823 & 0.809 & 0.815 & \underline{0.857} & 0.807 & 0.825 & 0.648 & \textbf{0.881} \\
       & Precision & 0.130 & 0.660 & 0.662 & 0.641 & \underline{0.799} & 0.647 & 0.724 & 0.120 & \textbf{0.818} \\
       & Recall    & 0.141 & 0.489 & 0.509 & 0.546 & \underline{0.623} & 0.542 & 0.560 & 0.140 & \textbf{0.661} \\
       & F1        & 0.122 & 0.537 & 0.554 & 0.561 & \underline{0.672} & 0.570 & 0.595 & 0.120 & \textbf{0.712} \\
    \midrule
    \multirow{4}{*}{logistic}
       & Accuracy  & 0.073 & 0.648 & 0.705 & 0.709 & \underline{0.750} & 0.709 & 0.740 & 0.094 & \textbf{0.838} \\
       & Precision & 0.147 & 0.486 & 0.544 & 0.538 & \underline{0.575} & 0.549 & 0.559 & 0.144 & \textbf{0.663} \\
       & Recall    & 0.165 & 0.668 & 0.680 & 0.678 & \underline{0.730} & 0.676 & 0.694 & 0.128 & \textbf{0.854} \\
       & F1        & 0.072 & 0.521 & 0.579 & 0.577 & \underline{0.617} & 0.580  & 0.597 & 0.076 & \textbf{0.714} \\
    \midrule
    \multirow{4}{*}{RF}
       & Accuracy  & 0.676 (0.000) & 0.846 (0.000) & 0.823 (0.001) & 0.820 (0.001) & \underline{0.862 (0.000)} & 0.826 (0.001) & 0.839 (0.001) & 0.676 (0.000) & \textbf{0.920 (0.001)} \\
       & Precision & 0.111 (0.043) & \underline{0.799 (0.006)} & 0.716 (0.007) & 0.697 (0.010) & 0.750 (0.056) & 0.707 (0.002) & 0.693 (0.004) & 0.097 (0.000) & \textbf{0.866 (0.003)} \\
       & Recall    & 0.143 (0.000) & 0.526 (0.002) & 0.514 (0.002) & 0.520 (0.004) & \underline{0.597 (0.002)} & 0.548 (0.001) & 0.561 (0.003) & 0.143 (0.000) & \textbf{0.773 (0.004)} \\
       & F1        & 0.115 (0.000) & 0.580 (0.002) & 0.557 (0.002) & 0.556 (0.003) & \underline{0.638 (0.002)} & 0.582 (0.001) & 0.598 (0.003) & 0.115 (0.000) & \textbf{0.804 (0.003)} \\
    \midrule
    \multirow{4}{*}{MLP}
       & Accuracy  & 0.446 (0.147) & 0.627 (0.037) & 0.717 (0.031) & 0.735 (0.019) & \underline{0.783 (0.022)} & 0.721 (0.022) & 0.758 (0.025) & 0.308 (0.203) & \textbf{0.833 (0.022)} \\
       & Precision & 0.154 (0.006) & 0.453 (0.019) & 0.546 (0.022) & 0.560 (0.021) & \underline{0.589 (0.025)} & 0.553 (0.026) & 0.579 (0.024) & 0.137 (0.006) & \textbf{0.672 (0.025)} \\
       & Recall    & 0.165 (0.003) & 0.627 (0.020) & 0.722 (0.006) & 0.717 (0.006) & \underline{0.758 (0.006)} & 0.696 (0.013) &  0.741 (0.010) & 0.145 (0.002) & \textbf{0.851 (0.009)} \\
       & F1        & 0.138 (0.020) & 0.470 (0.023) & 0.594 (0.019) & 0.603 (0.015) & \underline{0.643 (0.023)} & 0.589 (0.015) & 0.627 (0.019) & 0.094 (0.044) & \textbf{0.723 (0.027)} \\
    \bottomrule
  \end{tabular}}
  \vspace{-1em}
  \label{tab:classification_results_all}
\end{table*}

\blue{In Table~\ref{table:clustering-results-combined} we show that TSFMs generally perform well: (i) the first or second best ranked model in each metric comes from the \texttt{Chronos} models; (ii) both \texttt{Chronos} models always outperform both \texttt{Moirai} models; 
and (iii) The \texttt{Chronos} models and hand-crafted features are highly competitive, frequently alternating between the best and second-best performance. We find the domain-specific \texttt{Astromer} models generally yield poor performance, notably always worse than the TSFMs which are not trained on light curves. We further analyze the poor performance of \texttt{Astromer-1} in Appendix~\ref{app:astromer_1} and find that its embeddings collapse to similar directions. Appendix~\ref{app:astromer_1} further discusses how this is expected for both \texttt{Astromer} models based on results from previous studies.}
\blue{Finally,  hand-crafted features achieve the highest global separability (top NMI under both K-means and Ward), reflecting coarse class alignment. In contrast, \texttt{Chronos} leads on pairwise consistency (best ARI for K-means, and ARI is more sensitive to pairwise correctness), suggesting that its embeddings form small, pure neighborhoods rather than single, class-wide clusters.}

\subsection{Supervised Classification}
\label{sec:results_supervised_classification}

\begin{figure*}
  \centering
  \includegraphics[width=0.90\linewidth]{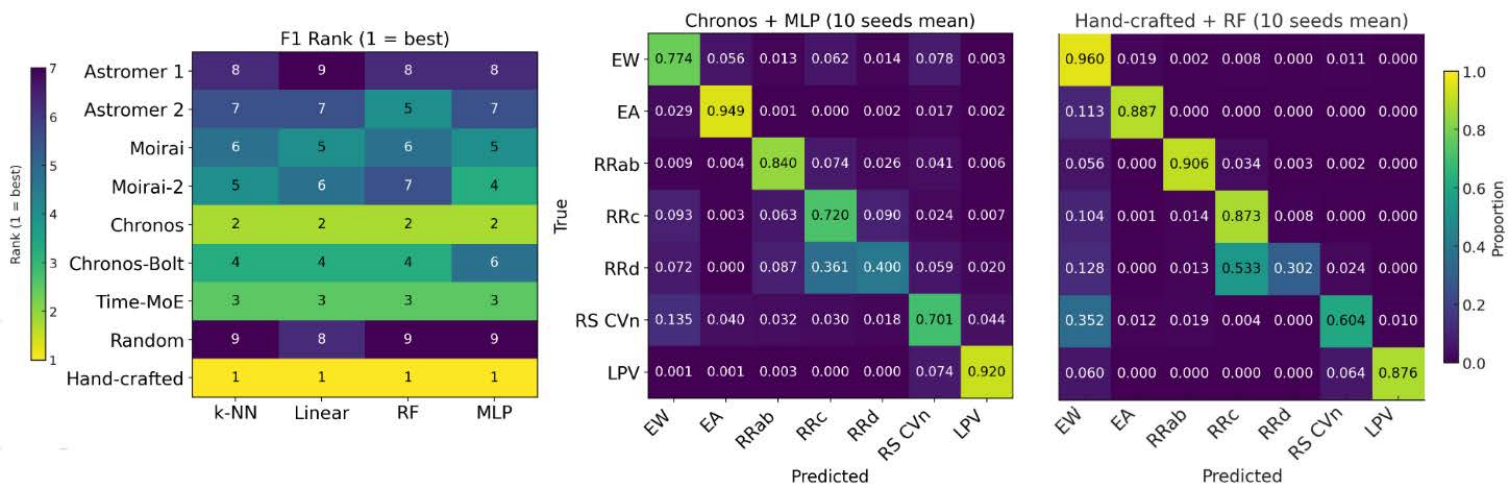}
  \vspace{-0.7em}
  \caption{
  \blue{\textit{Left:} 
  F1 Ranking of baselines across classifier heads. \texttt{Chronos} consistently outperforms other TSFMs and the domain-specific \texttt{Astromer} models, the hand-crafted (HC) features perform the best overall.
  \textit{Right:}
  Confusion matrix of \texttt{Chronos} + MLP, the best performing TSFM-classifier combination, and of HC features + RF classification, the SOTA baseline in astrophysics.
  \texttt{Chronos} yields better performance on most classes (EA, RRd, RS~CVn, and LPV), indicating TSFM's efficacy in extracting information for classification.
  }}
  \label{fig:summary}
  \vspace{-0.8em}
\end{figure*}

Table~\ref{tab:classification_results_all} and the left panel of Figure~\ref{fig:summary} show that (i) the \texttt{Chronos} models perform very well compared to the \texttt{Moirai} and \texttt{Astromer} models; (ii) \texttt{Astromer-2} performs better than the \texttt{Moirai} models in some metrics; and (iii) unlike in the clustering results, \texttt{Chronos} outperforms \texttt{Chronos-Bolt} and achieves second-best performance in nearly all metrics. The hand-crafted features remain superior, yielding an F1 score of \newblue{$0.804\pm0.003$} with the RF classifier. We also find that the RF classifier generally performs better than others, although this is somewhat model-dependent. \newblue{We conduct a feature importance analysis on the hand-crafted features in Appendix~\ref{app:HCF_feat_importance} and find that the periods of the variability in either passband are the most important features.} 

The two confusion matrices in \newblue{Figure~\ref{fig:summary}} show that (i) both \texttt{Chronos} and the hand-crafted features often confuse RRd sources as RRc; (ii) \texttt{Chronos} yields better performance on most classes (EA, RRd, RS~CVn, and LPV) although loses overall due to the larger margins in the classes where the hand-crafted features perform better (EW, RRab, RRc). These results show that, despite having never seen astronomical time series, \texttt{Chronos} clearly extracts useful information from the data for supervised classification. The complete set of confusion matrices for all embedding–classifier combinations can be found in Appendix~\ref{app:confusion}.
We further include recent TSFMs family designing for classification results in \cref{app:tsfm_classification}.

\vspace{-0.2em}
\subsection{\blue{Out-of-Distribution Source Detection}}
\label{sec:results_ood}

\begin{table}
\captionsetup{skip=1.5pt}
\centering
\setlength{\tabcolsep}{2pt}   %
\renewcommand{\arraystretch}{0.92}
\footnotesize
\caption{Purity of top N-percentile of sources selected by out-of-distribution source detection. Best results are highlighted in bold, and the second-best are underlined. Chronos-Bolt ranks first across all metrics with hand-crafted (HC) features being a distant second.}
\label{table:OOD}

\begin{tabular}{
    >{\raggedright\arraybackslash}p{0.24\columnwidth}   %
    >{\centering\arraybackslash}p{0.24\columnwidth}
    >{\centering\arraybackslash}p{0.24\columnwidth}
    >{\centering\arraybackslash}p{0.24\columnwidth}
}
\toprule
 & \textbf{Top 1\%} & \textbf{Top 5\%} & \textbf{Top 10\%} \\
\midrule
\texttt{Astromer-1}   & \newblue{0.017(0.016)} & \newblue{0.091(0.019)} & \newblue{0.120(0.001)} \\
\texttt{Astromer-2}   & \newblue{0.135(0.027)} & \newblue{0.126(0.006)} & \newblue{0.120(0.002)} \\
\texttt{Moirai}       & \newblue{0.169(0.024)} & \newblue{0.143(0.007)} & \newblue{0.150(0.004)} \\
\texttt{Moirai-2}     & \newblue{0.143(0.024)} & \newblue{0.204(0.007)} & \newblue{0.198(0.004)} \\
\texttt{Chronos}      & \newblue{0.139(0.054)} & \newblue{0.116(0.028)} & \newblue{0.149(0.021)} \\
\texttt{Chronos-Bolt} & \textbf{\newblue{0.569(0.060)}} & \textbf{\newblue{0.532(0.055)}} & \textbf{\newblue{0.519(0.038)}} \\
\texttt{Time-MoE}     & \newblue{0.137(0.013)} & \newblue{0.175(0.006)} & \newblue{0.176(0.004)} \\
Random & 0.116 & 0.116 & 0.116 \\
HC features  & \newblue{\underline{0.213(0.013)}} & \newblue{\underline{0.271(0.014)}} & \newblue{\underline{0.259(0.003)}} \\
\bottomrule
\end{tabular}
\end{table}

Table \ref{table:OOD} shows: (i) \texttt{Chronos-Bolt} is exceptional at isolating OOD sources from the inliers; (ii) the second-ranked hand-crafted features deliver considerably worse OOD purity; and (iii) every other model provides only marginal gain over random selection. Of all the stars evaluated,  $\sim11\%$ are OOD samples. This demonstrates that by applying the MCIF approach to the \texttt{Chronos-Bolt} embeddings, we are able to recover nearly half of the OOD sources by evaluating just $10\%$ of the data, a $\sim5\times$ improvement in search efficiency over the random embeddings. As these OOD events often correspond to astrophysically rare or anomalous sources, they are of great interest for the astrophysics community. We posit that \texttt{Chronos-Bolt}'s patch-based, multi-step generation is less sensitive to step-level variation than \texttt{Chronos}’s autoregressive next-token training, hence encouraging a tighter inlier manifold and larger off-manifold distances for rare morphologies. This yields weaker clustering of inliers but stronger OOD isolation.

\subsection{Attention Score Analysis}
\label{sec: attention_score_analysis}

\begin{figure}[t]
  \centering
  \includegraphics[width=\columnwidth]{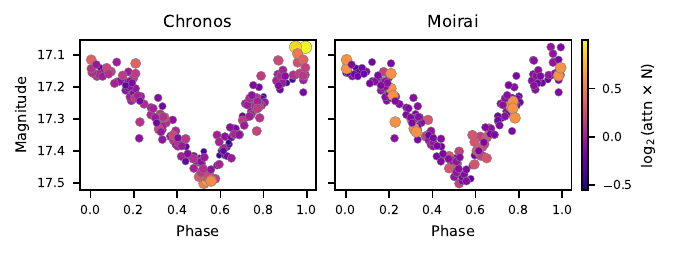}
  \vspace{-2em}
  \caption{\small Attention score for an RRc light curve, phase-folded with the catalog period. 
  Chronos concentrates attention on the brightness peak, while Moirai's attention is diffuse across the cycle. See detailed setup and more visualizations in \cref{app:attn_score}.}
  \label{fig:rrc_attention}
\end{figure}

We further investigate different TSFMs architectures' efficacy in capturing periodic information from the irregular time series. We hypothesize the key to be the tokenization method. \texttt{Chronos} uses point-wise tokenization, whereas \texttt{Moirai} uses patch-wise tokenization. Point-wise tokenization is less destructive when the raw sequence contains irregular gaps and non-uniform local periodic structure, because patching can conflate consecutive observations separated by very different time intervals.

Figure~\ref{fig:rrc_attention} visualizes the attention scores of \texttt{Chronos} and \texttt{Moirai} on one example light curve. \texttt{Chronos} concentrates attention on brightness extrema, thereby capturing the full range of variability in the light curve, including both rapid and slow changes in brightness. In contrast, \texttt{Moirai} groups neighboring measurements into patches. For irregularly sampled light curves, each patch can correspond to dramatically different time baselines, exacerbating the challenge of irregular sampling and making it difficult to learn the underlying variability.

\vspace{-0.5em}
\subsection{TSFM Failure Analysis}
\vspace{-0.5em}

\begin{figure}[t]
  \centering
  \includegraphics[width=0.95\columnwidth]{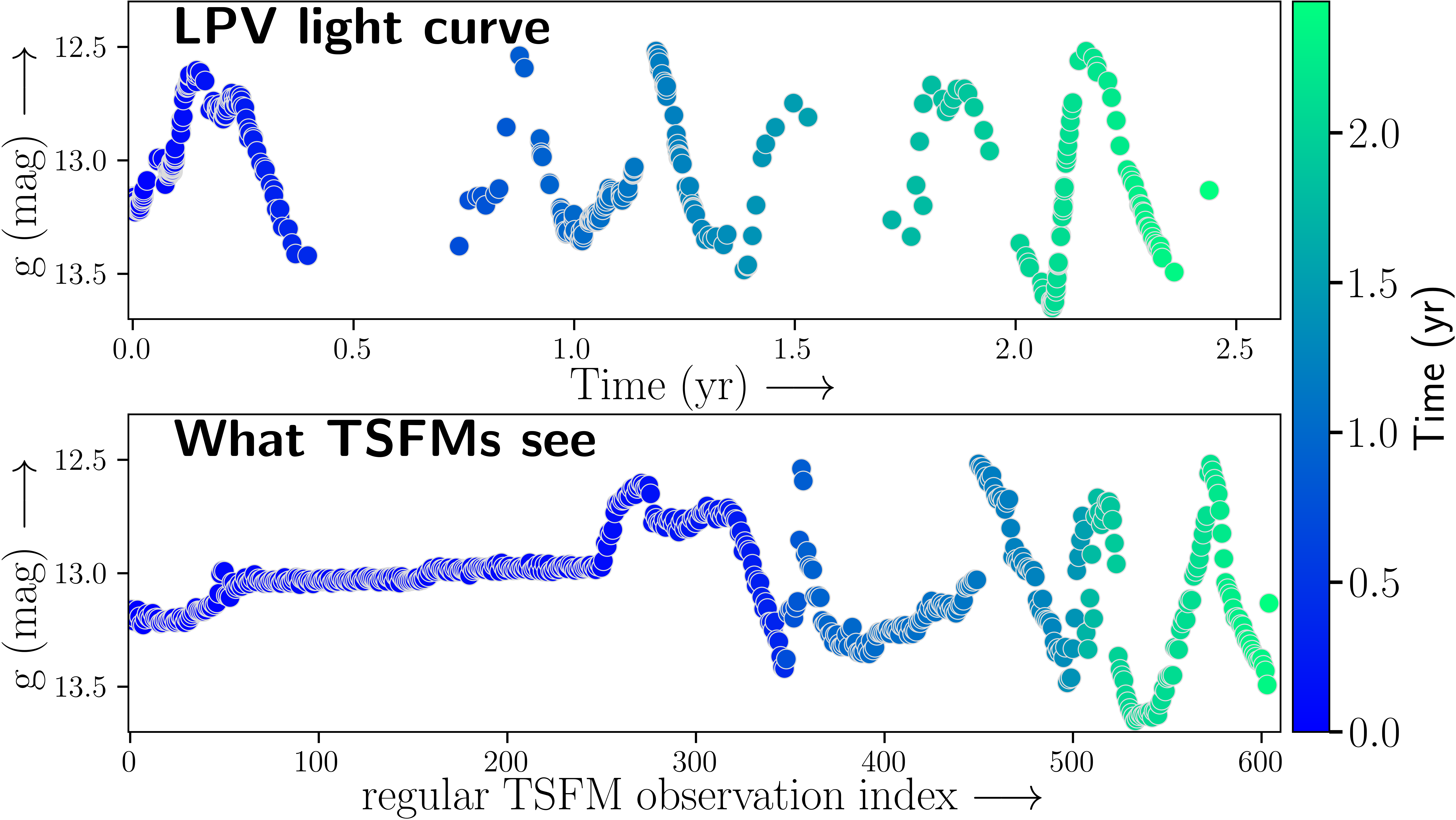}
  \caption{\newblue{A long period variable (LPV) star's light curve (\textit{top}: the real, physical index) and indexed by the order of the measurements (\textit{bottom}: a non-physical index). The coloring corresponds to the physical time index, showing that discarding relative timestamp information causes the TSFMs to receive a warped view of the light curve, limiting their interpretation of the data.}}
  \label{fig:tsfm_view}
\end{figure}

\cref{fig:tsfm_view} illustrates that the TSFMs evaluated here treat irregularly sampled data as regular, resulting in a warped view of the time series that obscures key physical structures like periodicity (\cref{app:HCF_feat_importance}). That TSFMs perform competitively despite this informational loss suggests that TSFMs with mechanisms specifically architected for irregular sampling and irregular pre-training data will likely unlock significant performance gains. Such a model could drive a paradigm shift from bespoke, fully-supervised pipelines toward off-the-shelf foundational representations for variable star analysis. Pushing analysis frontiers in this way is critical for the era of petabyte-scale light curve data sets. Complimentarily, astronomy has challenging, information-rich time series in game-changing quantities to offer the AI-for-time-series field, and making use of these time series would also help open new applications of TSFMs to other scientific domains with similarly challenging time series.

\vspace{-0.5em}
\subsection{TSFM Fine-tuning}
\vspace{-0.5em}
\label{sec:finetuning}

We also study TSFMs fine-tuning on astrophysical data. Since \texttt{Chronos} is the best-performing TSFM, we fine-tune it on StarEmbed and evaluate it on supervised classification.

\begin{table}[t]
\centering
\caption{Classification fine-tuning results for Chronos-tiny on ZTF light curves with an \textbf{MLP head}. Same HPO grid and reporting protocol as Table~\ref{tab:appx-finetune-linear}. Bold indicates the best in each row.}
\label{tab:finetune-mlp}
\begin{tabular}{@{}l@{\hskip 6pt}c@{\hskip 6pt}c@{\hskip 6pt}c@{}}
\toprule
Metric & Full FT & LoRA & LayerNorm \\
\midrule
\shortstack{Trainable\\Params} & 9.6M (100\%) & 1.3M (13.2\%) & 1.2M (12.4\%) \\
Macro F1 & 0.659 (0.026) & \textbf{0.661 (0.015)} & 0.648 (0.016) \\
Accuracy & \textbf{0.778 (0.028)} & 0.772 (0.020) & 0.774 (0.023) \\
Macro Prec & \textbf{0.613 (0.034)} & \textbf{0.613 (0.018)} & 0.605 (0.023) \\
Macro Rec & 0.766 (0.005) & \textbf{0.767 (0.006)} & 0.766 (0.006) \\
\bottomrule
\end{tabular}
\end{table}

\textbf{Classification Fine-tuning versus Forecasting Fine-tuning.} 
We compare two fine-tuning objectives. The first setting, denoted \texttt{Chronos-FTfcst}, uses the original \texttt{Chronos} forecasting objective. The second setting, denoted \texttt{Chronos-FTcls}, performs end-to-end classification fine-tuning with an MLP classification head.

\Cref{tab:appx-chronos-ft-wide} shows that \texttt{Chronos-FTfcst} does not improve downstream classification and is slightly worse than zero-shot \texttt{Chronos}. In contrast, \texttt{Chronos-FTcls} improves three of the four MLP-based classification metrics over zero-shot \texttt{Chronos}, suggesting that fine-tuning is helpful for \texttt{Chronos} when performed directly on the downstream classification task. The end-to-end classification fine-tuned \texttt{Chronos} still remains below hand-crafted features, however, which indicates that current TSFMs are not yet sufficient to replace even the conceptually simple domain-specific baseline on our challenging scientific dataset.

\textbf{Full Fine-tuning versus Parameter-Efficient \& Partial Fine-tuning.} Regarding fine-tuning efficiency, we also compare full fine-tuning, LoRA fine-tuning, and partial fine-tuning in which only the LayerNorm parameters are updated. We run each of these experiments with both a linear head and an MLP head.

\begin{table}[ht]
\centering
\caption{Classification fine-tuning results for \texttt{Chronos} on ZTF light curves with a \textbf{linear head}. Bold indicates the best in each row.}
\label{tab:appx-finetune-linear}
\begin{tabular}{@{}l@{\hskip 6pt}c@{\hskip 6pt}c@{\hskip 6pt}c@{}}
\toprule
Metric & Full FT & LoRA & LayerNorm \\
\midrule
\shortstack{Trainable\\Params} & 8.4M (100\%) & 102K (1.2\%) & 9.2K (0.1\%) \\
Macro F1   & \textbf{0.661 (0.010)} & 0.622 (0.022) & 0.622 (0.018) \\
Accuracy   & \textbf{0.777 (0.010)} & 0.746 (0.023) & 0.750 (0.029) \\
Macro Prec & \textbf{0.615 (0.012)} & 0.572 (0.026) & 0.571 (0.022) \\
Macro Rec  & \textbf{0.766 (0.005)} & 0.760 (0.006) & 0.758 (0.004) \\
\bottomrule
\end{tabular}
\end{table}

\Cref{tab:appx-finetune-linear} shows that with a linear head, full fine-tuning yields the best performance, indicating that when the classification head is linear, updating only a small number of parameters is not sufficient to match full fine-tuning. With an MLP head, full fine-tuning and LoRA yield comparable top performance, despite LoRA tuning fewer than one seventh as many parameters. LayerNorm-only tuning also improves under the MLP head, but remains below LoRA. These results suggest that when only a small number of parameters are updated during fine-tuning, using a more expressive classification head becomes important. Under a linear head, both LoRA and LayerNorm-only tuning appear to be similarly constrained by the limited expressiveness of the head, resulting in lower F1 scores. Under an MLP head, this bottleneck is alleviated by the nonlinear classifier. LoRA also shows a clearer advantage over LayerNorm-only tuning, possibly because LoRA adapts the attention projections and can therefore better focus on discriminative patterns in the time series, whereas LayerNorm-only tuning is limited to scale-and-shift updates.
See \cref{app:finetune} for full results.

\vspace{-0.5em}
\subsection{TSFM Architecture Improvement}
\label{sec:rope_irregular}
\vspace{-0.5em}

\begin{table}[t]
\centering
\caption{Pre-training TSFMs with a modified RoPE incorporating irregular time gaps and test on sinusoidal synthetic data. We evaluate: (1) RoPE + True timestamp, replacing sequential position indices with actual observation timestamps, (2) vanilla baseline that is the original \texttt{Moirai}. Each sample is generated following $x(t) = A\sin(2\pi f t + \varphi) + \varepsilon$.
The results indicate the importance of modeling irregular time gaps in the positional encoding. See prediction examples in \cref{fig:sin_prediction_1,fig:appx-sin-pred-rope-only}.}
\label{tab:sinusoidal_high_irreg}
\begin{tabular}{lcc}
\toprule
Metric & RoPE + True timestamp & Vanilla \\
\midrule
$R^2$ $\uparrow$ 
& $\mathbf{0.188 \pm 0.102}$ 
& $0.016 \pm 0.039$ \\

Pearson $\uparrow$ 
& $\mathbf{0.351 \pm 0.114}$ 
& $0.202 \pm 0.098$ \\

MSE $\downarrow$ 
& $\mathbf{0.041 \pm 0.006}$ 
& $0.050 \pm 0.002$ \\

MAE $\downarrow$ 
& $\mathbf{0.159 \pm 0.017}$ 
& $0.183 \pm 0.007$ \\
\bottomrule
\end{tabular}
\end{table}

\begin{figure}[t]
  \centering
  \includegraphics[width=\columnwidth]{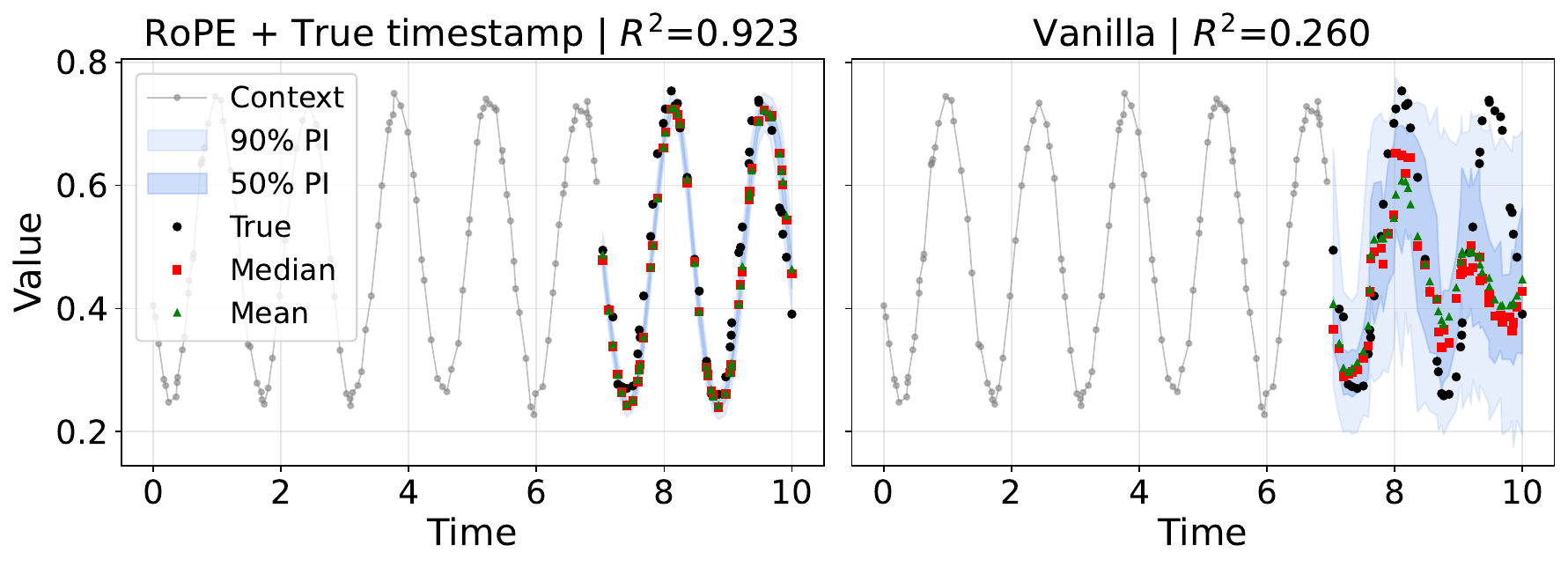}
  \vspace{-1.5em}
  \caption{\small Forecasting visualization of \cref{tab:sinusoidal_high_irreg}.  Encoding the time information correctly into \texttt{Moirai} through RoME yields perfect predictions on the test data, whereas vanilla \texttt{Moirai}'s predictions have the wrong frequency. }
  \label{fig:sin_prediction_1}
\end{figure}

We also study architectural improvements of TSFMs to identify directions for developing more capable TSFMs natively aligned with irregularly sampled time series. 
A central limitation observed throughout our experiments is that current TSFMs do not properly treat irregular sampling (see \cref{fig:tsfm_view}). Existing positional encodings are typically applied to regular token or patch indices, but in astronomical light curves, adjacent observations in the input sequence correspond to non-uniform time intervals.

We modify Rotary Positional Encoding (RoPE) in \texttt{Moirai} by incorporating the true observation timestamps instead of using the patch index. \Cref{tab:sinusoidal_high_irreg} and Figure~\ref{fig:sin_prediction_1} show that our modification outperforms the vanilla version which does not properly handle the irregular sampling. These observations isolate an important design option for future development and pre-training of TSFMs capable of properly treating irregular temporal sampling. See more experimental details and forecasting examples in \cref{app:rope_irregular}.  Closely related work by \citet{zivanovic2026rotary} also demonstrates the effectiveness of continuous-time RoPE within a masked-autoencoder framework for irregular time series.

\vspace{-0.5em}
\section{Discussion and Conclusions}
\vspace{-0.5em}
\label{sec:discussion_conclusions}

We introduced \texttt{StarEmbed}, a public benchmark of TSFMs on real multi-band stellar light curves, whose irregular sampling and heteroskedasticity differ substantially from typical TSFM pre-training corpora. By harmonizing expert-vetted labels with $\sim$40,000 ZTF light curves, we provide a rigorously curated seven-class dataset and conduct comprehensive benchmarking experiments targeting the following categories: (i) TSFM zero-shot performance on three key downstream astrophysics tasks (unsupervised clustering, supervised classification, and OOD source detection); (ii) interpretibility analysis (TSFM embedding clusterings and quality analysis, TSFM attention score analysis, failure analysis); (iii) study of TSFM training for irregular time series: TSFM fine-tuning, TSFM architecture design for better pre-training, and TSFM scaling on astrophysics data. The conclusions are as follows.

\textbf{TSFM Performance Benchmark (\cref{sec:results_unsupervised_clustering,sec:results_supervised_classification,sec:results_ood}). } We compare domain-specific embeddings, SOTA general-purpose TSFMs, and the long-standing and top-performing method in astrophysics, hand-crafted features. For clustering, \texttt{Chronos} matches hand-crafted features. For supervised classification, hand-crafted features perform best, with \texttt{Chronos} consistently second with a small margin. For OOD detection, \texttt{Chronos-Bolt} significantly outperforms hand-crafted features. Across all tasks, TSFMs generally outperform the domain-specific transformer, \texttt{Astromer}. 

\textbf{Interpretibility Analysis.} Both TSFMs' embedding clusterings and quality analysis show the efficacy of zero-shot TSFMs' embeddings in capturing the differences between different classes of astrophysics data. The failure analysis shows that all TSFMs fall short in capturing periodic information from the irregular time series compared to hand-crafted features. The attention score analysis further unveils that \texttt{Chronos}'s model architecture could retain much better periodic information compared to \texttt{Moirai}'s.

\textbf{Study of TSFM Development on Irregular Time Series.} Beyond benchmarking zero-shot TSFMs, we also study TSFM training, especially on our irregular time series data. Our fine-tuning experiment shows that fine-tuning with a classification head yield a better results on the downstream classification task compared to fine-tuning with a forecasting objective as in pre-training. In terms of pre-training TSFMs from scratch, we demonstrate that a modified RoPE encoding that incorporates the irregular time gaps significantly improves TSFMs ability in capturing periodic information in irregular time series. Last but not least, we also investigate the scaling dynamics of TSFMs, which shows that model performance in general scales with model size.

Taken together, we demonstrate both the capabilities and the boundaries of TSFM zero-shot generalization on astrophysical time series with a comprehensive benchmark and analysis. Furthermore, we also provide additional meaningful insights beyond zero-shot in training better TSFMs on astrophysics data and irregular time series. By releasing all data, code, and model wrappers, \texttt{StarEmbed} serves to forge the bridge between astronomy and AI for time series fields and to provide the community with a benchmark for foundation model advances on challenging time-series data.

\section*{Acknowledgments}

We gratefully acknowledge the support of the NSF-Simons AI-Institute for the Sky (SkAI) via grants NSF AST-2421845 and Simons Foundation MPS-AI-00010513.

Dennis Wu is supported by the Cognitive Science Advanced Research Fellowship. 

Zwicky Transient Facility access for all authors was supported by Northwestern University and the Center for Interdisciplinary Exploration and Research in Astrophysics (CIERA). A.A.M.~is supported by DoE award \#DE-SC0025599. A.A.M. is also supported by Cottrell Scholar Award \#CS-CSA-2025-059 from Research Corporation for Science Advancement. N.R. is supported by NSF award \#2421845 and a Northwestern University Presidential Fellowship award.

This research has made use of NASA's Astrophysics Data System.

Based on observations obtained with the Samuel Oschin Telescope 48-inch and the 60-inch Telescope at the Palomar Observatory as part of the Zwicky Transient Facility project. ZTF is supported by the National Science Foundation under Grants No. AST-1440341, AST-2034437, and currently Award \#2407588. ZTF receives additional funding from the ZTF partnership. Current members include Caltech, USA; Caltech/IPAC, USA; University of Maryland, USA; University of California, Berkeley, USA; University of Wisconsin at Milwaukee, USA; Cornell University, USA; Drexel University, USA; University of North Carolina at Chapel Hill, USA; Institute of Science and Technology, Austria; National Central University, Taiwan, and OKC, University of Stockholm, Sweden. Operations are conducted by Caltech's Optical Observatory (COO), Caltech/IPAC, and the University of Washington at Seattle, USA.

The CSS survey is funded by the National Aeronautics and Space
Administration under Grant No. NNG05GF22G issued through the Science Mission Directorate Near-Earth Objects Observations Program.  The CRTS survey is supported by the U.S.~National Science Foundation under grants AST-0909182 and AST-1313422.

This research was supported in part through the computational resources and staff contributions provided for the Quest high performance computing facility at Northwestern University which is jointly supported by the Office of the Provost, the Office for Research, and Northwestern University Information Technology.

This research used both the DeltaAI advanced computing and data resource, which is supported by the National Science Foundation (award OAC 2320345) and the State of Illinois, and the Delta advanced computing and data resource which is supported by the National Science Foundation (award OAC 2005572) and the State of Illinois.. Delta and DeltaAI are joint efforts of the University of Illinois Urbana-Champaign and its National Center for Supercomputing Applications.

This work was completed in part at the NCSA Open Hackathon, part of the Open Hackathons program. The authors would like to acknowledge OpenACC-Standard.org for their support.

\section*{Impact Statement}
This paper introduces \texttt{StarEmbed}, a benchmark for evaluating time series foundation models on astronomical observations of variable stars. Our goal is to advance scientific understanding of how general time series foundation models perform on complex real-world time series data. While improved machine learning methods have broad societal implications, we do not identify any specific ethical concerns or negative impacts arising directly from this benchmarking study beyond those commonly associated with machine learning research.

\bibliography{refs}

@article{debosscher07,
	adsurl = {http://adsabs.harvard.edu/abs/2007A%26A...475.1159D},
	archiveprefix = {arXiv},
	author = {{Debosscher}, J. and {Sarro}, L.~M. and {Aerts}, C. and {Cuypers}, J. and {Vandenbussche}, B. and {Garrido}, R. and {Solano}, E.},
	doi = {10.1051/0004-6361:20077638},
	eprint = {0711.0703},
	journal = {\aap},
	month = dec,
	pages = {1159-1183},
	title = {{Automated supervised classification of variable stars. I. Methodology}},
	volume = 475,
	year = 2007}

@article{pimentel2022deep,
  title={Deep attention-based supernovae classification of multiband light curves},
  author={Pimentel, {\'O}scar and Est{\'e}vez, Pablo A and F{\"o}rster, Francisco},
  journal={The Astronomical Journal},
  volume={165},
  number={1},
  pages={18},
  year={2022},
  publisher={IOP Publishing}
}

@article{naul2018recurrent,
  title={A recurrent neural network for classification of unevenly sampled variable stars},
  author={Naul, Brett and Bloom, Joshua S and P{\'e}rez, Fernando and Van Der Walt, St{\'e}fan},
  journal={Nature Astronomy},
  volume={2},
  number={2},
  pages={151--155},
  year={2018},
  publisher={Nature Publishing Group UK London}
}

@ARTICLE{Bellm+2019a,
       author = {{Bellm}, Eric C. and {Kulkarni}, Shrinivas R. and {Graham}, Matthew J. and {Dekany}, Richard and {Smith}, Roger M. and {Riddle}, Reed and {Masci}, Frank J. and {Helou}, George and {Prince}, Thomas A. and {Adams}, Scott M. and {Barbarino}, C. and {Barlow}, Tom and {Bauer}, James and {Beck}, Ron and {Belicki}, Justin and {Biswas}, Rahul and {Blagorodnova}, Nadejda and {Bodewits}, Dennis and {Bolin}, Bryce and {Brinnel}, Valery and {Brooke}, Tim and {Bue}, Brian and {Bulla}, Mattia and {Burruss}, Rick and {Cenko}, S. Bradley and {Chang}, Chan-Kao and {Connolly}, Andrew and {Coughlin}, Michael and {Cromer}, John and {Cunningham}, Virginia and {De}, Kishalay and {Delacroix}, Alex and {Desai}, Vandana and {Duev}, Dmitry A. and {Eadie}, Gwendolyn and {Farnham}, Tony L. and {Feeney}, Michael and {Feindt}, Ulrich and {Flynn}, David and {Franckowiak}, Anna and {Frederick}, S. and {Fremling}, C. and {Gal-Yam}, Avishay and {Gezari}, Suvi and {Giomi}, Matteo and {Goldstein}, Daniel A. and {Golkhou}, V. Zach and {Goobar}, Ariel and {Groom}, Steven and {Hacopians}, Eugean and {Hale}, David and {Henning}, John and {Ho}, Anna Y.~Q. and {Hover}, David and {Howell}, Justin and {Hung}, Tiara and {Huppenkothen}, Daniela and {Imel}, David and {Ip}, Wing-Huen and {Ivezi{\'c}}, {\v{Z}}eljko and {Jackson}, Edward and {Jones}, Lynne and {Juric}, Mario and {Kasliwal}, Mansi M. and {Kaspi}, S. and {Kaye}, Stephen and {Kelley}, Michael S.~P. and {Kowalski}, Marek and {Kramer}, Emily and {Kupfer}, Thomas and {Landry}, Walter and {Laher}, Russ R. and {Lee}, Chien-De and {Lin}, Hsing Wen and {Lin}, Zhong-Yi and {Lunnan}, Ragnhild and {Giomi}, Matteo and {Mahabal}, Ashish and {Mao}, Peter and {Miller}, Adam A. and {Monkewitz}, Serge and {Murphy}, Patrick and {Ngeow}, Chow-Choong and {Nordin}, Jakob and {Nugent}, Peter and {Ofek}, Eran and {Patterson}, Maria T. and {Penprase}, Bryan and {Porter}, Michael and {Rauch}, Ludwig and {Rebbapragada}, Umaa and {Reiley}, Dan and {Rigault}, Mickael and {Rodriguez}, Hector and {van Roestel}, Jan and {Rusholme}, Ben and {van Santen}, Jakob and {Schulze}, S. and {Shupe}, David L. and {Singer}, Leo P. and {Soumagnac}, Maayane T. and {Stein}, Robert and {Surace}, Jason and {Sollerman}, Jesper and {Szkody}, Paula and {Taddia}, F. and {Terek}, Scott and {Van Sistine}, Angela and {van Velzen}, Sjoert and {Vestrand}, W. Thomas and {Walters}, Richard and {Ward}, Charlotte and {Ye}, Quan-Zhi and {Yu}, Po-Chieh and {Yan}, Lin and {Zolkower}, Jeffry},
        title = "{The Zwicky Transient Facility: System Overview, Performance, and First Results}",
      journal = {pasp},
         year = 2019,
        month = jan,
       volume = {131},
       number = {995},
        pages = {018002},
          doi = {10.1088/1538-3873/aaecbe},
archivePrefix = {arXiv},
       eprint = {1902.01932},
 primaryClass = {astro-ph.IM},
       adsurl = {https://ui.adsabs.harvard.edu/abs/2019PASP..131a8002B}
}

@article{Ivezic+2019,
	adsurl = {https://ui.adsabs.harvard.edu/abs/2019ApJ...873..111I},
	archiveprefix = {arXiv},
	author = {{Ivezi{\'c}}, {\v{Z}}eljko and {Kahn}, Steven M. and {Tyson}, J. Anthony and {Abel}, Bob and {Acosta}, Emily and {Allsman}, Robyn and {Alonso}, David and {AlSayyad}, Yusra and {Anderson}, Scott F. and {Andrew}, John and {Angel}, James Roger P. and {Angeli}, George Z. and {Ansari}, Reza and {Antilogus}, Pierre and {Araujo}, Constanza and {Armstrong}, Robert and {Arndt}, Kirk T. and {Astier}, Pierre and {Aubourg}, {\'E}ric and {Auza}, Nicole and {Axelrod}, Tim S. and {Bard}, Deborah J. and {Barr}, Jeff D. and {Barrau}, Aurelian and {Bartlett}, James G. and {Bauer}, Amanda E. and {Bauman}, Brian J. and {Baumont}, Sylvain and {Bechtol}, Ellen and {Bechtol}, Keith and {Becker}, Andrew C. and {Becla}, Jacek and {Beldica}, Cristina and {Bellavia}, Steve and {Bianco}, Federica B. and {Biswas}, Rahul and {Blanc}, Guillaume and {Blazek}, Jonathan and {Bland ford}, Roger D. and {Bloom}, Josh S. and {Bogart}, Joanne and {Bond}, Tim W. and {Booth}, Michael T. and {Borgland}, Anders W. and {Borne}, Kirk and {Bosch}, James F. and {Boutigny}, Dominique and {Brackett}, Craig A. and {Bradshaw}, Andrew and {Brand t}, William Nielsen and {Brown}, Michael E. and {Bullock}, James S. and {Burchat}, Patricia and {Burke}, David L. and {Cagnoli}, Gianpietro and {Calabrese}, Daniel and {Callahan}, Shawn and {Callen}, Alice L. and {Carlin}, Jeffrey L. and {Carlson}, Erin L. and {Chand rasekharan}, Srinivasan and {Charles-Emerson}, Glenaver and {Chesley}, Steve and {Cheu}, Elliott C. and {Chiang}, Hsin-Fang and {Chiang}, James and {Chirino}, Carol and {Chow}, Derek and {Ciardi}, David R. and {Claver}, Charles F. and {Cohen-Tanugi}, Johann and {Cockrum}, Joseph J. and {Coles}, Rebecca and {Connolly}, Andrew J. and {Cook}, Kem H. and {Cooray}, Asantha and {Covey}, Kevin R. and {Cribbs}, Chris and {Cui}, Wei and {Cutri}, Roc and {Daly}, Philip N. and {Daniel}, Scott F. and {Daruich}, Felipe and {Daubard}, Guillaume and {Daues}, Greg and {Dawson}, William and {Delgado}, Francisco and {Dellapenna}, Alfred and {de Peyster}, Robert and {de Val-Borro}, Miguel and {Digel}, Seth W. and {Doherty}, Peter and {Dubois}, Richard and {Dubois-Felsmann}, Gregory P. and {Durech}, Josef and {Economou}, Frossie and {Eifler}, Tim and {Eracleous}, Michael and {Emmons}, Benjamin L. and {Fausti Neto}, Angelo and {Ferguson}, Henry and {Figueroa}, Enrique and {Fisher-Levine}, Merlin and {Focke}, Warren and {Foss}, Michael D. and {Frank}, James and {Freemon}, Michael D. and {Gangler}, Emmanuel and {Gawiser}, Eric and {Geary}, John C. and {Gee}, Perry and {Geha}, Marla and {Gessner}, Charles J.~B. and {Gibson}, Robert R. and {Gilmore}, D. Kirk and {Glanzman}, Thomas and {Glick}, William and {Goldina}, Tatiana and {Goldstein}, Daniel A. and {Goodenow}, Iain and {Graham}, Melissa L. and {Gressler}, William J. and {Gris}, Philippe and {Guy}, Leanne P. and {Guyonnet}, Augustin and {Haller}, Gunther and {Harris}, Ron and {Hascall}, Patrick A. and {Haupt}, Justine and {Hernand ez}, Fabio and {Herrmann}, Sven and {Hileman}, Edward and {Hoblitt}, Joshua and {Hodgson}, John A. and {Hogan}, Craig and {Howard}, James D. and {Huang}, Dajun and {Huffer}, Michael E. and {Ingraham}, Patrick and {Innes}, Walter R. and {Jacoby}, Suzanne H. and {Jain}, Bhuvnesh and {Jammes}, Fabrice and {Jee}, M. James and {Jenness}, Tim and {Jernigan}, Garrett and {Jevremovi{\'c}}, Darko and {Johns}, Kenneth and {Johnson}, Anthony S. and {Johnson}, Margaret W.~G. and {Jones}, R. Lynne and {Juramy-Gilles}, Claire and {Juri{\'c}}, Mario and {Kalirai}, Jason S. and {Kallivayalil}, Nitya J. and {Kalmbach}, Bryce and {Kantor}, Jeffrey P. and {Karst}, Pierre and {Kasliwal}, Mansi M. and {Kelly}, Heather and {Kessler}, Richard and {Kinnison}, Veronica and {Kirkby}, David and {Knox}, Lloyd and {Kotov}, Ivan V. and {Krabbendam}, Victor L. and {Krughoff}, K. Simon and {Kub{\'a}nek}, Petr and {Kuczewski}, John and {Kulkarni}, Shri and {Ku}, John and {Kurita}, Nadine R. and {Lage}, Craig S. and {Lambert}, Ron and {Lange}, Travis and {Langton}, J. Brian and {Le Guillou}, Laurent and {Levine}, Deborah and {Liang}, Ming and {Lim}, Kian-Tat and {Lintott}, Chris J. and {Long}, Kevin E. and {Lopez}, Margaux and {Lotz}, Paul J. and {Lupton}, Robert H. and {Lust}, Nate B. and {MacArthur}, Lauren A. and {Mahabal}, Ashish and {Mand elbaum}, Rachel and {Markiewicz}, Thomas W. and {Marsh}, Darren S. and {Marshall}, Philip J. and {Marshall}, Stuart and {May}, Morgan and {McKercher}, Robert and {McQueen}, Michelle and {Meyers}, Joshua and {Migliore}, Myriam and {Miller}, Michelle and {Mills}, David J. and {Miraval}, Connor and {Moeyens}, Joachim and {Moolekamp}, Fred E. and {Monet}, David G. and {Moniez}, Marc and {Monkewitz}, Serge and {Montgomery}, Christopher and {Morrison}, Christopher B. and {Mueller}, Fritz and {Muller}, Gary P. and {Mu{\~n}oz Arancibia}, Freddy and {Neill}, Douglas R. and {Newbry}, Scott P. and {Nief}, Jean-Yves and {Nomerotski}, Andrei and {Nordby}, Martin and {O'Connor}, Paul and {Oliver}, John and {Olivier}, Scot S. and {Olsen}, Knut and {O'Mullane}, William and {Ortiz}, Sandra and {Osier}, Shawn and {Owen}, Russell E. and {Pain}, Reynald and {Palecek}, Paul E. and {Parejko}, John K. and {Parsons}, James B. and {Pease}, Nathan M. and {Peterson}, J. Matt and {Peterson}, John R. and {Petravick}, Donald L. and {Libby Petrick}, M.~E. and {Petry}, Cathy E. and {Pierfederici}, Francesco and {Pietrowicz}, Stephen and {Pike}, Rob and {Pinto}, Philip A. and {Plante}, Raymond and {Plate}, Stephen and {Plutchak}, Joel P. and {Price}, Paul A. and {Prouza}, Michael and {Radeka}, Veljko and {Rajagopal}, Jayadev and {Rasmussen}, Andrew P. and {Regnault}, Nicolas and {Reil}, Kevin A. and {Reiss}, David J. and {Reuter}, Michael A. and {Ridgway}, Stephen T. and {Riot}, Vincent J. and {Ritz}, Steve and {Robinson}, Sean and {Roby}, William and {Roodman}, Aaron and {Rosing}, Wayne and {Roucelle}, Cecille and {Rumore}, Matthew R. and {Russo}, Stefano and {Saha}, Abhijit and {Sassolas}, Benoit and {Schalk}, Terry L. and {Schellart}, Pim and {Schindler}, Rafe H. and {Schmidt}, Samuel and {Schneider}, Donald P. and {Schneider}, Michael D. and {Schoening}, William and {Schumacher}, German and {Schwamb}, Megan E. and {Sebag}, Jacques and {Selvy}, Brian and {Sembroski}, Glenn H. and {Seppala}, Lynn G. and {Serio}, Andrew and {Serrano}, Eduardo and {Shaw}, Richard A. and {Shipsey}, Ian and {Sick}, Jonathan and {Silvestri}, Nicole and {Slater}, Colin T. and {Smith}, J. Allyn and {Smith}, R. Chris and {Sobhani}, Shahram and {Soldahl}, Christine and {Storrie-Lombardi}, Lisa and {Stover}, Edward and {Strauss}, Michael A. and {Street}, Rachel A. and {Stubbs}, Christopher W. and {Sullivan}, Ian S. and {Sweeney}, Donald and {Swinbank}, John D. and {Szalay}, Alexander and {Takacs}, Peter and {Tether}, Stephen A. and {Thaler}, Jon J. and {Thayer}, John Gregg and {Thomas}, Sandrine and {Thornton}, Adam J. and {Thukral}, Vaikunth and {Tice}, Jeffrey and {Trilling}, David E. and {Turri}, Max and {Van Berg}, Richard and {Vanden Berk}, Daniel and {Vetter}, Kurt and {Virieux}, Francoise and {Vucina}, Tomislav and {Wahl}, William and {Walkowicz}, Lucianne and {Walsh}, Brian and {Walter}, Christopher W. and {Wang}, Daniel L. and {Wang}, Shin-Yawn and {Warner}, Michael and {Wiecha}, Oliver and {Willman}, Beth and {Winters}, Scott E. and {Wittman}, David and {Wolff}, Sidney C. and {Wood-Vasey}, W. Michael and {Wu}, Xiuqin and {Xin}, Bo and {Yoachim}, Peter and {Zhan}, Hu},
	doi = {10.3847/1538-4357/ab042c},
	eid = {111},
	eprint = {0805.2366},
	journal = {apj},
	month = {Mar},
	number = {2},
	pages = {111},
	primaryclass = {astro-ph},
	title = {{LSST: From Science Drivers to Reference Design and Anticipated Data Products}},
	volume = {873},
	year = {2019}}

@ARTICLE{Dekany+2020,
       author = {{Dekany}, Richard and {Smith}, Roger M. and {Riddle}, Reed and {Feeney}, Michael and {Porter}, Michael and {Hale}, David and {Zolkower}, Jeffry and {Belicki}, Justin and {Kaye}, Stephen and {Henning}, John and {Walters}, Richard and {Cromer}, John and {Delacroix}, Alex and {Rodriguez}, Hector and {Reiley}, Daniel J. and {Mao}, Peter and {Hover}, David and {Murphy}, Patrick and {Burruss}, Rick and {Baker}, John and {Kowalski}, Marek and {Reif}, Klaus and {Mueller}, Phillip and {Bellm}, Eric and {Graham}, Matthew and {Kulkarni}, Shrinivas R.},
        title = "{The Zwicky Transient Facility: Observing System}",
      journal = {pasp},
         year = 2020,
        month = mar,
       volume = {132},
       number = {1009},
          eid = {038001},
        pages = {038001},
          doi = {10.1088/1538-3873/ab4ca2},
archivePrefix = {arXiv},
       eprint = {2008.04923},
 primaryClass = {astro-ph.IM},
       adsurl = {https://ui.adsabs.harvard.edu/abs/2020PASP..132c8001D}
}

@ARTICLE{Drake+2014,
       author = {{Drake}, A.~J. and {Graham}, M.~J. and {Djorgovski}, S.~G. and {Catelan}, M. and {Mahabal}, A.~A. and {Torrealba}, G. and {Garc{\'\i}a-{\'A}lvarez}, D. and {Donalek}, C. and {Prieto}, J.~L. and {Williams}, R. and {Larson}, S. and {Christen sen}, E. and {Belokurov}, V. and {Koposov}, S.~E. and {Beshore}, E. and {Boattini}, A. and {Gibbs}, A. and {Hill}, R. and {Kowalski}, R. and {Johnson}, J. and {Shelly}, F.},
        title = "{The Catalina Surveys Periodic Variable Star Catalog}",
      journal = {apjs},
         year = 2014,
        month = jul,
       volume = {213},
       number = {1},
          eid = {9},
        pages = {9},
          doi = {10.1088/0067-0049/213/1/9},
archivePrefix = {arXiv},
       eprint = {1405.4290},
 primaryClass = {astro-ph.SR},
       adsurl = {https://ui.adsabs.harvard.edu/abs/2014ApJS..213....9D}
}

@ARTICLE{Gupta+2025_MCIF,
       author = {{Gupta}, Rithwik and {Muthukrishna}, Daniel and {Lochner}, Michelle},
        title = "{A classifier-based approach to multiclass anomaly detection for astronomical transients}",
      journal = {RAS Techniques and Instruments},
         year = 2025,
        month = jan,
       volume = {4},
          eid = {rzae054},
        pages = {rzae054},
          doi = {10.1093/rasti/rzae054},
archivePrefix = {arXiv},
       eprint = {2403.14742},
 primaryClass = {astro-ph.IM},
       adsurl = {https://ui.adsabs.harvard.edu/abs/2025RASTI...4...54G}
}

@ARTICLE{Drake+2009,
       author = {{Drake}, A.~J. and {Djorgovski}, S.~G. and {Mahabal}, A. and {Beshore}, E. and {Larson}, S. and {Graham}, M.~J. and {Williams}, R. and {Christensen}, E. and {Catelan}, M. and {Boattini}, A. and {Gibbs}, A. and {Hill}, R. and {Kowalski}, R.},
        title = "{First Results from the Catalina Real-Time Transient Survey}",
      journal = {\apj},
         year = 2009,
        month = may,
       volume = {696},
       number = {1},
        pages = {870-884},
          doi = {10.1088/0004-637X/696/1/870},
archivePrefix = {arXiv},
       eprint = {0809.1394},
 primaryClass = {astro-ph},
       adsurl = {https://ui.adsabs.harvard.edu/abs/2009ApJ...696..870D}
}

@ARTICLE{Feast+1987,
       author = {{Feast}, M.~W. and {Walker}, A.~R.},
        title = "{Cepheids as distance indicators.}",
      journal = {\araa},
         year = 1987,
        month = jan,
       volume = {25},
        pages = {345-375},
          doi = {10.1146/annurev.aa.25.090187.002021},
       adsurl = {https://ui.adsabs.harvard.edu/abs/1987ARA&A..25..345F}
}

@BOOK{Catelan+2015,
       author = {{Catelan}, M. and {Smith}, H.~A.},
        title = "{Pulsating Stars}",
         year = 2015,
       publisher = "{Wiley-VCH}",
       adsurl = {https://ui.adsabs.harvard.edu/abs/2015pust.book.....C}
}

@ARTICLE{Clementini+2003,
       author = {{Clementini}, Gisella and {Gratton}, Raffaele and {Bragaglia}, Angela and {Carretta}, Eugenio and {Di Fabrizio}, Luca and {Maio}, Marcella},
        title = "{Distance to the Large Magellanic Cloud: The RR Lyrae Stars}",
      journal = {\aj},
         year = 2003,
        month = mar,
       volume = {125},
       number = {3},
        pages = {1309-1329},
          doi = {10.1086/367773},
       adsurl = {https://ui.adsabs.harvard.edu/abs/2003AJ....125.1309C}
}

@ARTICLE{Ripepi+2017,
       author = {{Ripepi}, Vincenzo and {Cioni}, Maria-Rosa L. and {Moretti}, Maria Ida and {Marconi}, Marcella and {Bekki}, Kenji and {Clementini}, Gisella and {de Grijs}, Richard and {Emerson}, Jim and {Groenewegen}, Martin A.~T. and {Ivanov}, Valentin D. and {Molinaro}, Roberto and {Muraveva}, Tatiana and {Oliveira}, Joana M. and {Piatti}, Andr{\'e}s E. and {Subramanian}, Smitha and {van Loon}, Jacco Th.},
        title = "{The VMC survey - XXV. The 3D structure of the Small Magellanic Cloud from Classical Cepheids}",
      journal = {\mnras},
         year = 2017,
        month = nov,
       volume = {472},
       number = {1},
        pages = {808-827},
          doi = {10.1093/mnras/stx2096},
archivePrefix = {arXiv},
       eprint = {1707.04500},
 primaryClass = {astro-ph.GA},
       adsurl = {https://ui.adsabs.harvard.edu/abs/2017MNRAS.472..808R}
}

@ARTICLE{Genovali+2014,
       author = {{Genovali}, K. and {Lemasle}, B. and {Bono}, G. and {Romaniello}, M. and {Fabrizio}, M. and {Ferraro}, I. and {Iannicola}, G. and {Laney}, C.~D. and {Nonino}, M. and {Bergemann}, M. and {Buonanno}, R. and {Fran{\c{c}}ois}, P. and {Inno}, L. and {Kudritzki}, R. -P. and {Matsunaga}, N. and {Pedicelli}, S. and {Primas}, F. and {Th{\'e}venin}, F.},
        title = "{On the fine structure of the Cepheid metallicity gradient in the Galactic thin disk}",
      journal = {\aap},
         year = 2014,
        month = jun,
       volume = {566},
          eid = {A37},
        pages = {A37},
          doi = {10.1051/0004-6361/201323198},
archivePrefix = {arXiv},
       eprint = {1403.6128},
 primaryClass = {astro-ph.GA},
       adsurl = {https://ui.adsabs.harvard.edu/abs/2014A&A...566A..37G}
}

@ARTICLE{Moreno-Cartagena+2025,
       author = {{Moreno-Cartagena}, D. and {Protopapas}, P. and {Cabrera-Vives}, G. and {C{\'a}diz-Leyton}, M. and {Becker}, I. and {Donoso-Oliva}, C.},
        title = "{Leveraging pre-trained vision Transformers for multi-band photometric light curve classification}",
      journal = {\aap},
         year = 2025,
        month = oct,
       volume = {703},
          eid = {A41},
        pages = {A41},
          doi = {10.1051/0004-6361/202554289},
archivePrefix = {arXiv},
       eprint = {2502.20479},
 primaryClass = {astro-ph.IM},
       adsurl = {https://ui.adsabs.harvard.edu/abs/2025A&A...703A..41M}
}

@ARTICLE{Malanchev+2021,
       author = {{Malanchev}, K.~L. and {Pruzhinskaya}, M.~V. and {Korolev}, V.~S. and {Aleo}, P.~D. and {Kornilov}, M.~V. and {Ishida}, E.~E.~O. and {Krushinsky}, V.~V. and {Mondon}, F. and {Sreejith}, S. and {Volnova}, A.~A. and {Belinski}, A.~A. and {Dodin}, A.~V. and {Tatarnikov}, A.~M. and {Zheltoukhov}, S.~G. and {(The SNAD Team)}},
        title = "{Anomaly detection in the Zwicky Transient Facility DR3}",
      journal = {\mnras},
         year = 2021,
        month = apr,
       volume = {502},
       number = {4},
        pages = {5147-5175},
          doi = {10.1093/mnras/stab316},
archivePrefix = {arXiv},
       eprint = {2012.01419},
 primaryClass = {astro-ph.IM},
       adsurl = {https://ui.adsabs.harvard.edu/abs/2021MNRAS.502.5147M}
}

@inproceedings{pan2024astromlab,
  title={AstroMLab 2: AstroLLaMA-2-70B Model and Benchmarking Specialised LLMs for Astronomy},
  author={Pan, Rui and Nguyen, Tuan Dung and Arora, Hardik and Accomazzi, Alberto and Ghosal, Tirthankar and Ting, Yuan-Sen},
  booktitle={SC24-W: Workshops of the International Conference for High Performance Computing, Networking, Storage and Analysis},
  pages={87--96},
  year={2024},
  organization={IEEE}
}

@ARTICLE{Rehemtulla+2024,
       author = {{Rehemtulla}, Nabeel and {Miller}, Adam A. and {Jegou Du Laz}, Theophile and {Coughlin}, Michael W. and {Fremling}, Christoffer and {Perley}, Daniel A. and {Qin}, Yu-Jing and {Sollerman}, Jesper and {Mahabal}, Ashish A. and {Laher}, Russ R. and {Riddle}, Reed and {Rusholme}, Ben and {Kulkarni}, Shrinivas R.},
        title = "{The Zwicky Transient Facility Bright Transient Survey. III. BTSbot: Automated Identification and Follow-up of Bright Transients with Deep Learning}",
      journal = {\apj},
         year = 2024,
        month = sep,
       volume = {972},
       number = {1},
          eid = {7},
        pages = {7},
          doi = {10.3847/1538-4357/ad5666},
archivePrefix = {arXiv},
       eprint = {2401.15167},
 primaryClass = {astro-ph.IM},
       adsurl = {https://ui.adsabs.harvard.edu/abs/2024ApJ...972....7R}
}

@ARTICLE{Gagliano+2023,
       author = {{Gagliano}, Alexander and {Contardo}, Gabriella and {Foreman-Mackey}, Daniel and {Malz}, Alex I. and {Aleo}, Patrick D.},
        title = "{First Impressions: Early-time Classification of Supernovae Using Host-galaxy Information and Shallow Learning}",
      journal = {\apj},
         year = 2023,
        month = sep,
       volume = {954},
       number = {1},
          eid = {6},
        pages = {6},
          doi = {10.3847/1538-4357/ace326},
archivePrefix = {arXiv},
       eprint = {2305.08894},
 primaryClass = {astro-ph.IM},
       adsurl = {https://ui.adsabs.harvard.edu/abs/2023ApJ...954....6G}
}

@ARTICLE{Carrasco-Davis+2021,
       author = {{Carrasco-Davis}, R. and {Reyes}, E. and {Valenzuela}, C. and {F{\"o}rster}, F. and {Est{\'e}vez}, P.~A. and {Pignata}, G. and {Bauer}, F.~E. and {Reyes}, I. and {S{\'a}nchez-S{\'a}ez}, P. and {Cabrera-Vives}, G. and {Eyheramendy}, S. and {Catelan}, M. and {Arredondo}, J. and {Castillo-Navarrete}, E. and {Rodr{\'\i}guez-Mancini}, D. and {Ruz-Mieres}, D. and {Moya}, A. and {Sabatini-Gacit{\'u}a}, L. and {Sep{\'u}lveda-Cobo}, C. and {Mahabal}, A.~A. and {Silva-Farf{\'a}n}, J. and {Camacho-I{\~n}iguez}, E. and {Galbany}, L.},
        title = "{Alert Classification for the ALeRCE Broker System: The Real-time Stamp Classifier}",
      journal = {\aj},
         year = 2021,
        month = dec,
       volume = {162},
       number = {6},
          eid = {231},
        pages = {231},
          doi = {10.3847/1538-3881/ac0ef1},
archivePrefix = {arXiv},
       eprint = {2008.03309},
 primaryClass = {astro-ph.IM},
       adsurl = {https://ui.adsabs.harvard.edu/abs/2021AJ....162..231C}
}

@ARTICLE{Duev+2021,
       author = {{Duev}, Dmitry A. and {van der Walt}, St{\'e}fan J.},
        title = "{Phenomenological classification of the Zwicky Transient Facility astronomical event alerts}",
      journal = {arXiv e-prints},
         year = 2021,
        month = nov,
          eid = {arXiv:2111.12142},
        pages = {arXiv:2111.12142},
          doi = {10.48550/arXiv.2111.12142},
archivePrefix = {arXiv},
       eprint = {2111.12142},
 primaryClass = {astro-ph.IM},
       adsurl = {https://ui.adsabs.harvard.edu/abs/2021arXiv211112142D}
}

@ARTICLE{zhang2024maven,
       author = {{Zhang}, Gemma and {Helfer}, Thomas and {Gagliano}, Alexander T. and {Mishra-Sharma}, Siddharth and {Ashley Villar}, V.},
        title = "{Maven: a multimodal foundation model for supernova science}",
      journal = {Machine Learning: Science and Technology},
         year = 2024,
        month = dec,
       volume = {5},
       number = {4},
          eid = {045069},
        pages = {045069},
          doi = {10.1088/2632-2153/ad990d},
archivePrefix = {arXiv},
       eprint = {2408.16829},
 primaryClass = {astro-ph.HE},
       adsurl = {https://ui.adsabs.harvard.edu/abs/2024MLS&T...5d5069Z}
}

@ARTICLE{Allam+2023,
       author = {{Allam}, Jr., Tarek and {Peloton}, Julien and {McEwen}, Jason D.},
        title = "{The Tiny Time-series Transformer: Low-latency High-throughput Classification of Astronomical Transients using Deep Model Compression}",
      journal = {arXiv e-prints},
         year = 2023,
        month = mar,
          eid = {arXiv:2303.08951},
        pages = {arXiv:2303.08951},
          doi = {10.48550/arXiv.2303.08951},
archivePrefix = {arXiv},
       eprint = {2303.08951},
 primaryClass = {astro-ph.IM},
       adsurl = {https://ui.adsabs.harvard.edu/abs/2023arXiv230308951A}
}

@ARTICLE{Alcock+2000,
       author = {{Alcock}, C. and {Allsman}, R.~A. and {Alves}, D.~R. and {Axelrod}, T.~S. and {Becker}, A.~C. and {Bennett}, D.~P. and {Cook}, K.~H. and {Dalal}, N. and {Drake}, A.~J. and {Freeman}, K.~C. and {Geha}, M. and {Griest}, K. and {Lehner}, M.~J. and {Marshall}, S.~L. and {Minniti}, D. and {Nelson}, C.~A. and {Peterson}, B.~A. and {Popowski}, P. and {Pratt}, M.~R. and {Quinn}, P.~J. and {Stubbs}, C.~W. and {Sutherland}, W. and {Tomaney}, A.~B. and {Vandehei}, T. and {Welch}, D.},
        title = "{The MACHO Project: Microlensing Results from 5.7 Years of Large Magellanic Cloud Observations}",
      journal = {\apj},
         year = 2000,
        month = oct,
       volume = {542},
       number = {1},
        pages = {281-307},
          doi = {10.1086/309512},
archivePrefix = {arXiv},
       eprint = {astro-ph/0001272},
 primaryClass = {astro-ph},
       adsurl = {https://ui.adsabs.harvard.edu/abs/2000ApJ...542..281A}
}

@inproceedings{woo2024unified,
  title = {Unified Training of Universal Time Series Forecasting Transformers},
  author = {Woo, Gerald and Liu, Chenghao and Kumar, Akshat and Xiong, Caiming and Savarese, Silvio and Sahoo, Doyen},
  booktitle = {Proceedings of the 41st International Conference on Machine Learning},
  pages = {53140--53164},
  year = {2024},
  volume = {235},
  series = {Proceedings of Machine Learning Research},
  publisher = {PMLR},
  url = {https://proceedings.mlr.press/v235/woo24a.html}
}

@article{donoso2023astromer,
  title={ASTROMER-A transformer-based embedding for the representation of light curves},
  author={Donoso-Oliva, C and Becker, I and Protopapas, Pavlos and Cabrera-Vives, Guillermo and Vishnu, M and Vardhan, Harsh},
  journal={Astronomy \& Astrophysics},
  volume={670},
  pages={A54},
  year={2023},
  publisher={EDP Sciences}
}

@ARTICLE{Lomb+1976,
       author = {{Lomb}, N.~R.},
        title = "{Least-Squares Frequency Analysis of Unequally Spaced Data}",
      journal = {\apss},
         year = 1976,
        month = feb,
       volume = {39},
       number = {2},
        pages = {447-462},
          doi = {10.1007/BF00648343},
       adsurl = {https://ui.adsabs.harvard.edu/abs/1976Ap&SS..39..447L}
}

@ARTICLE{Scargle+1982,
       author = {{Scargle}, J.~D.},
        title = "{Studies in astronomical time series analysis. II. Statistical aspects of spectral analysis of unevenly spaced data.}",
      journal = {\apj},
         year = 1982,
        month = dec,
       volume = {263},
        pages = {835-853},
          doi = {10.1086/160554},
       adsurl = {https://ui.adsabs.harvard.edu/abs/1982ApJ...263..835S}
}

@article{ansari2024chronos,
  title = {Chronos: Learning the Language of Time Series},
  author = {Ansari, Abdul Fatir and Stella, Lorenzo and Turkmen, Caner and Zhang, Xiyuan and Mercado, Pedro and Shen, Huibin and Shchur, Oleksandr and Rangapuram, Syama Sundar and Arango, Sebastian Pineda and Kapoor, Shubham and Zschiegner, Jasper and Maddix, Danielle C. and Wang, Hao and Mahoney, Michael W. and Torkkola, Kari and Wilson, Andrew Gordon and Bohlke-Schneider, Michael and Wang, Yuyang},
  journal = {Transactions on Machine Learning Research},
  year = {2024},
  url = {https://openreview.net/forum?id=gerNCVqqtR}
}

@inproceedings{yue2022ts2vec,
  title={Ts2vec: Towards universal representation of time series},
  author={Yue, Zhihan and Wang, Yujing and Duan, Juanyong and Yang, Tianmeng and Huang, Congrui and Tong, Yunhai and Xu, Bixiong},
  booktitle={Proceedings of the AAAI conference on artificial intelligence},
  volume={36},
  pages={8980--8987},
  year={2022}
}

@ARTICLE{shah2025oracle,
       author = {{Shah}, Ved G. and {Gagliano}, Alex and {Malanchev}, Konstantin and {Narayan}, Gautham and {Malz}, Alex I. and {the LSST Dark Energy Science Collaboration}},
        title = "{ORACLE: A Real-time, Hierarchical, Deep Learning Photometric Classifier for the LSST}",
      journal = {\apj},
         year = 2025,
        month = dec,
       volume = {995},
       number = {1},
          eid = {4},
        pages = {4},
          doi = {10.3847/1538-4357/ae1130},
archivePrefix = {arXiv},
       eprint = {2501.01496},
 primaryClass = {astro-ph.IM},
       adsurl = {https://ui.adsabs.harvard.edu/abs/2025ApJ...995....4S}
}

@article{cabrera2024atat,
  title={ATAT: Astronomical Transformer for time series and Tabular data},
  author={Cabrera-Vives, G and Moreno-Cartagena, D and Astorga, N and Reyes-Jainaga, I and F{\"o}rster, F and Huijse, P and Arredondo, J and Arancibia, AM Mu{\~n}oz and Bayo, A and Catelan, M and others},
  journal={Astronomy \& Astrophysics},
  volume={689},
  pages={A289},
  year={2024},
  publisher={EDP Sciences}
}

@ARTICLE{2020MNRAS.493.2981B,
       author = {{Becker}, I. and {Pichara}, K. and {Catelan}, M. and {Protopapas}, P. and {Aguirre}, C. and {Nikzat}, F.},
        title = "{Scalable end-to-end recurrent neural network for variable star classification}",
      journal = {mnras},
         year = 2020,
        month = apr,
       volume = {493},
       number = {2},
        pages = {2981-2995},
          doi = {10.1093/mnras/staa350},
archivePrefix = {arXiv},
       eprint = {2002.00994},
 primaryClass = {astro-ph.IM},
       adsurl = {https://ui.adsabs.harvard.edu/abs/2020MNRAS.493.2981B}
}

@ARTICLE{2019PASP..131k8002M,
       author = {{Muthukrishna}, Daniel and {Narayan}, Gautham and {Mandel}, Kaisey S. and {Biswas}, Rahul and {Hlo{\v{z}}ek}, Ren{\'e}e},
        title = "{RAPID: Early Classification of Explosive Transients Using Deep Learning}",
      journal = {pasp},
         year = 2019,
        month = nov,
       volume = {131},
       number = {1005},
        pages = {118002},
          doi = {10.1088/1538-3873/ab1609},
archivePrefix = {arXiv},
       eprint = {1904.00014},
 primaryClass = {astro-ph.IM},
       adsurl = {https://ui.adsabs.harvard.edu/abs/2019PASP..131k8002M}
}

@ARTICLE{2019AJ....158..257B,
       author = {{Boone}, Kyle},
        title = "{Avocado: Photometric Classification of Astronomical Transients with Gaussian Process Augmentation}",
      journal = {aj},
         year = 2019,
        month = dec,
       volume = {158},
       number = {6},
          eid = {257},
        pages = {257},
          doi = {10.3847/1538-3881/ab5182},
archivePrefix = {arXiv},
       eprint = {1907.04690},
 primaryClass = {astro-ph.IM},
       adsurl = {https://ui.adsabs.harvard.edu/abs/2019AJ....158..257B}
}

@ARTICLE{2020ApJ...905...94V,
       author = {{Villar}, V. Ashley and {Hosseinzadeh}, Griffin and {Berger}, Edo and {Ntampaka}, Michelle and {Jones}, David O. and {Challis}, Peter and {Chornock}, Ryan and {Drout}, Maria R. and {Foley}, Ryan J. and {Kirshner}, Robert P. and {Lunnan}, Ragnhild and {Margutti}, Raffaella and {Milisavljevic}, Dan and {Sanders}, Nathan and {Pan}, Yen-Chen and {Rest}, Armin and {Scolnic}, Daniel M. and {Magnier}, Eugene and {Metcalfe}, Nigel and {Wainscoat}, Richard and {Waters}, Christopher},
        title = "{SuperRAENN: A Semisupervised Supernova Photometric Classification Pipeline Trained on Pan-STARRS1 Medium-Deep Survey Supernovae}",
      journal = {apj},
         year = 2020,
        month = dec,
       volume = {905},
       number = {2},
          eid = {94},
        pages = {94},
          doi = {10.3847/1538-4357/abc6fd},
archivePrefix = {arXiv},
       eprint = {2008.04921},
 primaryClass = {astro-ph.HE},
       adsurl = {https://ui.adsabs.harvard.edu/abs/2020ApJ...905...94V}
}

@ARTICLE{2025MNRAS.537..931D,
       author = {{Dillmann}, Steven and {Mart{\'\i}nez-Galarza}, Juan Rafael and {Soria}, Roberto and {Stefano}, Rosanne Di and {Kashyap}, Vinay L.},
        title = "{Representation learning for time-domain high-energy astrophysics: Discovery of extragalactic fast X-ray transient XRT 200515}",
      journal = {mnras},
         year = 2025,
        month = feb,
       volume = {537},
       number = {2},
        pages = {931-955},
          doi = {10.1093/mnras/stae2808},
archivePrefix = {arXiv},
       eprint = {2412.01150},
 primaryClass = {astro-ph.HE},
       adsurl = {https://ui.adsabs.harvard.edu/abs/2025MNRAS.537..931D}
}

@ARTICLE{2021AJ....161..141S,
       author = {{S{\'a}nchez-S{\'a}ez}, P. and {Reyes}, I. and {Valenzuela}, C. and {F{\"o}rster}, F. and {Eyheramendy}, S. and {Elorrieta}, F. and {Bauer}, F.~E. and {Cabrera-Vives}, G. and {Est{\'e}vez}, P.~A. and {Catelan}, M. and {Pignata}, G. and {Huijse}, P. and {De Cicco}, D. and {Ar{\'e}valo}, P. and {Carrasco-Davis}, R. and {Abril}, J. and {Kurtev}, R. and {Borissova}, J. and {Arredondo}, J. and {Castillo-Navarrete}, E. and {Rodriguez}, D. and {Ruz-Mieres}, D. and {Moya}, A. and {Sabatini-Gacit{\'u}a}, L. and {Sep{\'u}lveda-Cobo}, C. and {Camacho-I{\~n}iguez}, E.},
        title = "{Alert Classification for the ALeRCE Broker System: The Light Curve Classifier}",
      journal = {aj},
         year = 2021,
        month = mar,
       volume = {161},
       number = {3},
          eid = {141},
        pages = {141},
          doi = {10.3847/1538-3881/abd5c1},
archivePrefix = {arXiv},
       eprint = {2008.03311},
 primaryClass = {astro-ph.IM},
       adsurl = {https://ui.adsabs.harvard.edu/abs/2021AJ....161..141S}
}

@ARTICLE{Nun+2015,
       author = {{Nun}, Isadora and {Protopapas}, Pavlos and {Sim}, Brandon and {Zhu}, Ming and {Dave}, Rahul and {Castro}, Nicolas and {Pichara}, Karim},
        title = "{FATS: Feature Analysis for Time Series}",
      journal = {arXiv e-prints},
         year = 2015,
        month = may,
          eid = {arXiv:1506.00010},
        pages = {arXiv:1506.00010},
          doi = {10.48550/arXiv.1506.00010},
archivePrefix = {arXiv},
       eprint = {1506.00010},
 primaryClass = {astro-ph.IM},
       adsurl = {https://ui.adsabs.harvard.edu/abs/2015arXiv150600010N}
}

@ARTICLE{Richards+2011,
       author = {{Richards}, Joseph W. and {Starr}, Dan L. and {Butler}, Nathaniel R. and {Bloom}, Joshua S. and {Brewer}, John M. and {Crellin-Quick}, Arien and {Higgins}, Justin and {Kennedy}, Rachel and {Rischard}, Maxime},
        title = "{On Machine-learned Classification of Variable Stars with Sparse and Noisy Time-series Data}",
      journal = {apj},
         year = 2011,
        month = may,
       volume = {733},
       number = {1},
          eid = {10},
        pages = {10},
          doi = {10.1088/0004-637X/733/1/10},
archivePrefix = {arXiv},
       eprint = {1101.1959},
 primaryClass = {astro-ph.IM},
       adsurl = {https://ui.adsabs.harvard.edu/abs/2011ApJ...733...10R}
}

@ARTICLE{Rucinski+2007,
       author = {{Rucinski}, Slavek M.},
        title = "{The short-period end of the contact binary period distribution based on the All-Sky Automated Survey}",
      journal = {\mnras},
         year = 2007,
        month = nov,
       volume = {382},
       number = {1},
        pages = {393-396},
          doi = {10.1111/j.1365-2966.2007.12377.x},
archivePrefix = {arXiv},
       eprint = {0708.3020},
 primaryClass = {astro-ph},
       adsurl = {https://ui.adsabs.harvard.edu/abs/2007MNRAS.382..393R}
}

@ARTICLE{Tylenda+2011,
       author = {{Tylenda}, R. and {Hajduk}, M. and {Kami{\'n}ski}, T. and {Udalski}, A. and {Soszy{\'n}ski}, I. and {Szyma{\'n}ski}, M.~K. and {Kubiak}, M. and {Pietrzy{\'n}ski}, G. and {Poleski}, R. and {Wyrzykowski}, {\L}. and {Ulaczyk}, K.},
        title = "{V1309 Scorpii: merger of a contact binary}",
      journal = {\aap},
         year = 2011,
        month = apr,
       volume = {528},
          eid = {A114},
        pages = {A114},
          doi = {10.1051/0004-6361/201016221},
archivePrefix = {arXiv},
       eprint = {1012.0163},
 primaryClass = {astro-ph.SR},
       adsurl = {https://ui.adsabs.harvard.edu/abs/2011A&A...528A.114T}
}

@ARTICLE{astromer2,
       author = {{Donoso-Oliva}, Cristobal and {Becker}, Ignacio and {Protopapas}, Pavlos and {Cabrera-Vives}, Guillermo and {C{\'a}diz-Leyton}, Martina and {Moreno-Cartagena}, Daniel},
        title = "{Generalizing across astronomical surveys: Few-shot light curve classification with Astromer 2}",
      journal = {\aap},
         year = 2026,
        month = mar,
       volume = {707},
          eid = {A170},
        pages = {A170},
          doi = {10.1051/0004-6361/202554026},
archivePrefix = {arXiv},
       eprint = {2502.02717},
 primaryClass = {astro-ph.IM},
       adsurl = {https://ui.adsabs.harvard.edu/abs/2026A&A...707A.170D}
}

@ARTICLE{FALCO,
       author = {{Zuo}, Xiaoxiong and {Tao}, Yihan and {Huang}, Yang and {Kang}, Zhixuan and {Chen}, Huaxi and {Cui}, Chenzhou and {Pan}, Jiashu and {Kong}, Xiao and {Tang}, Xiaoyu and {Han}, Henggeng and {Mu}, Haiyang and {Xu}, Yunfei and {Fan}, Dongwei and {Xue}, Guirong and {Luo}, Ali and {Liu}, Jifeng},
        title = "{FALCO: a Foundation model of Astronomical Light Curves for time dOmain astronomy}",
      journal = {arXiv e-prints},
         year = 2025,
        month = apr,
          eid = {arXiv:2504.20290},
        pages = {arXiv:2504.20290},
          doi = {10.48550/arXiv.2504.20290},
archivePrefix = {arXiv},
       eprint = {2504.20290},
 primaryClass = {astro-ph.IM},
       adsurl = {https://ui.adsabs.harvard.edu/abs/2025arXiv250420290Z}
}

@inproceedings{sklearn_api,
  author    = {Lars Buitinck and Gilles Louppe and Mathieu Blondel and
                Fabian Pedregosa and Andreas Mueller and Olivier Grisel and
                Vlad Niculae and Peter Prettenhofer and Alexandre Gramfort
                and Jaques Grobler and Robert Layton and Jake VanderPlas and
                Arnaud Joly and Brian Holt and Ga{\"{e}}l Varoquaux},
  title     = {{API} design for machine learning software: experiences from the scikit-learn
                project},
  booktitle = {ECML PKDD Workshop: Languages for Data Mining and Machine Learning},
  year      = {2013},
  pages = {108--122},
}

@article{neelakantan2022text,
  title={Text and code embeddings by contrastive pre-training},
  author={Neelakantan, Arvind and Xu, Tao and Puri, Raul and Radford, Alec and Han, Jesse Michael and Tworek, Jerry and Yuan, Qiming and Tezak, Nikolas and Kim, Jong Wook and Hallacy, Chris and others},
  journal={arXiv preprint arXiv:2201.10005},
  year={2022}
}

@inproceedings{zhou2021ibot,
  title = {{iBOT}: Image {BERT} Pre-Training with Online Tokenizer},
  author = {Zhou, Jinghao and Wei, Chen and Wang, Huiyu and Shen, Wei and Xie, Cihang and Yuille, Alan and Kong, Tao},
  booktitle = {International Conference on Learning Representations},
  year = {2022},
  url = {https://openreview.net/forum?id=ydopy-e6Dg}
}

@inproceedings{caron2021emerging,
  title={Emerging properties in self-supervised vision transformers},
  author={Caron, Mathilde and Touvron, Hugo and Misra, Ishan and J{\'e}gou, Herv{\'e} and Mairal, Julien and Bojanowski, Piotr and Joulin, Armand},
  booktitle={Proceedings of the IEEE/CVF international conference on computer vision},
  pages={9650--9660},
  year={2021}
}

@inproceedings{nie2022time,
  title = {A Time Series is Worth 64 Words: Long-term Forecasting with Transformers},
  author = {Nie, Yuqi and Nguyen, Nam H. and Sinthong, Phanwadee and Kalagnanam, Jayant},
  booktitle = {International Conference on Learning Representations},
  year = {2023},
  url = {https://openreview.net/forum?id=Jbdc0vTOcol}
}

@inproceedings{zhou2021informer,
  title={Informer: Beyond efficient transformer for long sequence time-series forecasting},
  author={Zhou, Haoyi and Zhang, Shanghang and Peng, Jieqi and Zhang, Shuai and Li, Jianxin and Xiong, Hui and Zhang, Wancai},
  booktitle={Proceedings of the AAAI conference on artificial intelligence},
  volume={35},
  pages={11106--11115},
  year={2021}
}

@ARTICLE{Pan+2024,
       author = {{Pan}, Jia-Shu and {Ting}, Yuan-Sen and {Huang}, Yang and {Yu}, Jie and {Liu}, Ji-Feng},
        title = "{The Scaling Law in Stellar Light Curves}",
      journal = {arXiv e-prints},
         year = 2024,
        month = may,
          eid = {arXiv:2405.17156},
        pages = {arXiv:2405.17156},
          doi = {10.48550/arXiv.2405.17156},
archivePrefix = {arXiv},
       eprint = {2405.17156},
 primaryClass = {astro-ph.IM},
       adsurl = {https://ui.adsabs.harvard.edu/abs/2024arXiv240517156P}
}

@ARTICLE{Edwards+2024,
       author = {{Edwards}, Thomas D.~P. and {Alvey}, James and {Alsing}, Justin and {Nguyen}, Nam H. and {Wandelt}, Benjamin D.},
        title = "{Scaling-laws for Large Time-series Models}",
      journal = {arXiv e-prints},
         year = 2024,
        month = may,
          eid = {arXiv:2405.13867},
        pages = {arXiv:2405.13867},
          doi = {10.48550/arXiv.2405.13867},
archivePrefix = {arXiv},
       eprint = {2405.13867},
 primaryClass = {cs.LG},
       adsurl = {https://ui.adsabs.harvard.edu/abs/2024arXiv240513867E}
}

@ARTICLE{Dubath+2011,
       author = {{Dubath}, P. and {Rimoldini}, L. and {S{\"u}veges}, M. and {Blomme}, J. and {L{\'o}pez}, M. and {Sarro}, L.~M. and {De Ridder}, J. and {Cuypers}, J. and {Guy}, L. and {Lecoeur}, I. and {Nienartowicz}, K. and {Jan}, A. and {Beck}, M. and {Mowlavi}, N. and {De Cat}, P. and {Lebzelter}, T. and {Eyer}, L.},
        title = "{Random forest automated supervised classification of Hipparcos periodic variable stars}",
      journal = {\mnras},
         year = 2011,
        month = jul,
       volume = {414},
       number = {3},
        pages = {2602-2617},
          doi = {10.1111/j.1365-2966.2011.18575.x},
archivePrefix = {arXiv},
       eprint = {1101.2406},
 primaryClass = {astro-ph.SR},
       adsurl = {https://ui.adsabs.harvard.edu/abs/2011MNRAS.414.2602D}
}

@ARTICLE{Kim+2016,
       author = {{Kim}, Dae-Won and {Bailer-Jones}, Coryn A.~L.},
        title = "{A package for the automated classification of periodic variable stars}",
      journal = {\aap},
         year = 2016,
        month = mar,
       volume = {587},
          eid = {A18},
        pages = {A18},
          doi = {10.1051/0004-6361/201527188},
archivePrefix = {arXiv},
       eprint = {1512.01611},
 primaryClass = {astro-ph.IM},
       adsurl = {https://ui.adsabs.harvard.edu/abs/2016A&A...587A..18K}
}

@ARTICLE{Sesar+2017,
       author = {{Sesar}, Branimir and {Hernitschek}, Nina and {Mitrovi{\'c}}, Sandra and {Ivezi{\'c}}, {\v{Z}}eljko and {Rix}, Hans-Walter and {Cohen}, Judith G. and {Bernard}, Edouard J. and {Grebel}, Eva K. and {Martin}, Nicolas F. and {Schlafly}, Edward F. and {Burgett}, William S. and {Draper}, Peter W. and {Flewelling}, Heather and {Kaiser}, Nick and {Kudritzki}, Rolf P. and {Magnier}, Eugene A. and {Metcalfe}, Nigel and {Tonry}, John L. and {Waters}, Christopher},
        title = "{Machine-learned Identification of RR Lyrae Stars from Sparse, Multi-band Data: The PS1 Sample}",
      journal = {\aj},
         year = 2017,
        month = may,
       volume = {153},
       number = {5},
          eid = {204},
        pages = {204},
          doi = {10.3847/1538-3881/aa661b},
archivePrefix = {arXiv},
       eprint = {1611.08596},
 primaryClass = {astro-ph.GA},
       adsurl = {https://ui.adsabs.harvard.edu/abs/2017AJ....153..204S}
}

@article{yang2024research,
  title={Research on information leakage in time series prediction based on empirical mode decomposition},
  author={Yang, Xinyi and Li, Jingyi and Jiang, Xuchu},
  journal={Scientific Reports},
  volume={14},
  number={1},
  pages={28362},
  year={2024},
  publisher={Nature Publishing Group UK London}
}

@article{kuhn1955hungarian,
  title={The Hungarian method for the assignment problem},
  author={Kuhn, Harold W},
  journal={Naval research logistics quarterly},
  volume={2},
  number={1-2},
  pages={83--97},
  year={1955},
  publisher={Wiley Online Library}
}

@inproceedings{sun2024lsenet,
  title = {{LSE}net: {L}orentz Structural Entropy Neural Network for Deep Graph Clustering},
  author = {Sun, Li and Huang, Zhenhao and Peng, Hao and Wang, Yujie and Liu, Chunyang and Yu, Philip S.},
  booktitle = {Proceedings of the 41st International Conference on Machine Learning},
  pages = {47078--47104},
  year = {2024},
  volume = {235},
  series = {Proceedings of Machine Learning Research},
  publisher = {PMLR},
  url = {https://proceedings.mlr.press/v235/sun24g.html}
}

@inproceedings{
  li2024image,
  title={Image Clustering with External Guidance},
  author={Li, Yunfan and Hu, Peng and Peng, Dezhong and Lv, Jiancheng and Fan, Jianping and Peng, Xi},
  booktitle={Forty-first International Conference on Machine Learning},
  year={2024},
  url={https://openreview.net/forum?id=JSYN891WnB}
}

@article{monnier2020deep,
  title={Deep transformation-invariant clustering},
  author={Monnier, Tom and Groueix, Thibault and Aubry, Mathieu},
  journal={Advances in neural information processing systems},
  volume={33},
  pages={7945--7955},
  year={2020}
}

@article{huang2020partially,
  title={Partially view-aligned clustering},
  author={Huang, Zhenyu and Hu, Peng and Zhou, Joey Tianyi and Lv, Jiancheng and Peng, Xi},
  journal={Advances in Neural Information Processing Systems},
  volume={33},
  pages={2892--2902},
  year={2020}
}

@ARTICLE{Liu+2025_moirai2,
       author = {{Liu}, Chenghao and {Aksu}, Taha and {Liu}, Juncheng and {Liu}, Xu and {Yan}, Hanshu and {Pham}, Quang and {Savarese}, Silvio and {Sahoo}, Doyen and {Xiong}, Caiming and {Li}, Junnan},
        title = "{Moirai 2.0: When Less Is More for Time Series Forecasting}",
      journal = {arXiv e-prints},
         year = 2025,
        month = nov,
          eid = {arXiv:2511.11698},
        pages = {arXiv:2511.11698},
          doi = {10.48550/arXiv.2511.11698},
archivePrefix = {arXiv},
       eprint = {2511.11698},
 primaryClass = {stat.ML},
       adsurl = {https://ui.adsabs.harvard.edu/abs/2025arXiv251111698L}
}

@inproceedings{Shi+2024_TimeMoE,
  title = {Time-{MoE}: Billion-Scale Time Series Foundation Models with Mixture of Experts},
  author = {Shi, Xiaoming and Wang, Shiyu and Nie, Yuqi and Li, Dianqi and Ye, Zhou and Wen, Qingsong and Jin, Ming},
  booktitle = {The Thirteenth International Conference on Learning Representations},
  year = {2025},
  url = {https://openreview.net/forum?id=e1wDDFmlVu}
}

@inproceedings{abnar2020quantifying,
  title={Quantifying attention flow in transformers},
  author={Abnar, Samira and Zuidema, Willem},
  booktitle={Proceedings of the 58th annual meeting of the association for computational linguistics},
  pages={4190--4197},
  year={2020}
}

@ARTICLE{Rehemtulla+2026,
       author = {{Rehemtulla}, Nabeel and {Miller}, Adam A. and {Walmsley}, Mike and {Shah}, Ved G. and {Jegou du Laz}, Theophile and {Coughlin}, Michael W. and {Sasli}, Argyro and {Bloom}, Joshua and {Fremling}, Christoffer and {Graham}, Matthew J. and {Groom}, Steven L. and {Hale}, David and {Mahabal}, Ashish A. and {Perley}, Daniel A. and {Purdum}, Josiah and {Rusholme}, Ben and {Sollerman}, Jesper and {Kasliwal}, Mansi M.},
        title = "{Pre-training Vision Models for the Classification of Alerts from Wide-field Time-domain Surveys}",
      journal = {\pasp},
         year = 2026,
        month = mar,
       volume = {138},
       number = {3},
          eid = {034503},
        pages = {034503},
          doi = {10.1088/1538-3873/ae50bc},
archivePrefix = {arXiv},
       eprint = {2512.11957},
 primaryClass = {astro-ph.IM},
       adsurl = {https://ui.adsabs.harvard.edu/abs/2026PASP..138c4503R}
}

@ARTICLE{Fei+2024,
       author = {{Fei}, Ya and {Yu}, Ce and {Li}, Kun and {Chen}, Xiaodian and {Zhang}, Yajie and {Cui}, Chenzhou and {Xiao}, Jian and {Xu}, Yunfei and {Tao}, Yihan},
        title = "{LEAVES: An Expandable Light-curve Data Set for Automatic Classification of Variable Stars}",
      journal = {\apjs},
         year = 2024,
        month = nov,
       volume = {275},
       number = {1},
          eid = {10},
        pages = {10},
          doi = {10.3847/1538-4365/ad785b},
       adsurl = {https://ui.adsabs.harvard.edu/abs/2024ApJS..275...10F}
}

@article{feofanov2026mantisv2,
  title={MantisV2: Closing the Zero-Shot Gap in Time Series Classification with Synthetic Data and Test-Time Strategies},
  author={Feofanov, Vasilii and Wen, Songkang and Zhang, Jianfeng and Pan, Lujia and Redko, Ievgen},
  journal={arXiv preprint arXiv:2602.17868},
  year={2026}
}

@inproceedings{chen2024visionts,
  title = {{VisionTS}: Visual Masked Autoencoders Are Free-Lunch Zero-Shot Time Series Forecasters},
  author = {Chen, Mouxiang and Shen, Lefei and Li, Zhuo and Wang, Xiaoyun Joy and Sun, Jianling and Liu, Chenghao},
  booktitle = {Proceedings of the 42nd International Conference on Machine Learning},
  pages = {8979--9007},
  year = {2025},
  volume = {267},
  series = {Proceedings of Machine Learning Research},
  publisher = {PMLR},
  url = {https://proceedings.mlr.press/v267/chen25be.html}
}

@article{zivanovic2026rotary,
  title={Rotary Masked Autoencoders are Versatile Learners},
  author={Zivanovic, Uros and Di Gioia, Serafina and Scaffidi, Andre and de los Rios, Mart{\'\i}n and Contardo, Gabriella and Trotta, Roberto},
  journal={Advances in Neural Information Processing Systems},
  volume={38},
  pages={133952--133987},
  year={2026}
}
\bibliographystyle{icml2026}

\newpage
\appendix
\onecolumn

\makeatletter
\let\addcontentsline\real@addcontentsline
\makeatother

\begingroup
\let\clearpage\relax  %
\renewcommand\contentsname{Supplementary Material}

\section*{Appendix}

\setcounter{tocdepth}{2}

\startcontents[appendix]
\printcontents[appendix]{l}{1}{\setcounter{tocdepth}{2}}
\endgroup

\clearpage

\section{Additional Zero-shot and Fine-tuning Experiments}
\label{app:additional_exps}

\subsection{Time Series Foundation Models for Classification}
\label{app:tsfm_classification}

In this section, we extend our benchmark to include two TSFMs designed for classification.

\paragraph{Mantis+ and MantisV2.}
\texttt{Mantis+} and \texttt{MantisV2}~\citep{feofanov2026mantisv2}
are two pre-trained TSFMs designed specifically for time series
classification. Both share the same architecture: a 3-branch
tokenizer followed by a 6-layer Transformer encoder. The three
branches apply a 1D convolution to the raw signal, a convolution to
the first-order temporal difference, and a per-patch mean/std
encoding, respectively. They are pre-trained with a
contrastive self-supervised objective on $2$M synthetic time series.
The two models differ only in their internal modules.
\texttt{Mantis+} uses sinusoidal positional encoding, whereas
\texttt{MantisV2} uses rotary positional encoding together with
RMSNorm and SwiGLU activations. As a result, \texttt{MantisV2} has
a more compact backbone ($4.19$M vs.\ $8.91$M parameters). The
contrastive objective on synthetic data is designed to produce
embeddings that transfer well to unseen domains.

\paragraph{Experiments Settings and Hyperparameters.}
For zero-shot embedding, we follow the same per-band-then-concat
pipeline used for the other TSFMs in our paper, with a context window
of $192 = 32 \times 6$ (the largest multiple of $32$ below $200$, as
required by \texttt{Mantis}'s tokenizer) and mean-pooled tokens from
the third Transformer layer following from \citep{feofanov2026mantisv2}. 
For fine-tuning, we unfreeze the entire encoder and append the same
3-layer MLP head used in our zero-shot MLP benchmark. The $g$ and $r$
bands are fed as two input channels into the final Transformer layer.
We use AdamW under a cosine learning-rate schedule, using
the same class-weighted cross-entropy loss as the
MLP benchmark. Hyperparameters (batch size, learning rate, dropout)
are tuned with the same grid search protocol as the zero-shot MLP. We
then run the best configuration with $10$ seeds and report the mean and
standard deviation in Table~\ref{tab:mantis_results}.

\paragraph{Results.}
\cref{tab:clustering_new_models,tab:classification_results_mantis} contains the zero-shot clustering and classification results, respectively.
Table~\ref{tab:mantis_results} compares both the zero-shot and the
fine-tuned classification results. 
In Table~\ref{tab:classification_results_all}, both \texttt{Mantis+} and \texttt{MantisV2} perform on par
with \texttt{Moirai} but score below \texttt{Chronos} and the
hand-crafted feature baseline, which we attribute to their
patch-based tokenization. 
For Table~\ref{tab:mantis_results}, we observe that fine-tuning improves both models substantially over their frozen-encoder baselines: with the same MLP
head, full fine-tuning raises macro-F1 from $0.570$ to $0.638$ for
\texttt{MantisV2} and from $0.600$ to $0.646$ for \texttt{Mantis+}.
Even with this gain, fine-tuned Mantis remains below the hand-crafted
feature baseline (macro-F1~$0.723$).

\begin{table}[htbp]
\captionsetup{skip=1.5pt}
  \caption{\blue{Unsupervised clustering results for classification-specific TSFMs. Both Mantis models exhibit weaker clustering performance compared to forecasting-based TSFMs.}}
  \centering
  \resizebox{\columnwidth}{!}{
  \begin{tabular}{@{}lcccccc@{}}
    \toprule
    \textbf{Model} & \textbf{NMI (K-Means)} & \textbf{ARI (K-Means)} & \textbf{F1 (K-Means)} & \textbf{NMI (Ward)} & \textbf{ARI (Ward)} & \textbf{F1 (Ward)} \\
    \midrule
    \texttt{Astromer-1}        & 0.0041(0.0001) & 0.0017(0.0011) & 0.1660(0.0014) & 0.0041 & 0.0001 & 0.1652 \\
    \texttt{Astromer-2}        & 0.0082(0.0010) & 0.0192(0.0078) & 0.1590(0.0042) & 0.0091 & 0.0310 & 0.1600 \\
    \texttt{Moirai}            & 0.1749(0.0017) & 0.0981(0.0028) & 0.2831(0.0034) & 0.1476 & 0.0828 & 0.2612 \\
    \texttt{\newblue{Moirai-2}} & 0.2017(0.0019) & 0.0967(0.0060) & 0.3007(0.0095) & 0.1850 & 0.0603 & 0.2655 \\
    \texttt{Chronos}           & 0.2374(0.0082) & \textbf{0.1596(0.0029)} & 0.3110(0.0362) & 0.1890 & 0.1217 & \textbf{0.3671} \\
    \texttt{Chronos-Bolt}      & 0.2120(0.0033) & 0.1306(0.0125) & 0.3128(0.0027) & \underline{0.2273} & \textbf{0.1553} & \underline{0.3662} \\
    \texttt{\newblue{Time-MoE}} & 0.2069(0.0108) & 0.0913(0.0090) & 0.3101(0.0227) & 0.1946 & 0.0944 & 0.3385 \\
    Hand-crafted Features      & \textbf{0.2700(0.0058)} & 0.1197(0.0092) & \textbf{0.3960(0.0271)} & \textbf{0.2508} & \underline{0.1319} & 0.3323 \\
    \midrule
    \texttt{\newblue{MantisV2}}   & 0.0566(0.0006) & 0.0096(0.0032) & 0.1679(0.0064) & 0.0838 & 0.0346 & 0.2276 \\
    \texttt{\newblue{Mantis+}}    & 0.1012(0.0137) & 0.0619(0.0068) & 0.2217(0.0183) & 0.1172 & 0.0568 & 0.2117 \\
    \bottomrule
  \end{tabular}
  }
  \label{tab:clustering_new_models}
\end{table}

\begin{table*}[htbp]
\large
\caption{\blue{Classification results including classification-specific TSFMs. 
We use frozen embeddings from layer $3$ follow the paper setting. They show layer $3$ yields the best feature-extraction performance. Other setting for embedding generation is the same as in our paper.
\texttt{MantisV2} and \texttt{Mantis+} have comparable performance as \texttt{Moirai}.
Unlike the point-wise tokenization of \texttt{Chronos}/\texttt{Time-MoE}, \texttt{Mantis} family uses patching like \texttt{Moirai}. We hypothesize this limit fine-grained temporal resolution on irregularly sampled light curves. See \cref{fig:attention_socre_1} for more evidence.
(Each metrics is averaged separately over 10 runs. Best results in \textbf{bold}, second-best \underline{underlined}.) }}
  \resizebox{\textwidth}{!}{
  \begin{tabular}{@{}llcccccc|ccc@{}}
    \toprule
    \textbf{Classifier} & \textbf{Metric} & \textbf{\texttt{Moirai}} & \textbf{\texttt{\newblue{Moirai-2}}} & \textbf{\texttt{Chronos}} & \textbf{\texttt{Chronos-Bolt}} & \textbf{\texttt{\newblue{Time-MoE}}} & \textbf{HC features} & \textbf{\texttt{\newblue{MantisV2}}} & \textbf{\texttt{\newblue{Mantis+}}} \\
    \midrule
    \multirow{4}{*}{$k$-NN}
       & Accuracy  & 0.809 & 0.815 & \underline{0.857} & 0.807 & 0.825 & \textbf{0.881} & 0.814 & 0.820 \\
       & Precision & 0.662 & 0.641 & \underline{0.799} & 0.647 & 0.724 & \textbf{0.818} & 0.657 & 0.667 \\
       & Recall    & 0.509 & 0.546 & \underline{0.623} & 0.542 & 0.560 & \textbf{0.661} & 0.501 & 0.509 \\
       & F1        & 0.554 & 0.561 & \underline{0.672} & 0.570 & 0.595 & \textbf{0.712} & 0.550 & 0.557 \\
    \midrule
    \multirow{4}{*}{logistic}
       & Accuracy  & 0.705 & 0.709 & \underline{0.750} & 0.709 & 0.740 & \textbf{0.838} & 0.743 & 0.739 \\
       & Precision & 0.544 & 0.538 & \underline{0.575} & 0.549 & 0.559 & \textbf{0.663} & 0.568 & 0.555 \\
       & Recall    & 0.680 & 0.678 & \underline{0.730} & 0.676 & 0.694 & \textbf{0.854} & 0.711 & 0.737 \\
       & F1        & 0.579 & 0.577 & \underline{0.617} & 0.580 & 0.597 & \textbf{0.714} & 0.603 & 0.595 \\
    \midrule
    \multirow{4}{*}{RF}
       & Accuracy  & 0.823 (0.001) & 0.820 (0.001) & \underline{0.862 (0.000)} & 0.826 (0.001) & 0.839 (0.001) & \textbf{0.920 (0.001)} & 0.840 (0.001) & 0.845 (0.001) \\
       & Precision & 0.716 (0.007) & 0.697 (0.010) & \underline{0.750 (0.056)} & 0.707 (0.002) & 0.693 (0.004) & \textbf{0.866 (0.003)} & 0.736 (0.003) & 0.731 (0.004) \\
       & Recall    & 0.514 (0.002) & 0.520 (0.004) & \underline{0.597 (0.002)} & 0.548 (0.001) & 0.561 (0.003) & \textbf{0.773 (0.004)} & 0.536 (0.002) & 0.533 (0.003) \\
       & F1        & 0.557 (0.002) & 0.556 (0.003) & \underline{0.638 (0.002)} & 0.582 (0.001) & 0.598 (0.003) & \textbf{0.804 (0.003)} & 0.588 (0.002) & 0.585 (0.003) \\
    \midrule
    \multirow{4}{*}{MLP}
       & Accuracy  & 0.717 (0.031) & 0.735 (0.019) & \underline{0.783 (0.022)} & 0.721 (0.022) & 0.758 (0.025) & \textbf{0.833 (0.022)} & 0.686 (0.036) & 0.731 (0.034) \\
       & Precision & 0.546 (0.022) & 0.560 (0.021) & \underline{0.589 (0.025)} & 0.553 (0.026) & 0.579 (0.024) & \textbf{0.672 (0.025)} & 0.551 (0.029) & 0.568 (0.019) \\
       & Recall    & 0.722 (0.006) & 0.717 (0.006) & \underline{0.758 (0.006)} & 0.696 (0.013) & 0.741 (0.010) & \textbf{0.851 (0.009)} & 0.714 (0.009) & 0.728 (0.010) \\
       & F1        & 0.594 (0.019) & 0.603 (0.015) & \underline{0.643 (0.023)} & 0.589 (0.015) & 0.627 (0.019) & \textbf{0.723 (0.027)} & 0.570 (0.019) & 0.600 (0.021) \\
    \bottomrule
  \end{tabular}}
\label{tab:classification_results_mantis}
\end{table*}

\begin{table}[htbp]
\caption{\blue{Comparison of zero-shot vs.\ fine-tuned classification for Mantis models. 
We append the same MLP head to the final layer of Mantis. The loss function is also the same weighted loss we use in MLP training. 
Hyperparameters were selected via the same grid search as the benchmark MLP. 
Best configuration: BS=256, LR=$10^{-5}$, dropout=0.1 for \texttt{MantisV2}; BS=256, LR=$10^{-5}$, dropout=0.0 for \texttt{Mantis+}. 
Full fine-tuning yields improvements, but still underperforms hand-crafted features. (Each metric is averaged separately over 10 runs. Best results in \textbf{bold}, second-best \underline{underlined}.)
}}

  \centering
  \begin{tabular}{@{}lcc|cc|c@{}}
    \toprule
    & \multicolumn{2}{c|}{\textbf{Zero-shot}} & \multicolumn{2}{c|}{\textbf{Fine-tuned}} & \\
    \cmidrule(lr){2-3} \cmidrule(lr){4-5}
    \textbf{Metric}
      & \textbf{\texttt{MantisV2}} & \textbf{\texttt{Mantis+}}
      & \textbf{\texttt{MantisV2-FT}} & \textbf{\texttt{Mantis+-FT}}
      & \textbf{HC features} \\
    \midrule
    Accuracy  & 0.686 (0.036) & 0.731 (0.034) & \underline{0.773 (0.020)} & 0.770 (0.018) & \textbf{0.833 (0.022)} \\
    Precision & 0.551 (0.029) & 0.568 (0.019) & 0.605 (0.017) & \underline{0.619 (0.016)} & \textbf{0.672 (0.025)} \\
    Recall    & 0.714 (0.009) & 0.728 (0.010) & 0.780 (0.005) & \underline{0.790 (0.014)} & \textbf{0.851 (0.009)} \\
    F1        & 0.570 (0.019) & 0.600 (0.021) & 0.638 (0.015) & \underline{0.646 (0.018)} & \textbf{0.723 (0.027)} \\
    \bottomrule
  \end{tabular}
  \label{tab:mantis_results}
\end{table}

\subsection{Fine-tuning Chronos}
\label{app:finetune}

This subsection expands on the fine-tuning study in \cref{sec:finetuning}.
We organize the experiments along two axes:
(i) \emph{fine-tuning objective} (forecasting vs.\ classification, \cref{tab:appx-chronos-ft-wide}), and
(ii) \emph{parameter-efficient/partial fine-tuning} (full fine-tuning vs.\ LoRA vs.\ LayerNorm-only, with either a linear or an MLP classification head, \cref{tab:appx-finetune-linear,tab:appx-finetune-mlp}).

\subsubsection{Classification vs.\ Forecasting Fine-tuning}

\paragraph{Experiments Settings and Hyperparameters.}
We compare two fine-tuning objectives for \texttt{Chronos-tiny}.
\begin{itemize}
    \item \textbf{\texttt{Chronos-FT\textsubscript{fcst}}.} 
    We fine-tune on ZTF light curves with the Chronos forecasting objective (context=512, prediction=64), using the default Chronos fine-tuning configuration (LR=$10^{-3}$ with linear decay, and BS  $=64$). We finetune on ${\sim}57$k univariate time series from g+r bands) with early stopping setting patience 10 (eval every 200 steps).
    \item \textbf{\texttt{Chronos-FT\textsubscript{cls}}.} We performs end-to-end classification fine-tuning. The $g$- and $r$-band light curves are fed through the shared encoder and the resulting $256$-dim averaged embeddings are concatenated, then passed to the same MLP head and weighted cross-entropy loss as in our benchmark. Hyperparameters are selected via the same grid search as the benchmark MLP (BS $\in \{32, 64, 128, 256\}$, LR $\in \{10^{-3}, 10^{-4}, 10^{-5}\}$, dropout $\in \{0.0, 0.1\}$); best configuration: BS=32, LR=$10^{-4}$, dropout=0.1.
\end{itemize}

\begin{table*}[htbp]
\large
\captionsetup{skip=1.5pt}
  \caption{\blue{Benchmark supervised classification including the two Chronos fine-tuning variants (\texttt{Chronos-FT\textsubscript{fcst}} and \texttt{Chronos-FT\textsubscript{cls}}). Best in \textbf{bold}, second-best \underline{underlined}. \texttt{Chronos-FT\textsubscript{cls}} improves 3 of 4 MLP metrics over zero-shot Chronos, while \texttt{Chronos-FT\textsubscript{fcst}} does not improve downstream classification and is slightly worse than zero-shot Chronos. 
  Since \texttt{Chronos-FT\textsubscript{cls}} jointly trains a classification head with the encoder rather than producing reusable embeddings, it only applies to the MLP row.
  For RF and MLP, each metrics is averaged separately over 10 runs.}}
  \resizebox{\textwidth}{!}{
  \begin{tabular}{@{}llccccccc|cc@{}}
    \toprule
    \textbf{Classifier} & \textbf{Metric} 
    & \textbf{\texttt{Moirai}} 
    & \textbf{\texttt{\newblue{Moirai-2}}} 
    & \textbf{\texttt{Chronos-Bolt}} 
    & \textbf{\texttt{\newblue{Time-MoE}}} 
    & \textbf{HC features} 
    & \textbf{\texttt{Chronos}} 
    & \textbf{\texttt{\newblue{Chronos-FT\textsubscript{fcst}}}} 
    & \textbf{\texttt{\newblue{Chronos-FT\textsubscript{cls}}}} \\
    \midrule
    \multirow{4}{*}{$k$-NN}
       & Accuracy   & 0.809 & 0.815 & 0.807 & 0.825 & \textbf{0.881} & \underline{0.857} & 0.850 & --- \\
       & Precision & 0.662 & 0.641 & 0.647 & 0.724 & \textbf{0.818} & \underline{0.799} & 0.712 & --- \\
       & Recall     & 0.509 & 0.546 & 0.542 & 0.560 & \textbf{0.661} & \underline{0.623} & 0.600 & --- \\
       & F1         & 0.554 & 0.561 & 0.570 & 0.595 & \textbf{0.712} & \underline{0.672} & 0.630 & --- \\
    \midrule
    \multirow{4}{*}{logistic}
       & Accuracy   & 0.705 & 0.709 & 0.709 & 0.740 & \textbf{0.838} & \underline{0.750} & 0.738 & --- \\
       & Precision  & 0.544 & 0.538 & 0.549 & 0.559 & \textbf{0.663} & \underline{0.575} & 0.549 & --- \\
       & Recall     & 0.680 & 0.678 & 0.676 & 0.694 & \textbf{0.854} & \underline{0.730} & 0.713 & --- \\
       & F1         & 0.579 & 0.577 & 0.580 & 0.597 & \textbf{0.714} & \underline{0.617} & 0.593 & --- \\
    \midrule
    \multirow{4}{*}{RF}
       & Accuracy   & 0.823 (0.001) & 0.820 (0.001) & 0.826 (0.001) & 0.839 (0.001) & \textbf{0.920 (0.001)} & \underline{0.862 (0.000)} & 0.861 (0.001) & --- \\
       & Precision  & 0.716 (0.007) & 0.697 (0.010) & 0.707 (0.002) & 0.693 (0.004) & \textbf{0.866 (0.003)} & \underline{0.750} (0.056) & 0.733 (0.041) & --- \\
       & Recall     & 0.514 (0.002) & 0.520 (0.004) & 0.548 (0.001) & 0.561 (0.003) & \textbf{0.773 (0.004)} & \underline{0.597 (0.002)} & 0.593 (0.003) & --- \\
       & F1         & 0.557 (0.002) & 0.556 (0.003) & 0.582 (0.001) & 0.598 (0.003) & \textbf{0.804 (0.003)} & \underline{0.638 (0.002)} & 0.628 (0.002) & --- \\
    \midrule
    \multirow{4}{*}{MLP}
       & Accuracy   & 0.717 (0.031) & 0.735 (0.019) & 0.721 (0.022) & 0.758 (0.025) & \textbf{0.833 (0.022)} & \underline{0.783 (0.022)} & 0.746 (0.026) & 0.778 (0.029) \\
       & Precision  & 0.546 (0.022) & 0.560 (0.021) & 0.553 (0.026) & 0.579 (0.024) & \textbf{0.672 (0.025)} & 0.589 (0.025) & 0.586 (0.024) & \underline{0.613 (0.036)} \\
       & Recall     & 0.722 (0.006) & 0.717 (0.006) & 0.696 (0.013) & 0.741 (0.010) & \textbf{0.851 (0.009)} & 0.758 (0.006) & 0.739 (0.009) & \underline{0.766 (0.005)} \\
       & F1         & 0.594 (0.019) & 0.603 (0.015) & 0.589 (0.015) & 0.627 (0.019) & \textbf{0.723 (0.027)} & 0.643 (0.023) & 0.627 (0.020) & \underline{0.659 (0.027)} \\
    \bottomrule
  \end{tabular}}
  \label{tab:appx-chronos-ft-wide}
\end{table*}

\subsubsection{Full vs.\ PEFT vs.\ Partial Fine-tuning}

\begin{table}[ht]
\centering
\caption{Classification fine-tuning results for Chronos-tiny on ZTF light curves with an \textbf{MLP head}. Same setup as \cref{tab:appx-finetune-linear}. Extends the condensed version reported as \cref{tab:finetune-mlp} in the main text. Bold indicates the best in each column.}
\label{tab:appx-finetune-mlp}
\begin{tabular}{lccccc}
\toprule
Mode & Trainable Params & Macro F1 & Accuracy & Macro Prec & Macro Rec \\
\midrule
Full FT     & 9.6M (100\%)   & 0.659 $\pm$ 0.026          & \textbf{0.778 $\pm$ 0.028} & \textbf{0.613 $\pm$ 0.034} & 0.766 $\pm$ 0.005          \\
LoRA        & 1.3M (13.2\%)  & \textbf{0.661 $\pm$ 0.015} & 0.772 $\pm$ 0.020          & \textbf{0.613 $\pm$ 0.018} & \textbf{0.767 $\pm$ 0.006} \\
LayerNorm   & 1.2M (12.4\%)  & 0.648 $\pm$ 0.016          & 0.774 $\pm$ 0.023          & 0.605 $\pm$ 0.023          & 0.766 $\pm$ 0.006          \\
\bottomrule
\end{tabular}
\end{table}

\begin{table}[ht]
\centering
\caption{Forecasting PEFT/partial fine-tuning of Chronos-tiny on ZTF light curves, evaluated via downstream classification on the resulting embeddings. Bold indicates the best in each row. The result aligns with the full fine-tuning results in \cref{tab:appx-chronos-ft-wide}. The downstream classification on embeddings from the forecasting fine-tuned models still underperforms the zero-shot baseline. These confirm that the issue lies in the forecasting objective itself, which is unsuitable for a model that does not treat the irregular sampling of our time series properly (see \cref{fig:tsfm_view}). Also see \cref{app:rope_irregular} for possible improvements on this issue.}
\label{tab:appx-forecast-ft}
\begin{tabular}{@{}llcccc@{}}
\toprule
\textbf{Classifier} & \textbf{Metric} & \textbf{Zero-shot} & \textbf{Full FT} & \textbf{LoRA} & \textbf{LayerNorm} \\
\midrule
\multirow{4}{*}{$k$-NN}
  & Accuracy  & 0.857 & 0.849 & \textbf{0.859} & 0.852 \\
  & Precision & \textbf{0.799} & 0.735 & 0.783 & 0.762 \\
  & Recall    & 0.623 & 0.603 & \textbf{0.628} & 0.618 \\
  & F1        & \textbf{0.672} & 0.632 & \textbf{0.672} & 0.654 \\
\midrule
\multirow{4}{*}{Logistic}
  & Accuracy  & \textbf{0.750} & 0.740 & \textbf{0.750} & 0.743 \\
  & Precision & \textbf{0.575} & 0.564 & 0.569 & 0.568 \\
  & Recall    & \textbf{0.730} & 0.711 & 0.724 & 0.725 \\
  & F1        & \textbf{0.617} & 0.601 & 0.610 & 0.608 \\
\bottomrule
\end{tabular}
\end{table}

\paragraph{Experiments Settings and Hyperparameters.}
Beside full fine-tuning, we further compare it with two other adaptation modes: LoRA ($r{=}8$, adapting the Q and V projections), and partial fine-tuning (only tuning the LayerNorms, following \citet{chen2024visionts}). 
Same as full fine-tuning, we test both forecasting and classification objectives. 
\begin{itemize}
    \item \textbf{Classification.} We pair each mode with either a linear head or an MLP head.
    Hyperparameters are selected via grid search over LR $\in \{10^{-3}, 10^{-4}, 10^{-5}\}$ and BS $\in \{32, 128, 256\}$, choosing the configuration with the highest macro-F1. Training uses early stopping with patience $5$ on validation loss. All metrics are reported over $10$ random seeds.
    \item \textbf{Forecasting.} HPO over LR $\in \{10^{-4}, 5{\times}10^{-4}, 10^{-3}\}$ and effective BS $\in \{32, 64, 128\}$ (gradient accumulation steps $=2$), selecting the best checkpoint by forecasting validation loss (early stopping with patience $10$). The downstream classifiers ($k$-NN with $k{=}5$ and logistic regression) are trained on the embeddings from each best checkpoint.
\end{itemize}

\paragraph{Result.}
See \cref{tab:appx-finetune-linear,tab:appx-finetune-mlp,tab:appx-forecast-ft}.

\subsection{Effect of Model Scale on Zero-Shot Performance}
\label{app:model_size}

\begin{table*}
\captionsetup{skip=1.5pt}
  \caption{\blue{Effect of model scale on downstream classification. Moirai benefits from scaling on both classifiers, and Chronos improves on logistic regression with scale. The gap to hand-crafted features narrows with increasing model size, though a meaningful gap remains even at the largest available scales.}}
  \centering
\begin{tabular}{@{}llccccc|ccc|c@{}}
    \toprule
    & & \multicolumn{5}{c|}{\textbf{\texttt{Chronos}}} & \multicolumn{3}{c|}{\textbf{\texttt{Moirai}}} & \multicolumn{1}{c}{\textbf{\texttt{HC features}}}\\
    \cmidrule(lr){3-7} \cmidrule(lr){8-10} \cmidrule(l){11-11}
    \textbf{Classifier} & \textbf{Metric}
      & \textbf{\texttt{Tiny}} & \textbf{\texttt{Mini}} & \textbf{\texttt{Small}} & \textbf{\texttt{Base}} & \textbf{\texttt{Large}}
      & \textbf{\texttt{Small}} & \textbf{\texttt{Base}} & \textbf{\texttt{Large}} & --- \\
    & \# Params & 8M & 20M & 46M & 200M & 710M & 14M & 91M & 311M & --- \\
    & Emb.\ dim & 512 & 768 & 1024 & 1536 & 2048 & 768 & 1536 & 2048 & 138 \\
    \midrule
    \multirow{4}{*}{$k$-NN}
       & Accuracy  & 0.857 & 0.859 & 0.848 & 0.849 & 0.850 & 0.809 & 0.824 & 0.849 & 0.881 \\
       & Precision & 0.799 & 0.769 & 0.775 & 0.751 & 0.785 & 0.662 & 0.709 & 0.722 & 0.818 \\
       & Recall    & 0.623 & 0.617 & 0.597 & 0.597 & 0.612 & 0.509 & 0.546 & 0.590 & 0.661 \\
       & F1        & 0.672 & 0.663 & 0.647 & 0.634 & 0.646 & 0.554 & 0.576 & 0.612 & 0.712 \\
    \midrule
    \multirow{4}{*}{Logistic}
       & Accuracy  & 0.750 & 0.766 & 0.764 & 0.783 & 0.794 & 0.705 & 0.751 & 0.780 & 0.838 \\
       & Precision & 0.575 & 0.587 & 0.580 & 0.588 & 0.602 & 0.544 & 0.558 & 0.598 & 0.663 \\
       & Recall    & 0.730 & 0.731 & 0.706 & 0.695 & 0.698 & 0.680 & 0.658 & 0.673 & 0.854 \\
       & F1        & 0.617 & 0.628 & 0.618 & 0.624 & 0.636 & 0.579 & 0.592 & 0.624 & 0.714 \\
    \bottomrule
  \end{tabular}
  \label{tab:model_scale}
\end{table*}

\begin{figure*}
  \centering
  \includegraphics[width=0.8\linewidth]{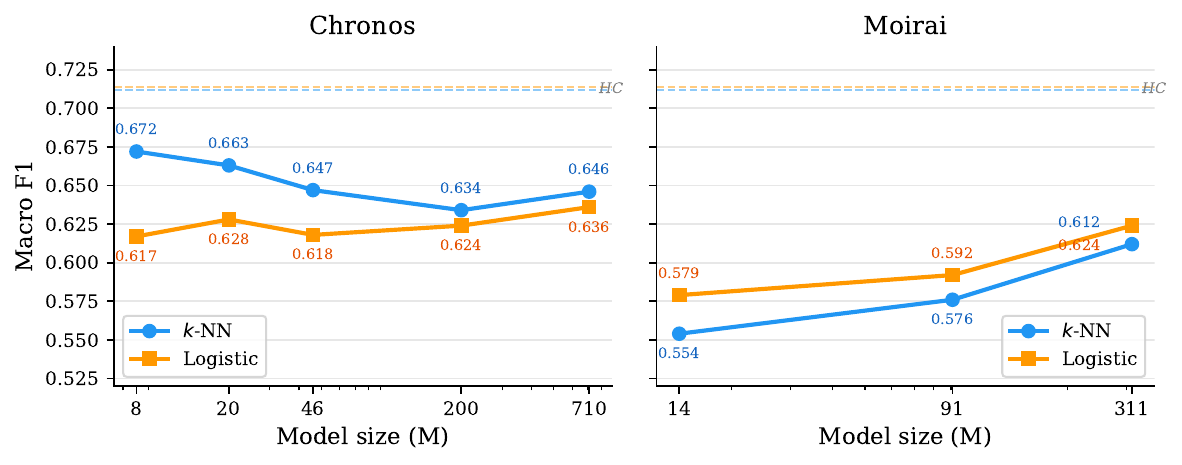}
  \caption{F1 score trend with increasing model size and hidden dimension.}
  \label{fig:scaling_f1}
\end{figure*}

\paragraph{Experiments Settings and Hyperparameters.}
We evaluate the full range of publicly available model sizes for both Chronos (Tiny/Mini/Small/Base/Large, $8$M--$710$M parameters) and Moirai (Small/Base/Large, $14$M--$311$M parameters) to test whether larger models narrow the gap to hand-crafted features. For each model size we extract zero-shot embeddings using the same protocol as our main benchmark and evaluate two deterministic downstream classifiers, $k$-NN and logistic regression. Hand-crafted features serve as the reference baseline.

\paragraph{Results.}
See \cref{tab:model_scale} and \cref{fig:scaling_f1}.

\subsection{Zero-Shot OOD Detection with Local Outlier Factor}
\label{app:ood_lof}

\begin{table*}[h]
\captionsetup{skip=1.5pt}
\centering
\setlength{\tabcolsep}{4pt}
\renewcommand{\arraystretch}{0.92}
\footnotesize
\caption{Purity of top N-percentile of sources selected by out-of-distribution source detection using Local Outlier Factor (LOF). We choose LOF because it has a different inductive bias from isolation forest. The best results are highlighted in \textbf{bold}, and the second-best are \underline{underlined}. We find that the chronos-bolt model continues to perform the best with chronos being a close second. With the LOF method, we also report much stronger performance for the best TSFM models when compared to the hand-crafted features, further strengthening the case to use TSFMs for OOD detection.}
\label{tab:OOD_LOF}
\begin{tabular}{
    >{\raggedright\arraybackslash}p{0.28\columnwidth}
    >{\centering\arraybackslash}p{0.18\columnwidth}
    >{\centering\arraybackslash}p{0.18\columnwidth}
    >{\centering\arraybackslash}p{0.18\columnwidth}
}
\toprule
 & \multicolumn{3}{c}{\textbf{Local Outlier Factor}}  \\
 & \textbf{Top 1\%} & \textbf{Top 5\%} & \textbf{Top 10\%} \\
\midrule
\texttt{Astromer-1}   & 0.108 & 0.167 & 0.203 \\
\texttt{Astromer-2}   & 0.236 & 0.220 & 0.187 \\
\texttt{Moirai}       & 0.281 & 0.232 & 0.184 \\
\texttt{Moirai-2}     & 0.079 & 0.122 & 0.126 \\
\texttt{Chronos}      & \textbf{\underline{0.989}} & \textbf{0.998} & \underline{0.992} \\
\texttt{Chronos-Bolt} & \textbf{\underline{0.989}} & \underline{0.996} & \textbf{0.997} \\
\texttt{Time-MoE}     & 0.157 & 0.198 & 0.179 \\
Random                & 0.116 & 0.116 & 0.116 \\
HC features           & 0.236 & 0.229 & 0.210 \\
\bottomrule
\end{tabular}
\end{table*}

\paragraph{Experiments Settings and Hyperparameters.}
To test the robustness of our OOD results, we introduce additional evaluations in this section. To complement the multi-class isolation forest method described in Section \ref{sec:results_ood}, we also test our embedding using Local Outlier Factor (LOF). We choose LOF because it imposes a different inductive bias than the isolation forest (IF) method. While IF is tree/partition-based, LOF is density-based and tests a different aspect of the embedding geometry. Our results in Table \ref{tab:OOD_LOF} confirm that, while the absolute purity values differ across IF and LOF, the relative trend is consistent across both OOD tests with the Chronos models performing the best on the OOD tests.

\subsection{TSFM Pre-training: Modified RoPE for Irregular Time Series}
\label{app:rope_irregular}

This subsection provides the full experimental setup and additional qualitative results supporting the modified-RoPE study reported in the main text (\cref{tab:sinusoidal_high_irreg}).

\paragraph{Experiments Settings and Hyperparameters.}
\textbf{Model.} We use \texttt{Moirai} as the base TSFM because it natively supports RoPE and input covariates.
\emph{Variants.} We evaluate three architectural variants of the positional treatment:
(1) \emph{RoPE + True timestamp} --- replacing sequential position indices with actual observation timestamps;
(2) \emph{$\Delta t$ as covariate} --- appending inter-observation time gaps as an additional input variate;
(3) \emph{Vanilla} --- neither modification.
\textbf{Data.} Each sample follows $x(t) = A\sin(2\pi f t + \varphi) + \varepsilon$ with $f \sim \mathcal{U}(0.5, 3.0)$, $A \sim \mathcal{U}(0.5, 2.0)$, $\varphi \sim \mathcal{U}(0, 2\pi)$, $\varepsilon \sim \mathcal{N}(0, 0.05)$. Each time series contains $160$ observations spanning $t \in [0, 10]$. The context window is $t < 7$ (${\sim}112$ points), the prediction window $t \geq 7$ (${\sim}48$ points), and the train/val/test split is $1{,}000/200/200$.
\textbf{Training.} LR $=5{\times}10^{-4}$ with cosine decay, batch size $128$, early stopping with patience $20$.
\textbf{Metrics.} Pearson correlation for shape recovery, plus $R^2$, MSE, and MAE for point-wise prediction accuracy.

\paragraph{Result.} See \cref{tab:appx-sinusoidal-high-irreg} and prediction examples in \cref{fig:appx-sin-pred-rope-only}.

\begin{table}[h]
\centering
\caption{Forecasting irregular sinusoidal time series with three architectural variants of MOIRAI's positional treatment. RoPE + True timestamp outperforms $\Delta t$ as covariate and Vanilla, indicating that preserving the shape of the irregular time series through positional encoding is the key. Extends the condensed comparison in \cref{tab:sinusoidal_high_irreg}; see prediction examples in \cref{fig:appx-sin-pred-rope-only}.}
\label{tab:appx-sinusoidal-high-irreg}
\begin{tabular}{lcccc}
\toprule
Variant & $R^2$ $\uparrow$ & Pearson $\uparrow$ & MSE $\downarrow$ & MAE $\downarrow$ \\
\midrule
RoPE + True timestamp    & $\mathbf{0.188 \pm 0.102}$ & $\mathbf{0.351 \pm 0.114}$ & $\mathbf{0.041 \pm 0.006}$ & $\mathbf{0.159 \pm 0.017}$ \\
$\Delta t$ as covariate  & $0.044 \pm 0.035$          & $0.277 \pm 0.058$          & $0.049 \pm 0.002$          & $0.177 \pm 0.006$ \\
Vanilla                  & $0.016 \pm 0.039$          & $0.202 \pm 0.098$          & $0.050 \pm 0.002$          & $0.183 \pm 0.007$ \\
\bottomrule
\end{tabular}
\end{table}

\begin{figure*}
  \centering
  \includegraphics[width=0.9\linewidth]{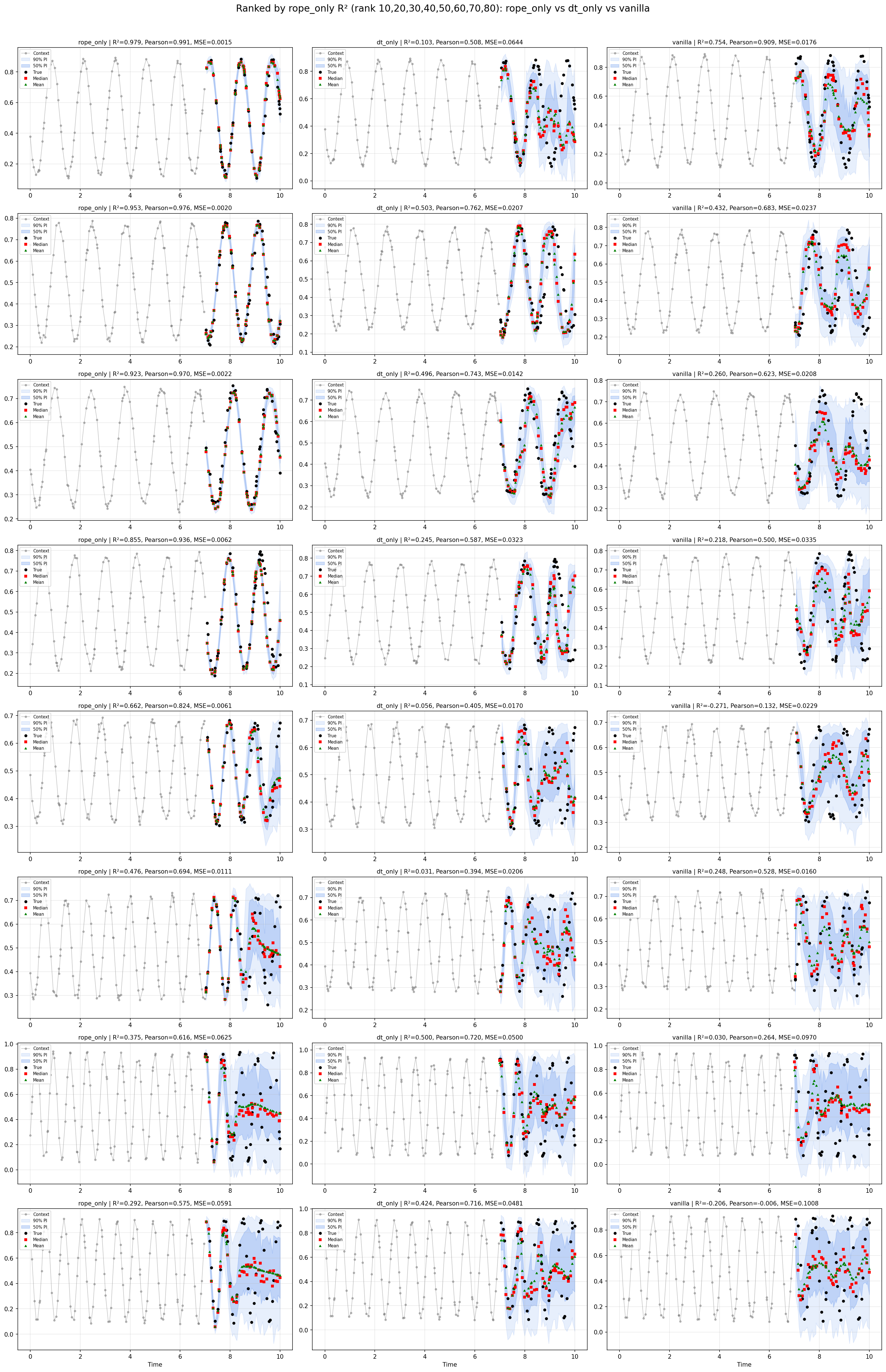}
  \caption{Predictions sorted by RoPE-only $R^2$ (ranks 10--80 out of 200), spanning from successful to failed predictions.}
  \label{fig:appx-sin-pred-rope-only}
\end{figure*}

\clearpage

\section{Extended Analysis of Results and Experiment Details}
\label{app:extended_analysis}

\subsection{Attention Analysis}
\label{app:attn_score}

\begin{figure*}[h]
  \centering
  \includegraphics[width=0.8\linewidth]{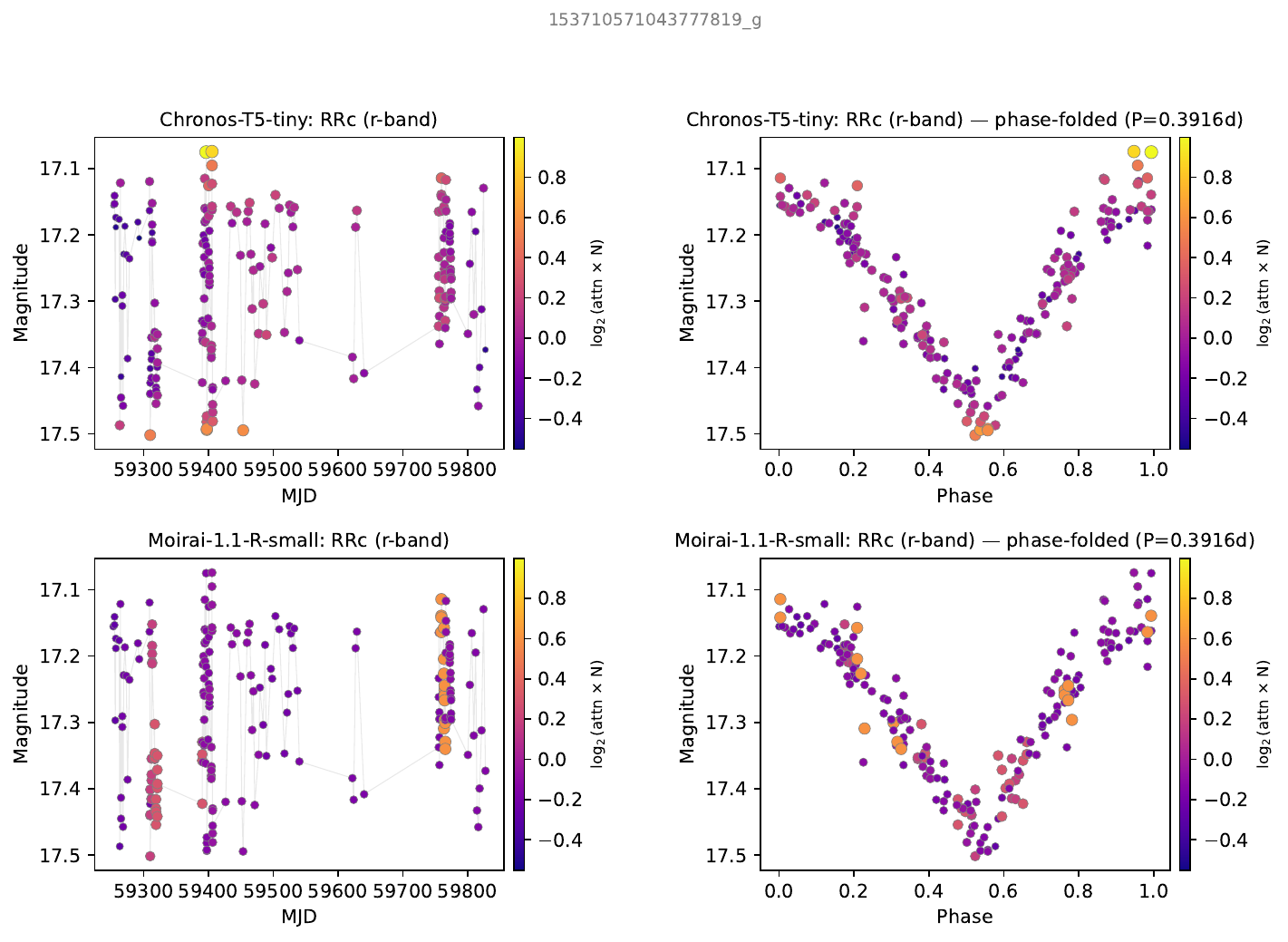}
  \caption{Chronos concentrates attention on brightness extrema in the phase-folded light curve, despite receiving only raw, irregularly sampled magnitudes without period information (we train a Chronos-embedding linear regressor and get $R^2 < 0.13$ for the period feature). \textbf{This indicates Chronos succeeds in attending to morphologically distinctive segments in the raw light curve (extreme values, rapid brightness changes) that coincide with physically meaningful phases.} Moirai, in contrast, distributes attention nearly uniformly or assigns high scores to non-distinctive segments. This stems from its patching mechanism: grouping consecutive observations into patches (16 time steps) conflates observations separated by irregular time gaps, destroying local morphological structure. See \cref{fig:attention_score_6} for other six classes.}
  \label{fig:attention_socre_1}
\end{figure*}

\begin{figure*}[h]
  \centering
  \includegraphics[width=0.48\textwidth]{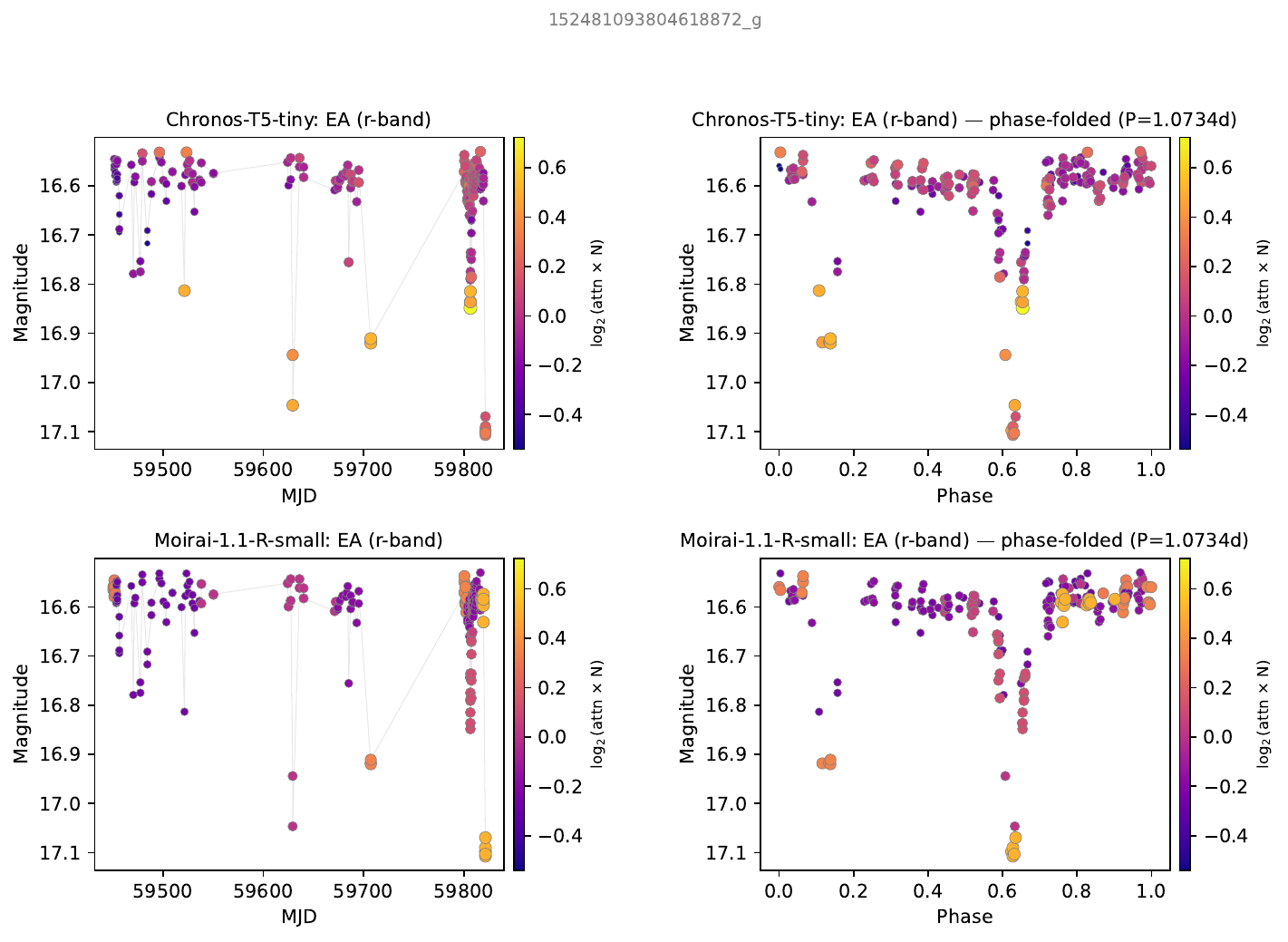}\hfill
  \includegraphics[width=0.48\textwidth]{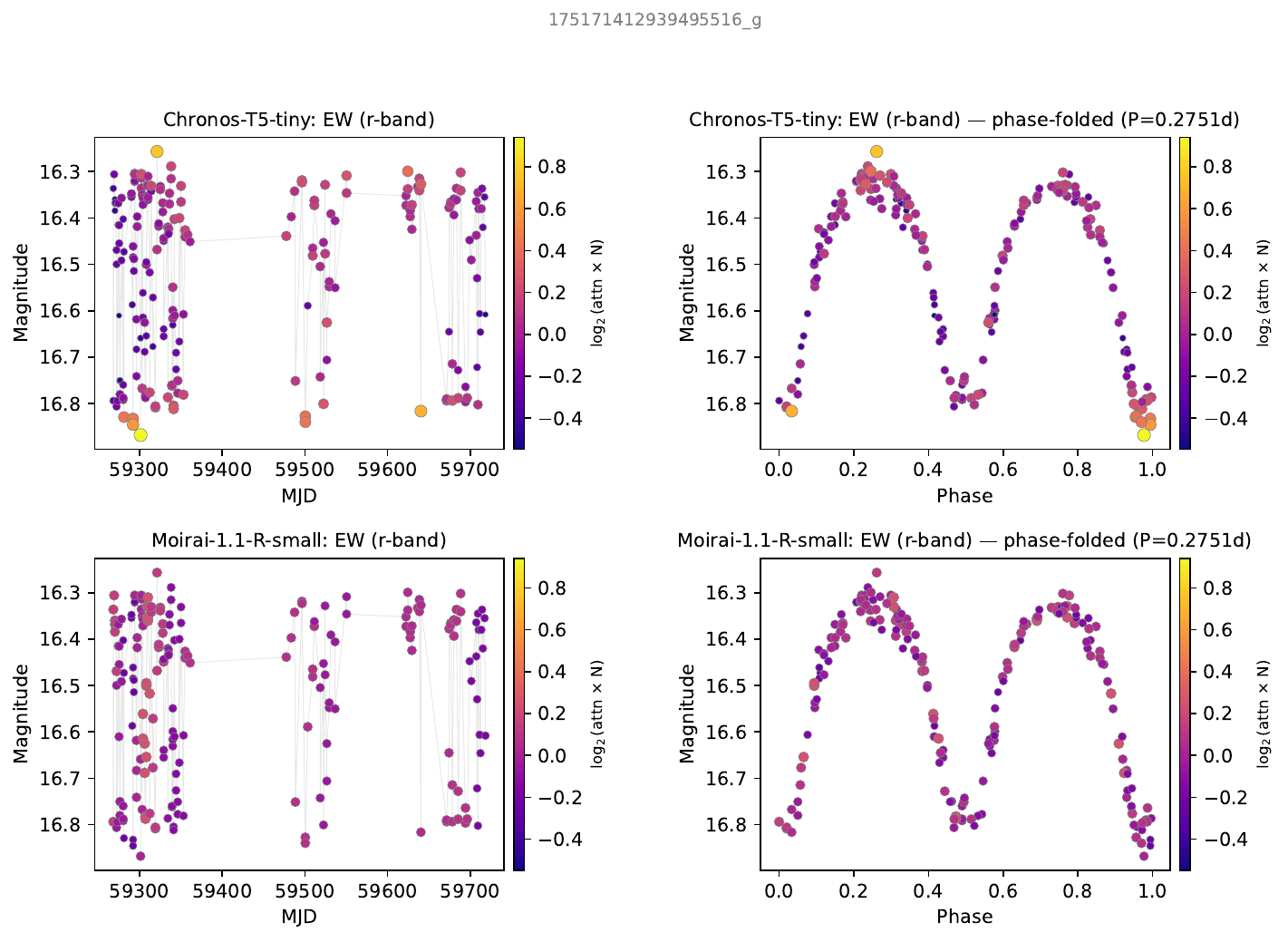}\\[0.5em]
  \includegraphics[width=0.48\textwidth]{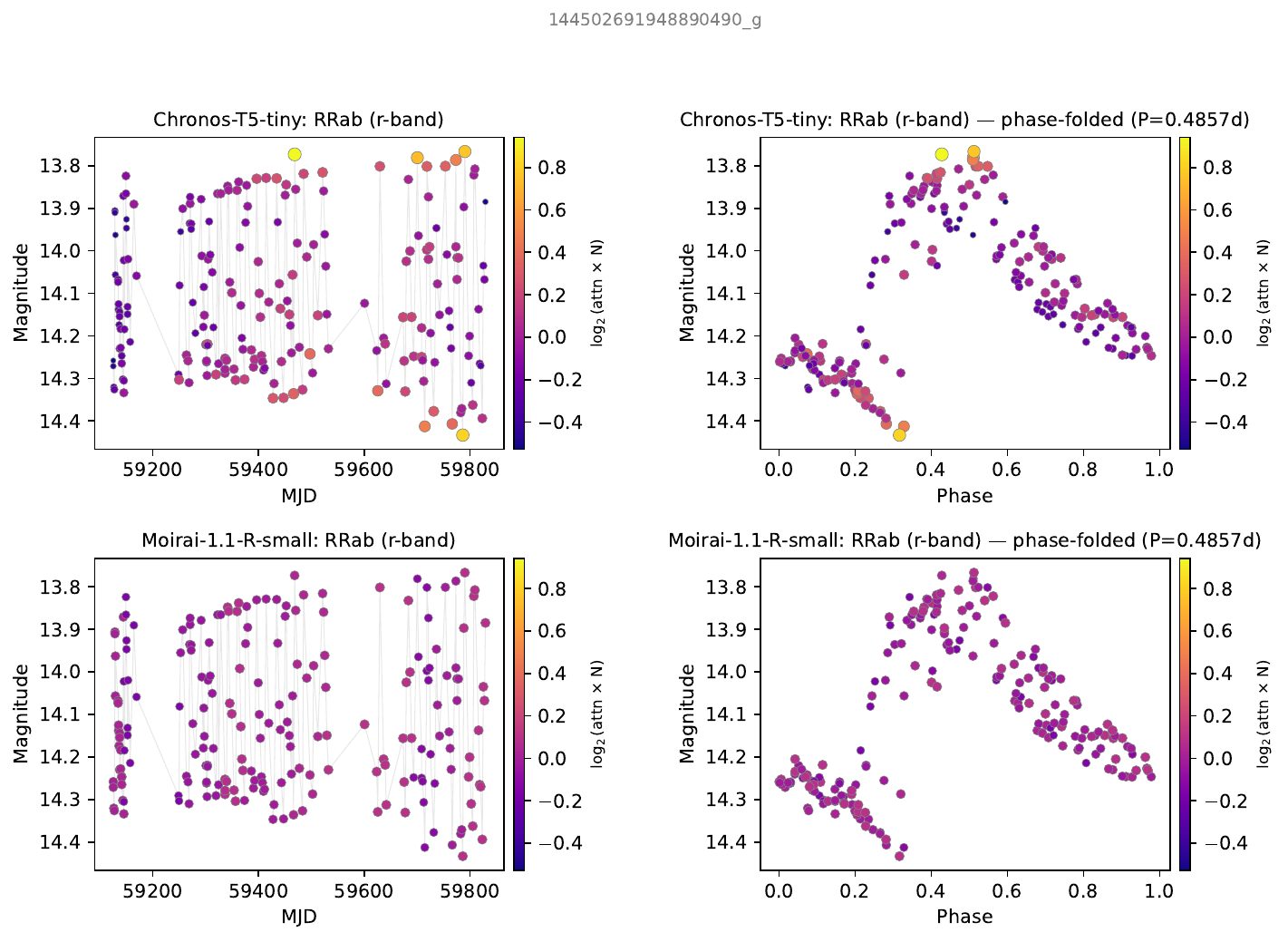}\hfill
  \includegraphics[width=0.48\textwidth]{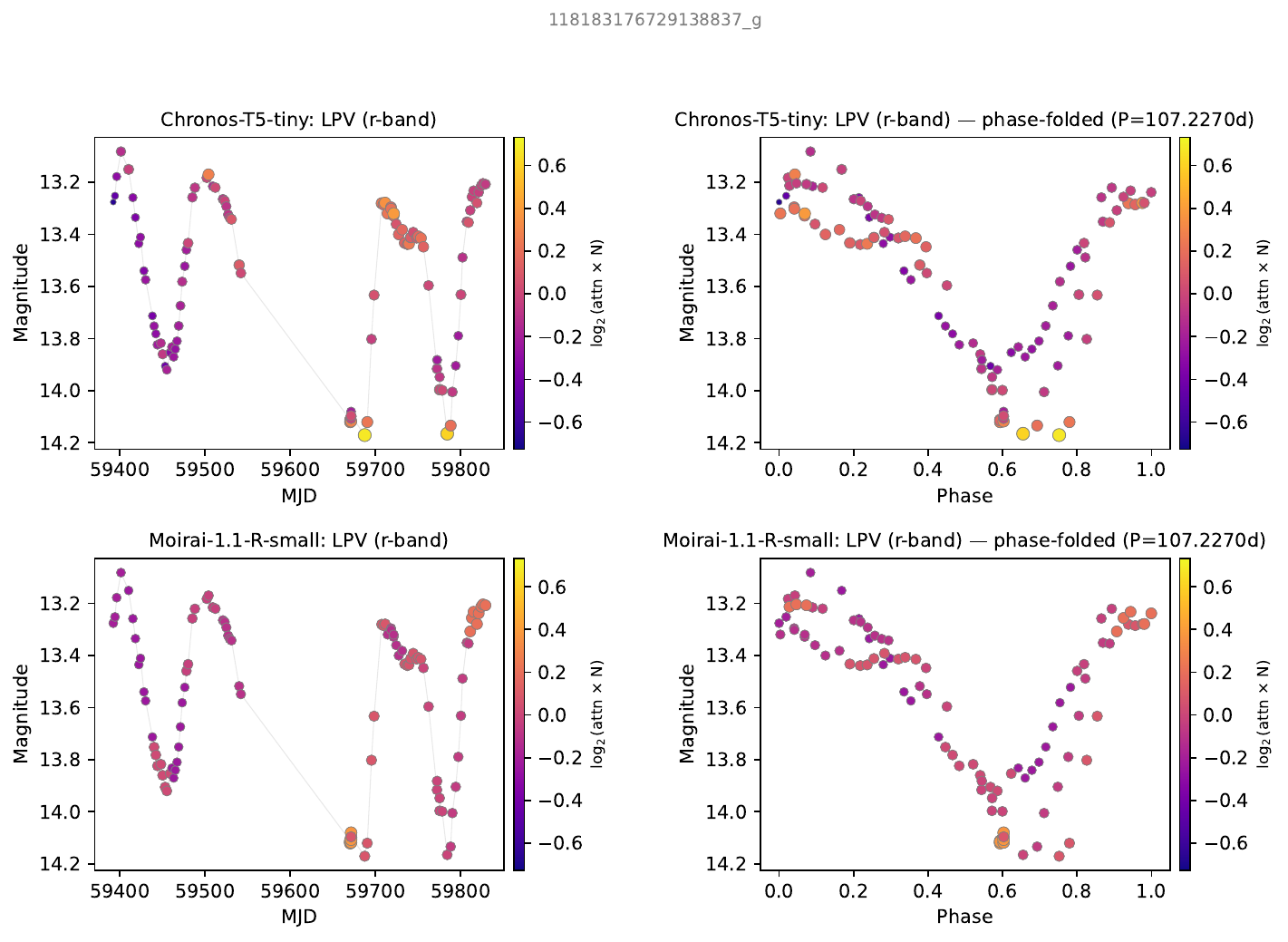}\\[0.5em]
  \includegraphics[width=0.48\textwidth]{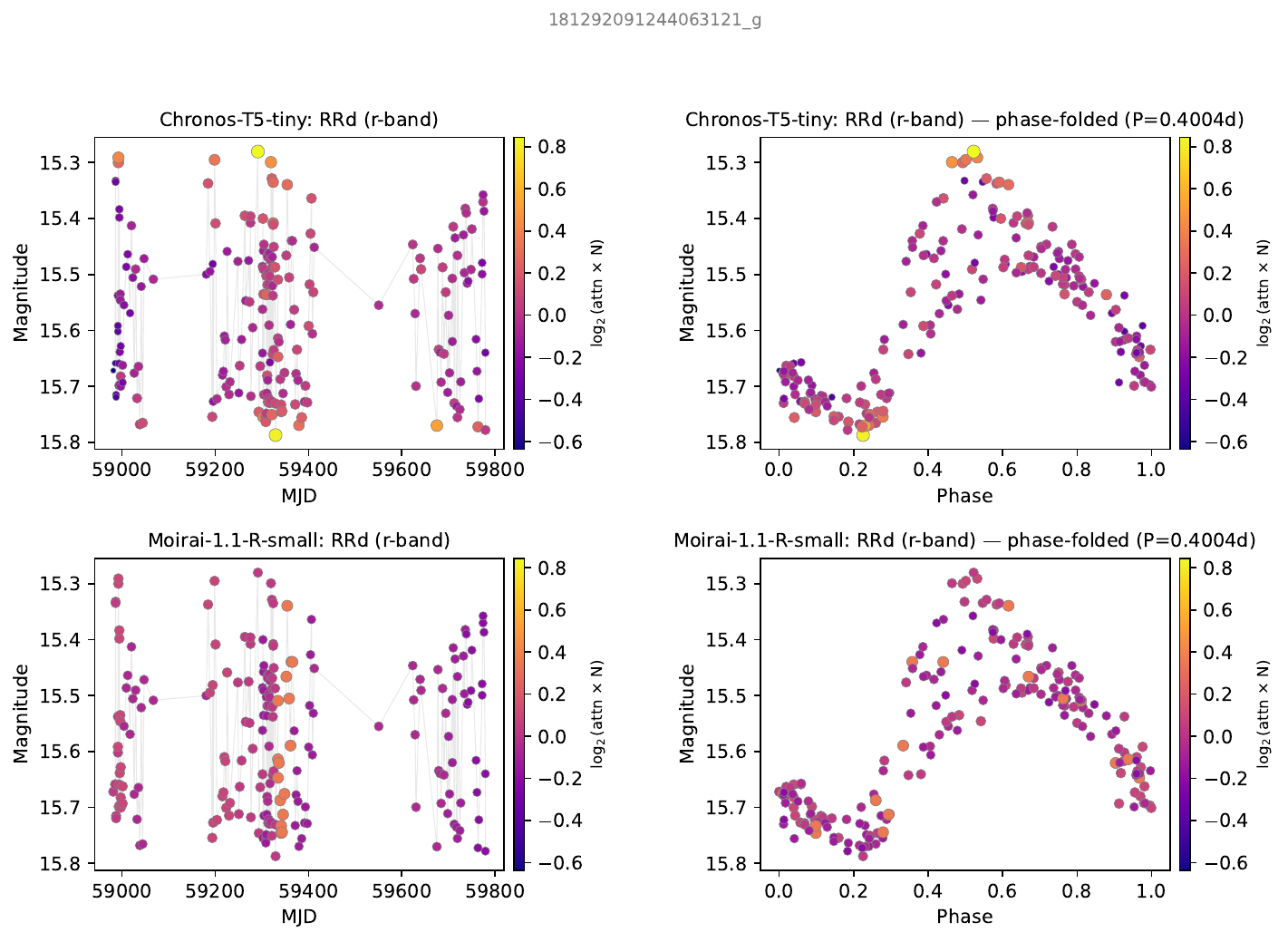}\hfill
  \includegraphics[width=0.48\textwidth]{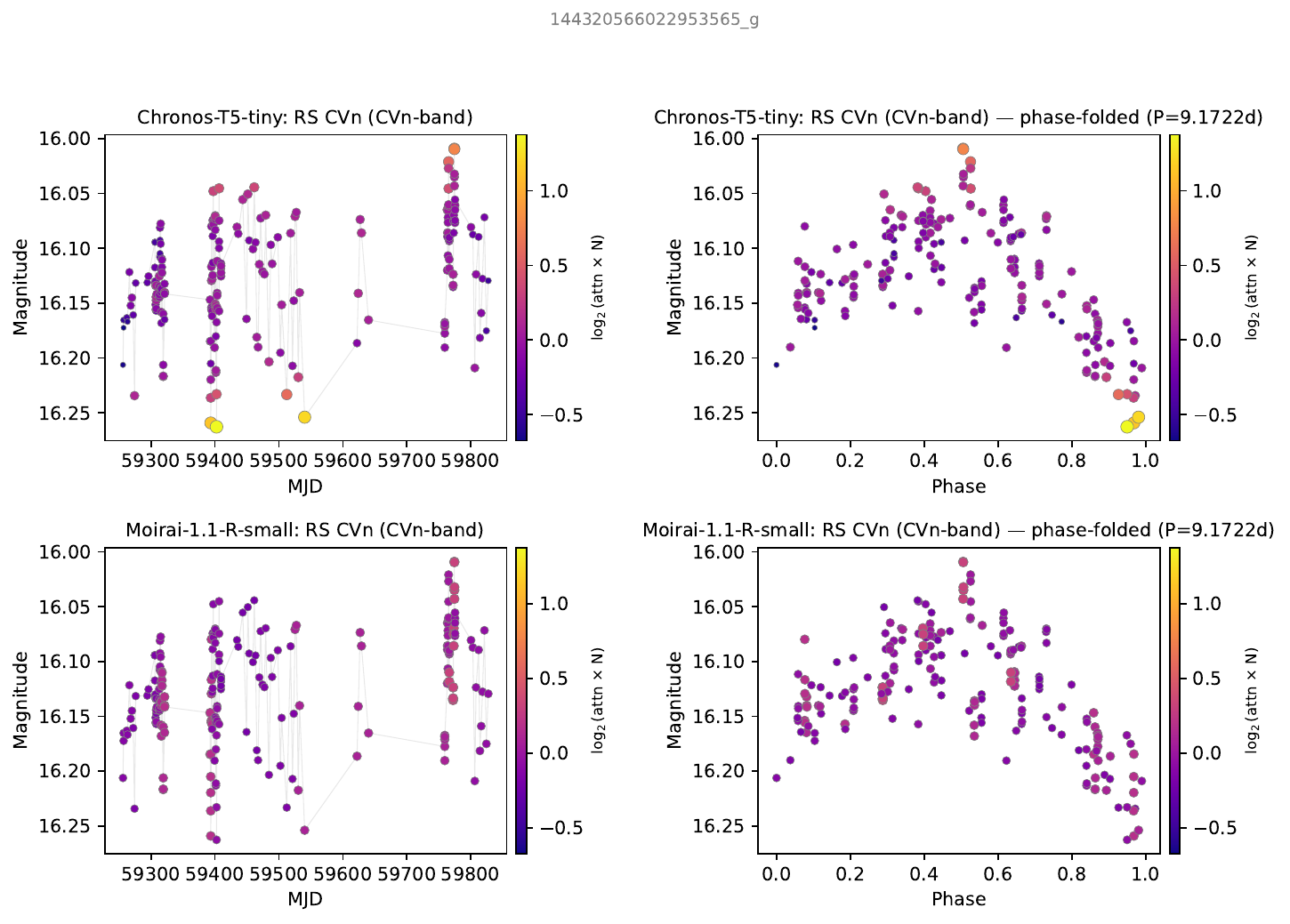}
  \caption{Attention rollout visualizations for the other six classes (EA, EW, RRab, LPV, RRd, RS~CVn). In each subplot, the upper row is \texttt{Chronos}, and the lower row is \texttt{Moirai}.}
  \label{fig:attention_score_6}
\end{figure*}

\paragraph{Attention Rollout Setup.}
We visualize attention patterns of \texttt{Chronos-tiny} and \texttt{Moirai-small} via attention rollout~\citep{abnar2020quantifying}. At the $\ell$-th layer, the attention matrix $A_\ell$ is averaged across heads and blended with the identity matrix ($0.5 \cdot A_\ell + 0.5 \cdot I$) to account for residual connections. The blended matrices are then multiplied across all layers to yield one importance score $attn$ per observation. We report $\log_2(N \cdot attn)$, where zero indicates averaged attention (every observation gets equal attention), positive values indicate above-average attention, and negative values indicate below-average attention.

\subsection{TSFMs Embedding Visualization (UMAP)}
\label{app:umap}

We include the UMAP visualizations for the embeddings from each embedding model to provide more intuition regarding the embedding space. As shown by \cref{fig:umap}, all time-series pre-trained models' embeddings show clear distinction and distribution of different clusters corresponding to different ground-truth classes. In comparison, as a baseline, the random embeddings show no clear clusters at all. \texttt{Astromer-1}'s embeddings and hand-crafted features do not show clear clusters either. \texttt{Astromer-2}'s embeddings show clearer cluster distribution but for some classes, the clusters are not distinctive from others either. These UMAP visualizations further demonstrate the promising potential of using time-series pre-trained models as light-curve embedding models.

\begin{figure}
    \centering
    \includegraphics[width=0.32\linewidth]{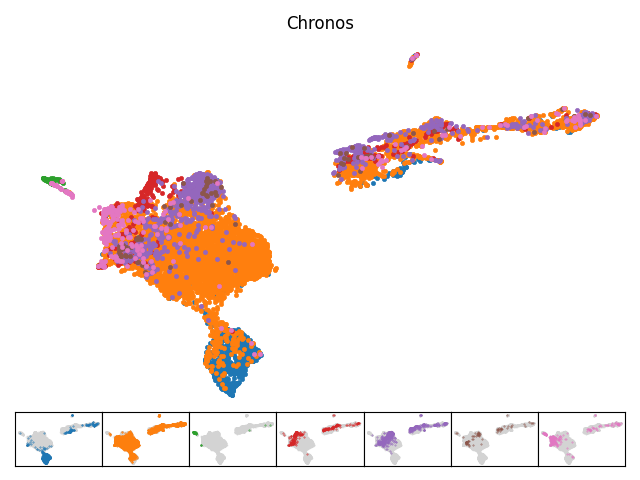}
    \includegraphics[width=0.32\linewidth]{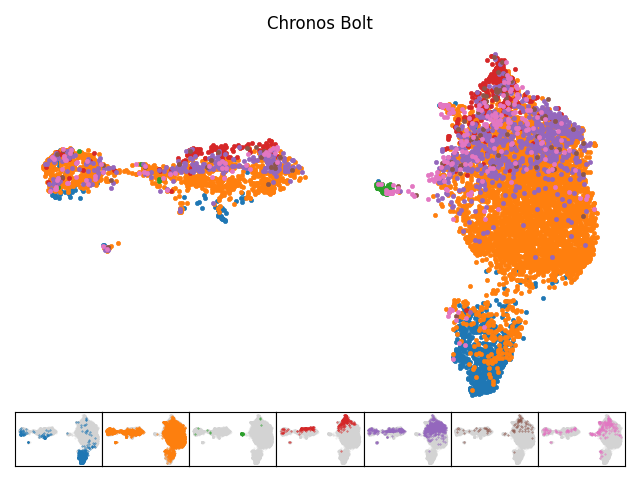}
    \includegraphics[width=0.32\linewidth]{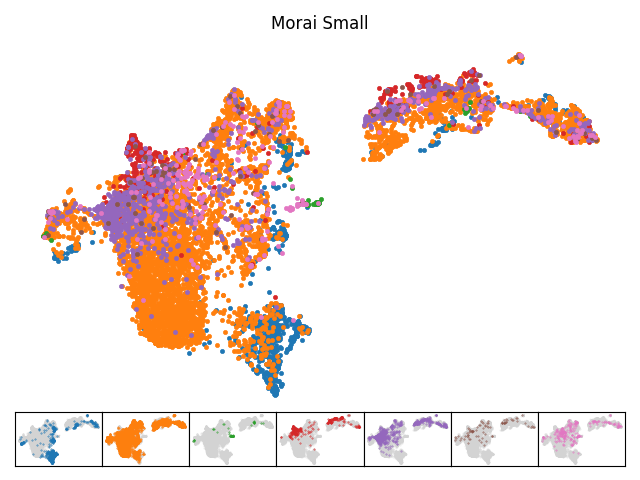}
    \includegraphics[width=0.32\linewidth]{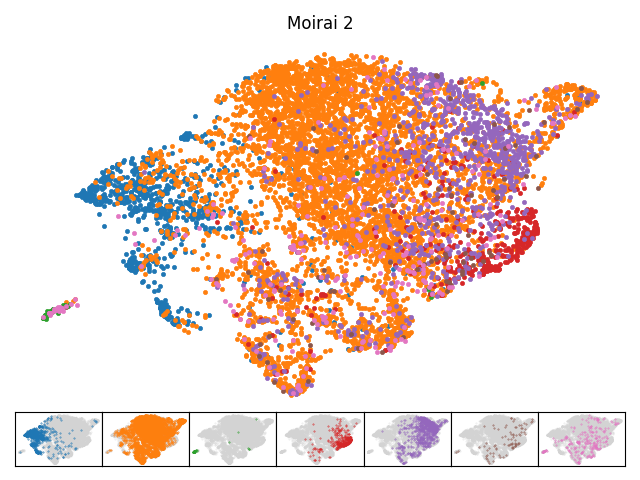}
    \includegraphics[width=0.32\linewidth]{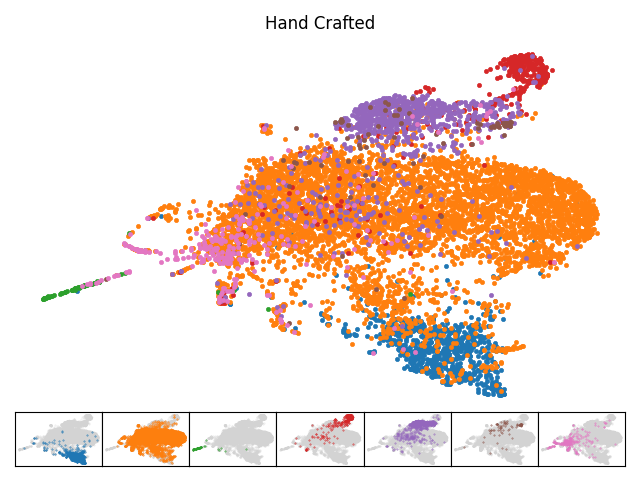}
    \includegraphics[width=0.32\linewidth]{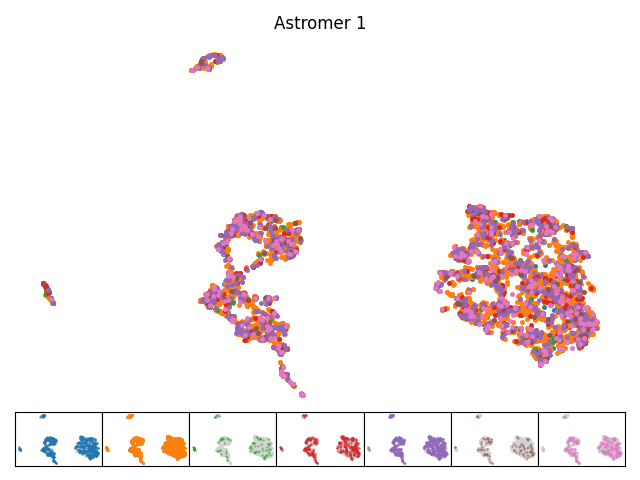}
    \includegraphics[width=0.32\linewidth]{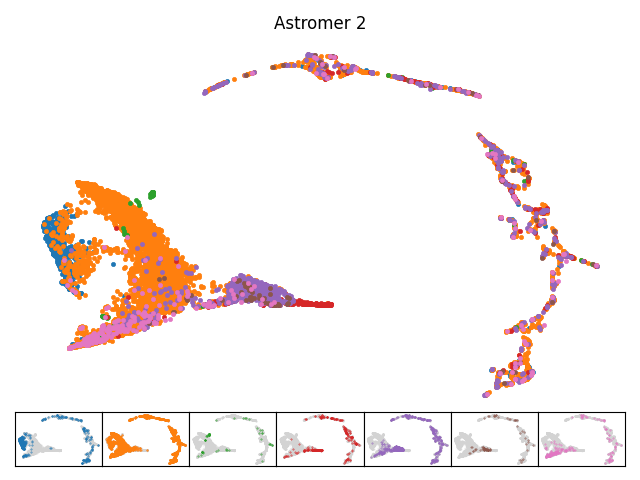}
    \includegraphics[width=0.32\linewidth]{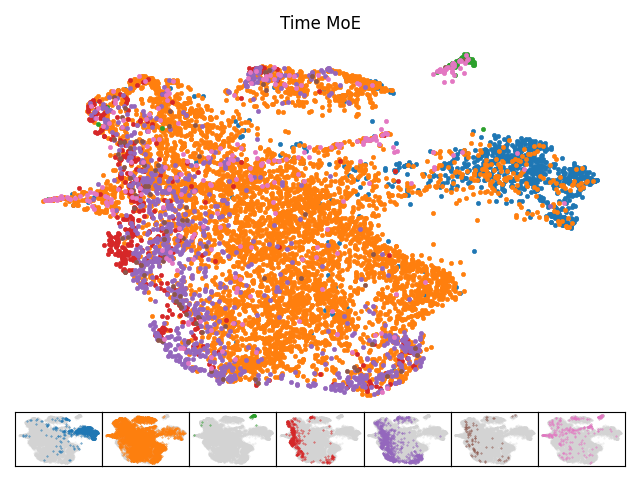}
    \includegraphics[width=0.27\linewidth]{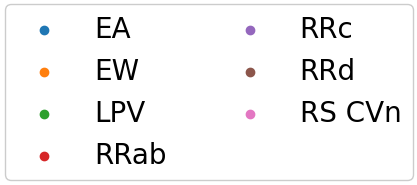}
    \caption{UMAP projections for each embedding model included in our analysis using the test set. Inset plots at the bottom of each figure show clustering of different classes.}
    \label{fig:umap}
\end{figure}

\subsection{\texttt{Astromer} Embedding Analysis}
\label{app:astromer_1}

\begin{figure}[t]
    \centering
    \minipage{0.49\textwidth}
        \includegraphics[width=\linewidth]{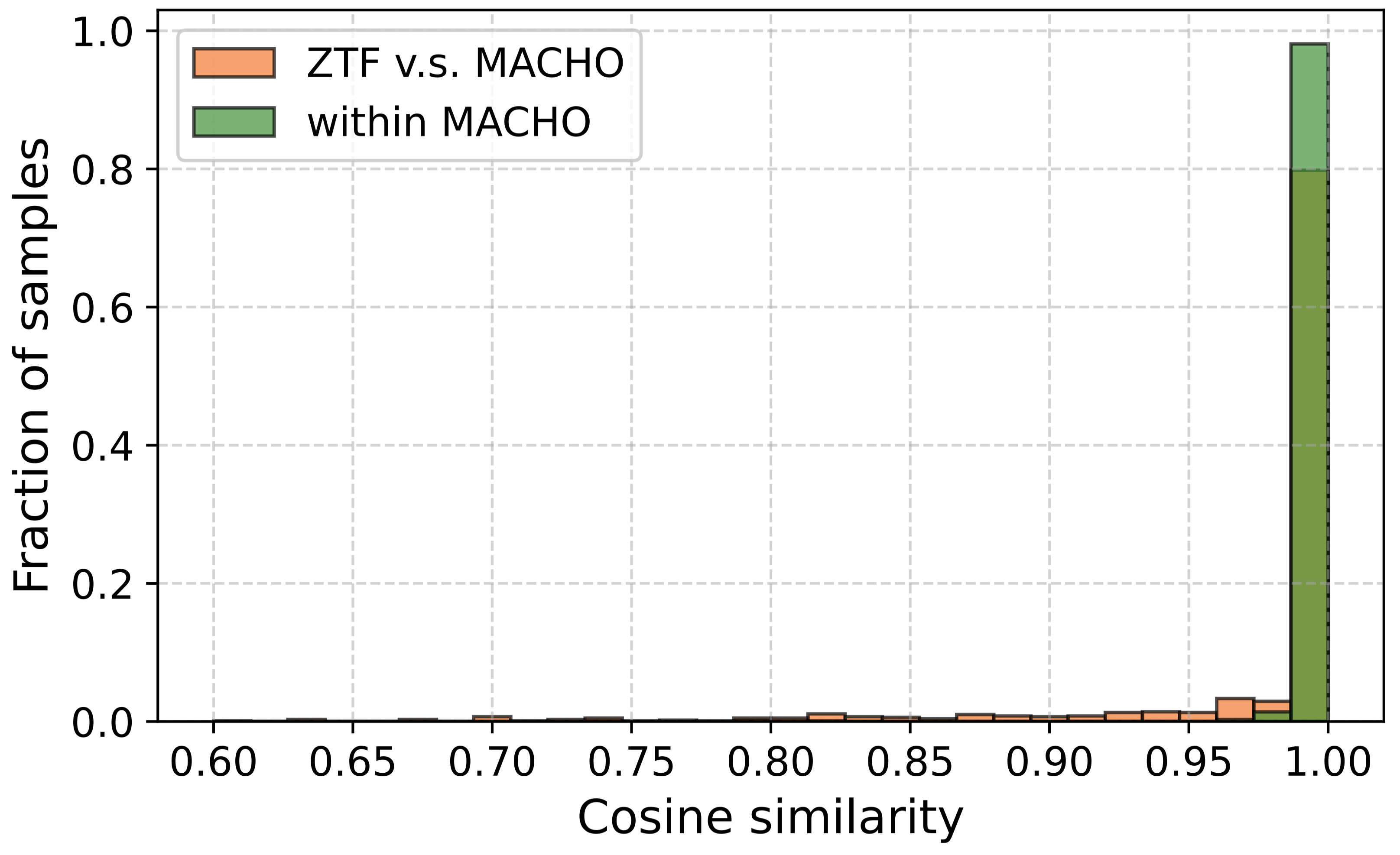}
    \endminipage
    \minipage{0.49\textwidth}
        \includegraphics[width=\linewidth]{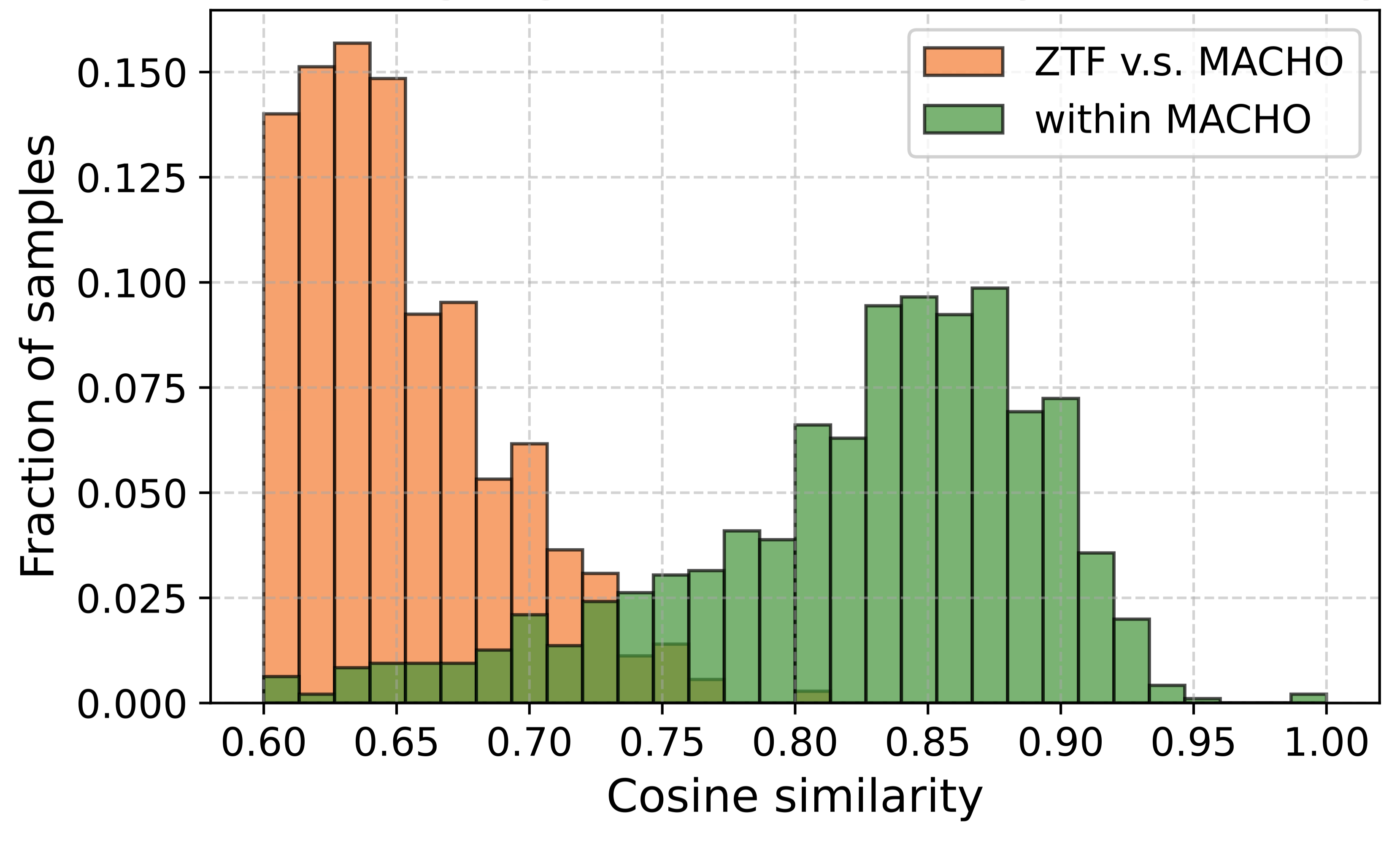}
    \endminipage
    \caption{\blue{\textbf{Comparison of embedding alignment for \texttt{Astromer-1} (left) and \texttt{Chronos} (right).}
    The plots show the distribution of cosine similarities between light-curve embeddings, both within the MACHO survey (green) and across surveys (ZTF vs.\ MACHO, orange).
    \texttt{Astromer-1} exhibits embedding collapse, with cosine similarities approaching $1.0$ no matter within- or cross-survey. This indicates that the model encodes little discriminative structure.
    In contrast, \texttt{Chronos} produces more meaningful embeddings: the wider distribution of cosine similarities preserves structural information, and the clear separation between within-survey and cross-survey pairs demonstrates its ability to capture the domain shift between datasets.}}
    \label{fig:cos_sim}
\end{figure}

\blue{We provide detailed analysis and additional experiments on the issue of the poor performance of \texttt{Astromer-1}.
First, \texttt{Astromer-1} needs further fine-tuning on the downstream-task dataset to achieve good performance on the variable-star classification task, according to \citet{donoso2023astromer}.
This is evidenced in their Fig.~11(a), which shows a clear increase of F1 score from $\sim$0.25 to $\sim$0.6 when fine-tuned on 20 to 500 variable stars per class of the MACHO dataset.
A similar trend is observed for other datasets, including OGLE-III and ATLAS. \texttt{Astromer-2}~\citep{astromer2} also shows improvement with fine-tuning, though its performance starts higher (around 0.65) even with just 20 samples per class. Please refer to Figures~11 and 12 of \citet{astromer2} for details.
These results indicate that \texttt{Astromer}'s low zero-shot performance in our benchmark is expected, as we intentionally evaluate the pre-trained checkpoints without any task-specific tuning.}

\blue{Second, we provide empirical analysis to show that \texttt{Astromer-1} embeddings collapse into similar directions.
Specifically, we randomly sample 1000 pairs of ZTF--MACHO data in the r-band, and compute the cosine similarity of embeddings from two models, \texttt{Astromer-1} and \texttt{Chronos}.
For comparison, we do the same for within-MACHO pairs.
The result is in \cref{fig:cos_sim}.
For \texttt{Astromer-1}, it is clear that its embeddings collapse to one direction. Even the 10th percentile of cosine similarity of embeddings within MACHO is 0.995, meaning every star is almost parallel to every other.
Even under the survey-data shift, the 10th percentile is still 0.948.
From the above we conclude that the embeddings of frozen \texttt{Astromer-1} encode very little discriminative structure. This explains why a downstream classifier has a hard time distinguishing different classes (low F1 score).
For \texttt{Chronos} (right figure), the cosine similarity within MACHO (green) shows a wide range of angles, indicating the embeddings preserve class information.
Furthermore, the embedding has a clear domain shift: the cosine similarity of ZTF--MACHO pairs (orange) is lower than that of within-MACHO pairs.
Unlike frozen \texttt{Astromer-1}, \texttt{Chronos} does not collapse everything into a single direction.
It still has room to spread out unseen patterns instead of forcing them into the old manifold.}

\subsection{Importance Analysis on Hand-crafted Features}
\label{app:HCF_feat_importance}
In this appendix we investigate the feature importance of the individual hand-crafted features to investigate the source of their excellent performance. We perform this analysis on the random forest trained for supervised classification with the hand-crafted features because it offers built-in functionality for quantitatively measuring the importance of each feature. The feature importance is computed using the mean decrease in impurity. In this framework, each time a feature is used to split a node, the associated reduction in the Gini impurity metric is recorded and weighted by the number of samples reaching that node. These weighted impurity decreases are then summed over all nodes and all trees in the ensemble, and finally normalized to yield a global, unit-scaled measure of how strongly each feature contributes to improving class separation within the forest.

\begin{figure}
    \centering
    \includegraphics[width=0.8\linewidth]{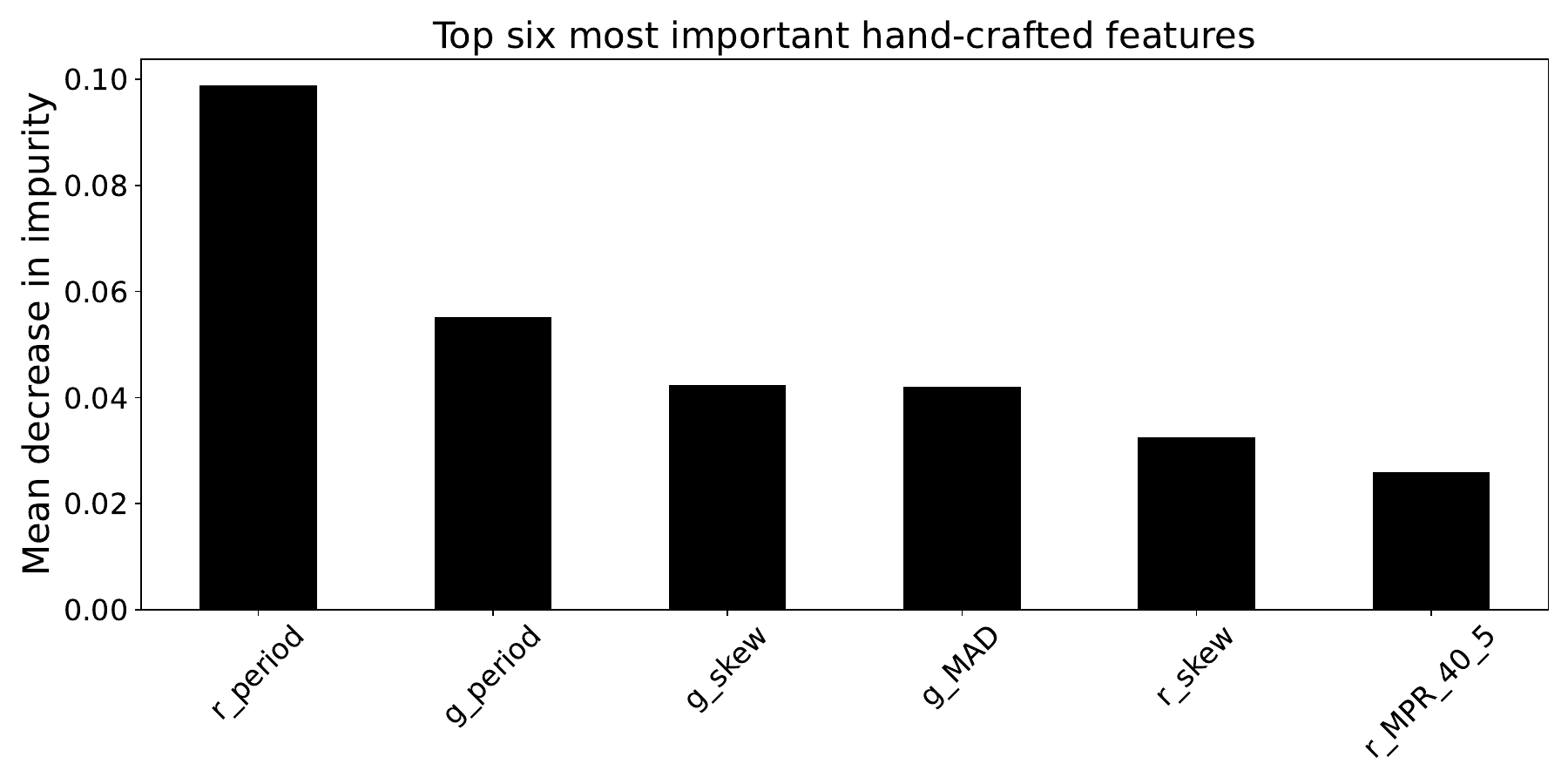}
    \caption{Top six most important hand-crafted features for random forest classification. The period of variability computed for either passband rank the highest. Other important metrics also include the skewness of the distribution of magnitudes, the median absolute deviation, and the magnitude percentage ratio for the 40th and 5th percentiles.}
    \label{fig:hcf_importance}
\end{figure}

Figure~\ref{fig:hcf_importance} shows the results of our feature importance analysis. We find that the period inferred from the $g$ and $r$ bands of the time series are the most important features driving the performance of the hand-crafted features. This aligns with expectations as numerous other studies performing variable star classification with tree-based classifiers come to a similar conclusion \cite{Richards+2011, Dubath+2011, Kim+2016}. We also find that the skewness, median absolute deviation, and the magnitude percentage ratio also rank very highly. These features encode information about the spread of the magnitude measurements, indirectly representing the morphology of the periodicity of the star. All features are described in Appendix~\ref{app:handcrafted}.

\subsection{Hyperparameters for Classification}
\label{app:hyperparameters}
\begin{table*}
  \caption{Best hyper-parameters for each model with the MLP classifier}
  \resizebox{\textwidth}{!}{%
  \begin{tabular}{@{}l l p{0.18\textwidth}@{}}
    \toprule
    \textbf{Method} & \textbf{Hyperparameters} & \textbf{Training epochs} \\
    \midrule
    \texttt{Astromer-1}           & \texttt{batch\_size=32, learning\_rate=0.0001, dropout=0.0} & 17 \\
    \texttt{Astromer-2}           & \texttt{batch\_size=32, learning\_rate=0.0001, dropout=0.0} & 23 \\
    \texttt{Moirai}        & \texttt{batch\_size=64, learning\_rate=0.001, dropout=0.0} & 11 \\
    \texttt{Moirai-2}        & \texttt{batch\_size=32, learning\_rate=0.0001, dropout=0.1} & 5 \\
    \texttt{Chronos}        & \texttt{batch\_size=32, learning\_rate=0.0001, dropout=0.0} & 26 \\
    \texttt{Chronos-Bolt}   & \texttt{batch\_size=128, learning\_rate=0.0001, dropout=0.1} & 17 \\
    \texttt{Time-MoE}   & \texttt{batch\_size=256, learning\_rate=0.0001, dropout=0.1} & 14 \\
    Random Embeddings   & \texttt{batch\_size=128, learning\_rate=0.0001, dropout=0.0}   & 5 \\
    Hand-crafted features  & \texttt{batch\_size=32, learning\_rate=0.0001, dropout=0.1}   & 30 \\
    \bottomrule
  \end{tabular}}%
  \label{tab:mlp_hyperparams}
\end{table*}

\begin{table*}
  \caption{Best hyperparameters for each model with the random forest classifier}
  \resizebox{\textwidth}{!}{
  \begin{tabular}{@{}l l p{0.18\textwidth}@{}}
    \toprule
    \textbf{Method} & \textbf{Hyper-parameters} & \textbf{Training time (s)} \\
    \midrule
    \texttt{Astromer-1}            & \texttt{max\_depth=10, min\_samples\_split=2, n\_estimators=200} & 84 \\
    \texttt{Astromer-2}           & \texttt{max\_depth=30, min\_samples\_split=10, n\_estimators=500} & 456 \\
    \texttt{Moirai}        & \texttt{max\_depth=None, min\_samples\_split=2, n\_estimators=500} & 738 \\
    \texttt{Moirai-2}        & \texttt{max\_depth=None, min\_samples\_split=2, n\_estimators=100} & 138 \\
    \texttt{Chronos}        & \texttt{max\_depth=None, min\_samples\_split=5, n\_estimators=100} & 126 \\
    \texttt{Chronos-Bolt}   & \texttt{max\_depth=30, min\_samples\_split=2, n\_estimators=500} & 450 \\
    \texttt{Time-MoE}   & \texttt{max\_depth=30, min\_samples\_split=5, n\_estimators=500} & 252 \\
    Random Embeddings   & \texttt{max\_depth=None, min\_samples\_split=2, n\_estimators=100}   & 198 \\
    Hand-crafted features  & \texttt{max\_depth=None, min\_samples\_split=10, n\_estimators=100}   & 36.4 \\
    \bottomrule
  \end{tabular}}
  \label{tab:rf_hyperparams}
\end{table*}

\blue{
We report the hyperparameter tuning process and summary for all classifiers in this section.
To fairly compare the different embeddings, we conduct a hyperparameter search on each model when training the downstream MLP and random forest classifier.
We use MLP with three hidden layers of sizes 1024, 512 and 256 and an output layer for class predictions.
we search over batch size $B \in \{128,256,512,1024\}$, learning rate $lr \in \{0.01, 0.001, 0.0001\}$ and dropout rate $\in \{0.0, 0.1\}$.
Every hyperparameter triple runs once on NVIDIA H100 4 GPUs.
The training process is at most 50 epochs, and stops early if the validation loss fails to improve for 3 epochs. In 
practice this training takes less than 30 epochs for all models before the early stopping is triggered.
For random forest, we search over maximum depth of the tree $\in \{\text{None}, 10, 20, 30 \}$, the minimum number of samples to split an internal node $\in \{2,5,10\}$, and number of estimator $\in \{100,200,500\}$.
We summarize the hyperparameters of the MLP and random forest classifiers in   Table~\ref{tab:mlp_hyperparams} and ~\ref{tab:rf_hyperparams}.}
\blue{For linear classifier, since the current training set is relatively small, we use the \texttt{LogisticRegression} (L-BFGS) from Scikit-learn library \citep{sklearn_api}, with \texttt{max\_iter = 5000}, \texttt{class\_weight = "balanced"}, and all other with default settings. 
L-BFGS is a deterministic full-batch method that converges to the global minimizer without learning rate tuning.
For $k$-NN, we use \texttt{KNeighborsClassifier} from Scikit-learn with default settings. Since $k$-NN and the default solver of logistic regression are both deterministic, we only report the 1 run result. }

\subsection{Computational Analysis}
\label{app:flops}

\begin{table}
\centering
\caption{FLOPs per forward pass (batch size = 1), with $L{=}200$ tokens and patch size $P{=}1$ applied uniformly to all models. Each multiply--add is counted as 2 FLOPs. For encoder--decoder models (Chronos-tiny, Chronos-Bolt-tiny), attention linear and quadratic columns aggregate self-attention in both encoder and decoder plus cross-attention in the decoder; the FFN column aggregates both stacks. For Time-MoE-50M, only \emph{active} FLOPs are reported (3 of 9 experts activated per layer); the total dense-equivalent would be $\approx 61$\,GFLOPs. Quadratic attention FLOPs scale as $O(L^2 d)$; FFN FLOPs scale as $O(L \cdot d \cdot d_\text{ff})$. ASTROMER-1/2 already operated at $L{=}200$, $P{=}1$ and are unchanged. Under these conditions, Chronos-Bolt-tiny and Chronos-tiny share the same transformer backbone ($d{=}256$, $d_\text{ff}{=}1024$, $4+4$ layers) and thus yield identical FLOPs; the sole original difference---input patching---is eliminated when $P{=}1$. Totals reflect main transformer operations only (attention $+$ FFN); auxiliary components such as positional encodings and normalisation layers contribute negligibly.}
\label{tab:flops_L200}
\setlength{\tabcolsep}{6pt}
\begin{tabular}{l r r r r}
\toprule
\multirow{2}{*}{\textbf{Model}} &
  \multicolumn{2}{c}{\textbf{Attention (MFLOPs)}} &
  \multirow{2}{*}{\shortstack{\textbf{FFN}\\\textbf{(MFLOPs)}}} &
  \multirow{2}{*}{\shortstack{\textbf{Total}\\\textbf{(GFLOPs)}}} \\
\cmidrule(lr){2-3}
 & Linear & Quadratic & & \\
\midrule
ASTROMER-1                          &     209.7 &    41.0 &      52.4 & $\mathbf{0.30}$  \\
ASTROMER-2                          &     629.1 &   123.0 &     157.3 & $\mathbf{0.91}$  \\
Chronos-tiny$^\dagger$              & 1{,}258.2 &   246.0 &  1{,}677.8 & $\mathbf{3.18}$  \\
Chronos-Bolt-tiny$^\dagger$         & 1{,}258.2 &   246.0 &  1{,}677.8 & $\mathbf{3.18}$  \\
Moirai-2.0-small                    & 1{,}415.7 &   184.3 &  1{,}887.4 & $\mathbf{3.49}$  \\
Moirai-small                        & 1{,}415.7 &   184.3 &  2{,}831.2 & $\mathbf{4.43}$  \\
Time-MoE-50M$^\ddagger$             & 2{,}831.5 &   368.6 & 16{,}986.9 & $\mathbf{20.19}$ \\
\bottomrule
\multicolumn{5}{l}{$^\dagger$ Identical FLOPs: both models share the same backbone at $P{=}1$.} \\
\multicolumn{5}{l}{$^\ddagger$ Active FLOPs only (3 of 9 experts per MoE layer).}
\end{tabular}
\end{table}

\begin{table}
\centering
\caption{FLOPs formulas used for each component (1 multiply-add = 2 FLOPs). $N$ = number of layers, $L$ = sequence length, $d$ = model dimension, $d_\text{ff}$ = FFN width, $E_a$ = active experts per token.}
\label{tab:flops_formulas}
\setlength{\tabcolsep}{8pt}
\begin{tabular}{l l}
\toprule
\textbf{Component}                & \textbf{FLOPs formula} \\
\midrule
Attn projections (Q, K, V, O)     & $N \times 4 \times 2 \times L \times d^2$ \\
Attn scores + weighted sum        & $N \times 2 \times 2 \times L^2 \times d$ \\
FFN (two linear layers)           & $N \times 2 \times 2 \times L \times d \times d_\text{ff}$ \\
Cross-attn linear                 & $N \times (2L_\text{dec} + 2L_\text{enc}) \times 2 \times d^2$ \\
Cross-attn scores + weighted sum  & $N \times 2 \times 2 \times L_\text{dec} \times L_\text{enc} \times d$ \\
MoE FFN (active)                  & $E_a \times N \times 2 \times 2 \times L \times d \times d_\text{ff}$ \\
\bottomrule
\end{tabular}
\end{table}

\begin{table*}
\centering
\caption{Model architectures, scales, parameter counts, embedding dimensions and forward-pass FLOPs. FLOPs are reported in MFLOPs except Total FLOPs, which is in GFLOPs. FLOPs numbers are from \cref{tab:flops_L200}.}
\scriptsize
\renewcommand{\arraystretch}{1.1}
\setlength{\tabcolsep}{4pt}
\resizebox{\textwidth}{!}{%
\begin{tabular}{l l l l c r r r r}
\toprule
Model & Architecture family & Exact scale used & Parameter count & Emb.\ dim & Attn.\ Linear & Attn.\ Quad. & FFN & Total FLOPs \\
\midrule
Astromer-1 & astronomy-specific transformer embedding model & Astromer-1 & 0.66M & 256 & 209.7 & 41.0 & 52.4 & 0.30 \\
Astromer-2 & astronomy-specific transformer embedding model & Astromer-2 & 5.4M & 256 & 629.1 & 123.0 & 157.3 & 0.91 \\
Moirai     & masked encoder-based universal forecasting transformer & Moirai-small & 14M & 384 & 1415.7 & 184.3 & 2831.2 & 4.43 \\
Moirai-2   & decoder-only universal forecasting transformer & Moirai-2.0-small & 11.4M & 384 & 1415.7 & 184.3 & 1887.4 & 3.49 \\
Chronos    & T5-style quantized time-series language model & Chronos-tiny & 8M & 256 & 1258.2 & 246.0 & 1677.8 & 3.18 \\
Chronos-Bolt & patch-based direct multi-step T5 encoder-decoder & Chronos-Bolt-tiny & 9M & 256 & 1258.2 & 246.0 & 1677.8 & 3.18 \\
Time-MoE   & decoder-only mixture-of-experts TSFM & Time-MoE-50M & 50M activated / 113M total & 384 & 2831.5 & 368.6 & 16986.9 & 20.19 \\
Hand-crafted features & domain-engineered feature baseline & --- & --- & 138 features & --- & --- & --- & --- \\
\bottomrule
\end{tabular}}
\label{tab:model_flops}
\end{table*}

Here we compare the computational cost of the models we evaluate.
To compare the computational cost for \emph{inference}, we calculate the FLOPs of the forward pass of each model. The results are in \cref{tab:flops_L200}. Model FLOPs are referenced from \cref{table:clustering-results-combined} in the main text. \Cref{tab:flops_formulas} lists the formulas used to compute FLOPs for each component, and \cref{tab:model_flops} summarizes architectures, parameter counts, embedding dimensions and total FLOPs side-by-side.

\clearpage

\clearpage

\section{Full List of Hand-crafted Features}
\label{app:handcrafted}
We select hand-crafted features from the libraries of established software packages: \texttt{FATS} \citep{Nun+2015} and \texttt{light\_curve} \citep{Malanchev+2021}. Each feature described here is computed for each passband individually and the embeddings are formed by concatenating the feature lists of the $g$ and $r$ embeddings. Tables~\ref{tab:FATS_feats_desc} and \ref{tab:LC_feats_desc} show the full list of features and descriptions from \texttt{FATS} and \texttt{light\_curve}, respectively.

\begin{table}[ht]
    \centering
    \begin{threeparttable}
        \caption{\texttt{light\_curve} features and plain-language descriptions}
        \label{tab:LC_feats_desc}
        \begin{tabular}{l p{0.64\linewidth}}
            \toprule
            \textbf{Feature} & \textbf{Intuitive one-sentence description} \\
            \midrule
            Amplitude & Half the peak-to-peak range—how far the light curve swings between brightest and faintest points. \\
            AndersonDarlingNormal & Scores how strongly the magnitude distribution departs from an ideal bell curve. \\
            BeyondNStd & Proportion of data points that sit more than $N$ standard deviations away from the mean, flagging outliers. \\
            Cusum & Total vertical span of the running cumulative sum, revealing slow drifts or trends. \\
            Eta & Von Neumann ratio: compares successive-point differences to overall scatter to catch rapid variability. \\
            EtaE & Eta re-weighted by time gaps so uneven sampling doesn’t skew the variability estimate. \\
            InterPercentileRange($p$) & Distance between the $p$ and $(1-p)$ quantiles—a robust width such as the IQR (when $p{=}0.25$). \\
            Kurtosis & Indicates whether the distribution is more peaked or heavy-tailed than a normal curve. \\
            LinearFit & Slope, error, and fit quality for a straight line that accounts for measurement uncertainties. \\
            LinearTrend & Slope and error of a simple least-squares line that ignores the error bars. \\
            MagnitudePercentageRatio & Ratio of inner to outer percentile widths, contrasting core spread with overall spread. \\
            MaximumSlope & Steepest single-step change in magnitude per unit time between consecutive points. \\
            Mean & Ordinary average magnitude. \\
            Median & Mid-point magnitude that splits the data into equal halves. \\
            MedianAbsoluteDeviation & Typical absolute distance from the median—a robust scatter measure. \\
            MedianBufferRangePercentage & Fraction of points that fall inside a narrow buffer zone around the median. \\
            OtsuSplit & Statistics describing the two groups produced by Otsu’s automatic thresholding of magnitudes. \\
            PercentAmplitude & Largest absolute deviation of any point from the median magnitude. \\
            ReducedChi2 & Reduced $\chi^{2}$ showing how well the data match their (weighted) mean given the quoted errors. \\
            Skew & Tells whether the distribution leans toward brighter or fainter extremes (positive or negative tail). \\
            StandardDeviation & Classical root-mean-square scatter of the magnitudes. \\
            StetsonK & Error-weighted “peakedness” measure that is robust to outliers in light-curve shape. \\
            WeightedMean & Average magnitude that gives greater weight to points with smaller measurement errors. \\
            \bottomrule
        \end{tabular}
        \begin{tablenotes}
            \item[See here for a detailed description: \url{https://github.com/light-curve/light-curve-python}]
        \end{tablenotes}
    \end{threeparttable}
\end{table}

\begin{table}[h]
    \centering
    \begin{threeparttable}
    \caption{\texttt{FATS} features and plain-language descriptions.}
    \label{tab:FATS_feats_desc}
    \begin{tabular}{l p{0.56\linewidth}}
    \toprule
        \textbf{Feature} & \textbf{Intuitive one-sentence description} \\
    \midrule
        PeriodLS & Best-fit period of the light curve using the Lomb–Scargle method. \\
        Period\_fit & The false alarm probability of the largest Lomb–Scargle periodogram value. \\
        Psi\_CS & The range of a cumulative sum metric computed on the phase-folded light curve. \\
        Psi\_eta & The variability index $\eta^e$ computed on the phase-folded light curve. \\
        Autocor\_length & The cross-correlation of the light curve with itself. \\
        PairSlopeTrend & The fraction of increasing first differences subtracted from the fraction of decreasing first differences, computed on the 30 most recent magnitude measurements. \\
        Freq\{N\}\_harmonics\_amplitude\_\{M\} & Amplitude of the $M$th harmonic of the $N$th dominant frequency. \\
        Freq\{N\}\_harmonics\_rel\_phase\_\{M\} & Relative phase of the $M$th harmonic of the $N$th dominant frequency. \\
        CAR\_sigma & Short-term variability amplitude in a continuous auto-regressive (CAR) model. \\
        CAR\_tau & Characteristic timescale of correlations in the CAR model. \\
        CAR\_mean & Long-term mean magnitude level in the CAR model. \\
    \bottomrule
    \end{tabular}
    \begin{tablenotes}
            \item[See here for a detailed description: \url{http://isadoranun.github.io/tsfeat/FeaturesDocumentation.html}]
        \end{tablenotes}
    \end{threeparttable}
\end{table}

\section{Astrophysical Description of Classes}
\label{app:classes_intro}
Our dataset contains seven total classes of periodic variable stars: EW, EA, RRab, RRc, RRd, RS CVn, and LPV. Here, we provide a high-level astrophysical description of each of these classes, including each class' observational characteristics and utility. In some cases, multiple classes are closely related so we describe them together. We also include descriptions of the classes which, due to their rarity, are considered out-of-distribution in this work: $\beta$-Lyrae, Blazhko, Anomalous Cepheids, Cepheid-II, HADS, LADS, ELL, Hump, PCEB, and EAup.

\subsection{Eclipsing Binaries (EW, EA)}
Eclipsing binary stars are pairs of stars orbiting each other and aligned with the observer in such a way that either star periodically blocks the light from the other. When neither star is eclipsed, the system is at maximum brightness, but when one star is eclipsed by the other, the total flux received from the system is suppressed, giving the binary star system periodic light curve behavior. This type of variability is described as extrinsic because it is not due to astrophysical properties of the stars themselves. Sub-categorization of eclipsing binaries is based on the configuration of the stars in the pair.

EW-type eclipsing binaries (also called W Ursae Majoris-type after the original EW system) are contact binaries. In this case, the two stars, typically dwarf stars, share a common outer envelope which entirely encapsulates them. This common envelope allows for the exchange of mass and energy between the pair, equilibrating their temperature. Their light curves exhibit constant and smooth variations where the dips from either star being eclipsed are of similar or identical depth. EW-type variables typically have short periods ($0.2 \lesssim P \textrm{ [d]} \lesssim 0.5$), with a notably unsolved period cut-off at $\sim0.2$ days \citep{Rucinski+2007, Drake+2014}. These systems are astrophysically valuable, in part, because they are expected to emit gravitational waves due to their tight orbits, and they also have the possibility of merging and triggering transient events \citep{Tylenda+2011}. 

EA-type eclipsing binaries (also called Algol-type binaries after the original EA system) are detached binaries. In this case, the two stars are not in contact and thus can have different temperatures and more varied orbits, manifesting as different light curve properties. The ellipticity of the orbit and the brightnesses of the stars affect the spacing and depths of the brightness dips. Multiple additional factors can affect their light curves, e.g., the presence of an accretion disk. EA systems tend to have longer periods than EW systems due to their wider orbits: $0.3 \lesssim P \textrm{ [d]} \lesssim 100$. EA systems are key for studying binary stellar evolution and populations, especially the exchange of mass between stars in a binary. 

\subsection{Active Binaries (RS CVn)}
RS Canum Venaticorum (RS CVn) stars are also stars in binary systems and are characterized by one of the stars exhibiting large magnetic spots on its surface. This manifests as observable variability as the RS CVn stars also have rapid rotational velocities. The resulting light curve affect also depends on the difference between the star's rotational period and the systems orbital period, and most RS CVn systems are tidally locked, meaning the two periods are closely matched. Some RS CVn are also eclipsing binaries, so their light curves can also show variability due to eclipses. Their periods tend to be $3 \lesssim P \textrm{ [d]} \lesssim 14$. RS CVn systems serve as extreme testbeds for studying stellar magnetic phenomena and evolution.

\subsection{RR Lyrae (RRab, RRc, RRd)}
RR Lyrae stars are low-mass stars exhibiting pulsations, cyclically expanding and contracting radially due to internal changes in opacity. Because RR Lyrae occur with only a small range of intrinsic brightnesses, their distances can be easily measured from their observed brightness. Among other utilities, this allows RR Lyrae to be used for measuring distances within the Milky Way and to nearby Galaxies. Their periods are also related to their chemical composition, so they can provide crucial information about Galactic structure and formation.

RR Lyrae can occur in different pulsation modes which define the various RR Lyrae subclasses. RRab stars pulsate in the fundamental mode; RRc in the first-overtone; and RRd are double-mode pulsators. RRab stars have light curves with a rapid brightening episode followed by a gradual fading, producing a sawtooth-like pattern. RRab typically have periods $0.4 \lesssim P \textrm{ [d]} \lesssim 1.0$. RRc stars exhibit sinusoid-like variability with typical periods of $0.2 \lesssim P \textrm{ [d]} \lesssim 0.5$. They also tend to exhibit a constant, slow drift in their periods. RRd stars pulsate in both the fundamental mode and the first-overtone and thus show a combination of two periodic signals in their light curves, typically with the first-overtone dominating.

\subsection{Long Period Variables (LPVs)}
LPVs are giant stars exhibiting pulsations with periods of $3 \lesssim P \textrm{ [d]} \lesssim 1000$. They include multiple subtypes each with different period-luminosity relations but we, as in \cite{Drake+2014}, consider these a single class. They pulsate with a similar mechanism to the RR Lyrae but have dramatically larger radii. Their outer layers are not very tightly bound to the rest of the star, which can lead to the star expunging mass and polluting the surrounding environment with gas. Thus, studying LPVs provides insights into the cycles of gas into and out of stars. Their period-luminosity relations also allow LPVs to act as distance measures.

\subsection{Beta-Lyrae}
Beta-Lyrae ($\beta$-Lyrae) stars are close binary star systems in which the outer gaseous layers of both stars have expanded to the point that the pair is enveloped in a shared gaseous envelope. At this stage, gaseous material can be transferred from one star to another, altering either star's evolution. Their periods are typically a few days, and their light curves display continuous variations in brightness rather than flat maxima or minima like EA-type binaries, for example. Their dynamics provide direct information on gaseous mass transfer between binary stars, stellar structure under extreme tidal distortion, and the role of binarity in late stellar evolution.

\subsection{Blazhko}
Blazhko variables are a sub-class of RR Lyrae variables, which exhibit the rare Blazhko effect in which their light curve amplitudes and phases are modulated over long time periods (tens to hundreds of days). In part owing to its rarity relative to normal RR Lyrae stars, there is not yet a consensus to the physical mechanism driving the Blazhko effect. The effect may be explained with magnetic fields or resonances within the star, so these stars are useful for studying exotic phenomena which can occur in stars.

\subsection{Type II Cepheids (Cepheid-II) and Anomalous Cepheids (ACEP)}
Cepheid-II stars have old, low-mass stars with periods typically of tens of days and can produce a variety of light curve morphologies. They deviate from classical Cepheids because they are much fainter, but they do follow their own period-luminosity relation, enabling them to also be used as distance measures.

ACEPs have periods and luminosities inbetween those of RR Lyrae and classical Cepheids (roughly 0.3–2 days) with amplitudes of about 0.3–1.0 magnitudes, and their light curves typically resemble those of RRab stars. Their physical nature is not very well understood, but they have be proposed to be a product of gaseous mass transfer in a binary star system. ACEPs provide special astrophysical insights into stellar evolution pathways involving binary interaction, as well as into the environments where they typically occur.

\subsection{High-amplitude Delta-Scutis (HADS) and Low-amplitude Delta-Scutis (LADS)}
Delta-Scuti ($\delta$-Scuti) stars are pulsating variables stars with short periods and are typically divided into the HADS and LADS subclasses based on the morphology of their light curves. HADS have simple, regular, sawtooth-like light curves with periods $<$0.3 days and amplitudes greater than $\sim$0.3 magnitudes. In contrast, LADS exhibit complex, multi-periodic light curves with amplitudes below $\sim$0.1 magnitudes. HADS provide clean tests of stellar pulsation theory and scaling relations, while LADS are testbeds for asteroseismology. 

\subsection{Ellipsoidal Binaries (ELL)}
ELLs are close binary star systems in which the stars are tidally distorted into ellipsoidal shapes, producing photometric variability without eclipses. ELL light curves have smooth, nearly sinusoidal variations with two unequal minima per cycle, arising from the elongated parts rotating in and out of view. They are astrophysically important because they reveal details of binary star evolution, stellar shapes, and the presence of companion objects such as white dwarfs, neutron stars, or black holes. 

\subsection{Hump variables}
The Hump class is used as a catch-all for the small amount of periodic variables which \cite{Drake+2014} were unable to classify into any other known classes but do show clear periodic variability. Some of these objects exhibit vaguely sawtooth-like variability, like what is seen in RRab stars, but others have smoother variability. 

\subsection{Post-Common-Envelope Binaries (PCEB)}
PCEBs are close binary stars that have recently emerged from a common-envelope evolutionary phase, in which one star expanded and engulfed its companion star inside its expanded outer layers (``envelope"). Their light curves show a wide range of morphologies, including eclipses and ellipsoidal modulations, depending on the physical properties of the system like the inclination of the orbit relative to our line-of-sight and the nature of either of the stars in the binary system. Their periods tend to be very short as the common-envelope phase drives angular momentum out of the orbit. The smaller star in these systems are often a white dwarf, so these systems are an opportunity to study potential progenitors of Type Ia supernovae.

\subsection{EA with unknown period (EAup)}
EAup are EA-type stars where \cite{Drake+2014} were unable to determine their periods for any reason.

\clearpage
\section{Complete Set of Confusion Matrices for Classification}
\label{app:confusion}
\begin{figure*}[htp]
  \centering
  \vspace{-0.5em}
  \includegraphics[width=\linewidth]{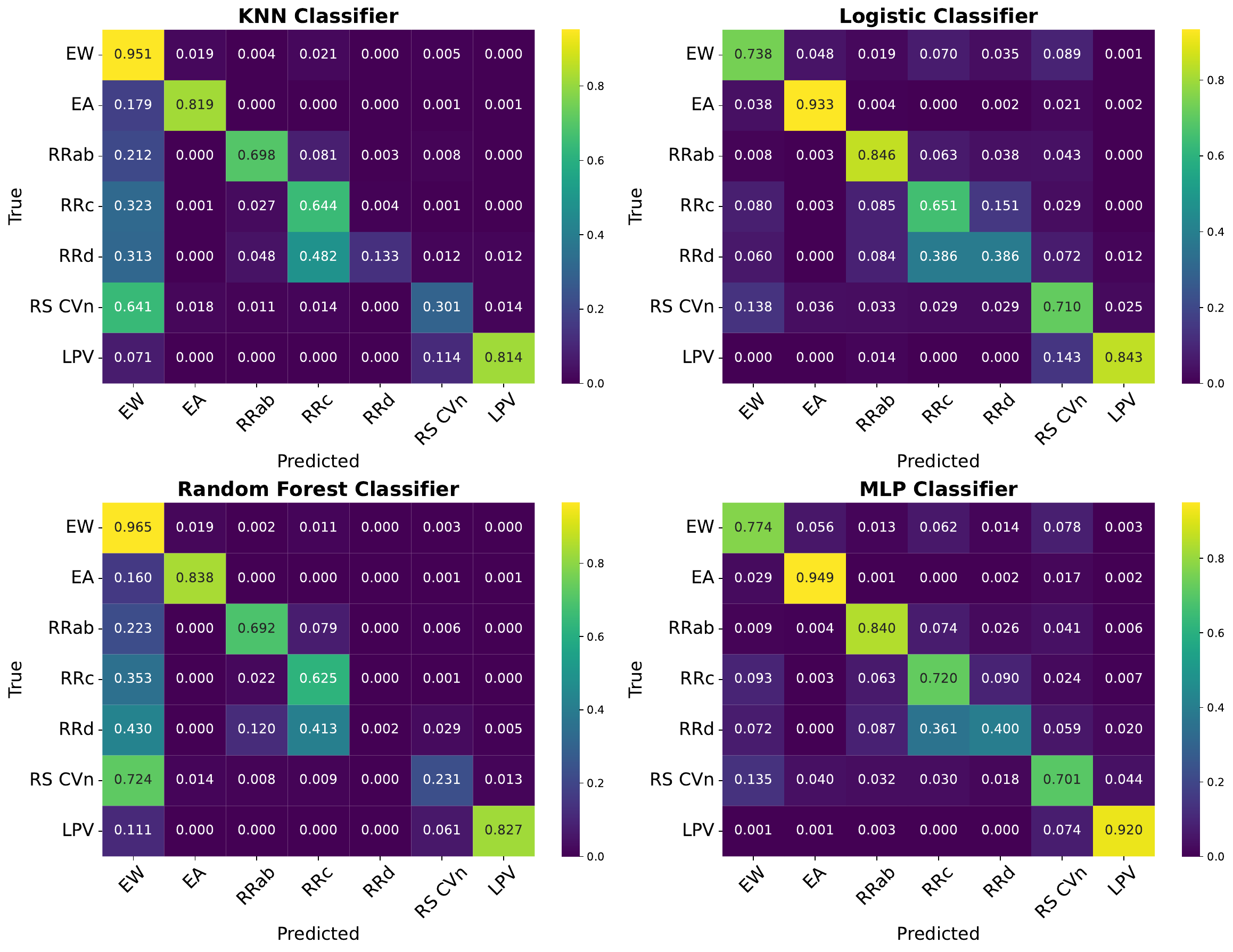}
  \caption{
  Confusion matrix of \texttt{Chronos} on four classifiers.
  }
  \label{fig:confusion_chronos_tiny}
  \vspace{-0.5em}
\end{figure*}

\begin{figure*}
  \centering
  \includegraphics[width=\linewidth]{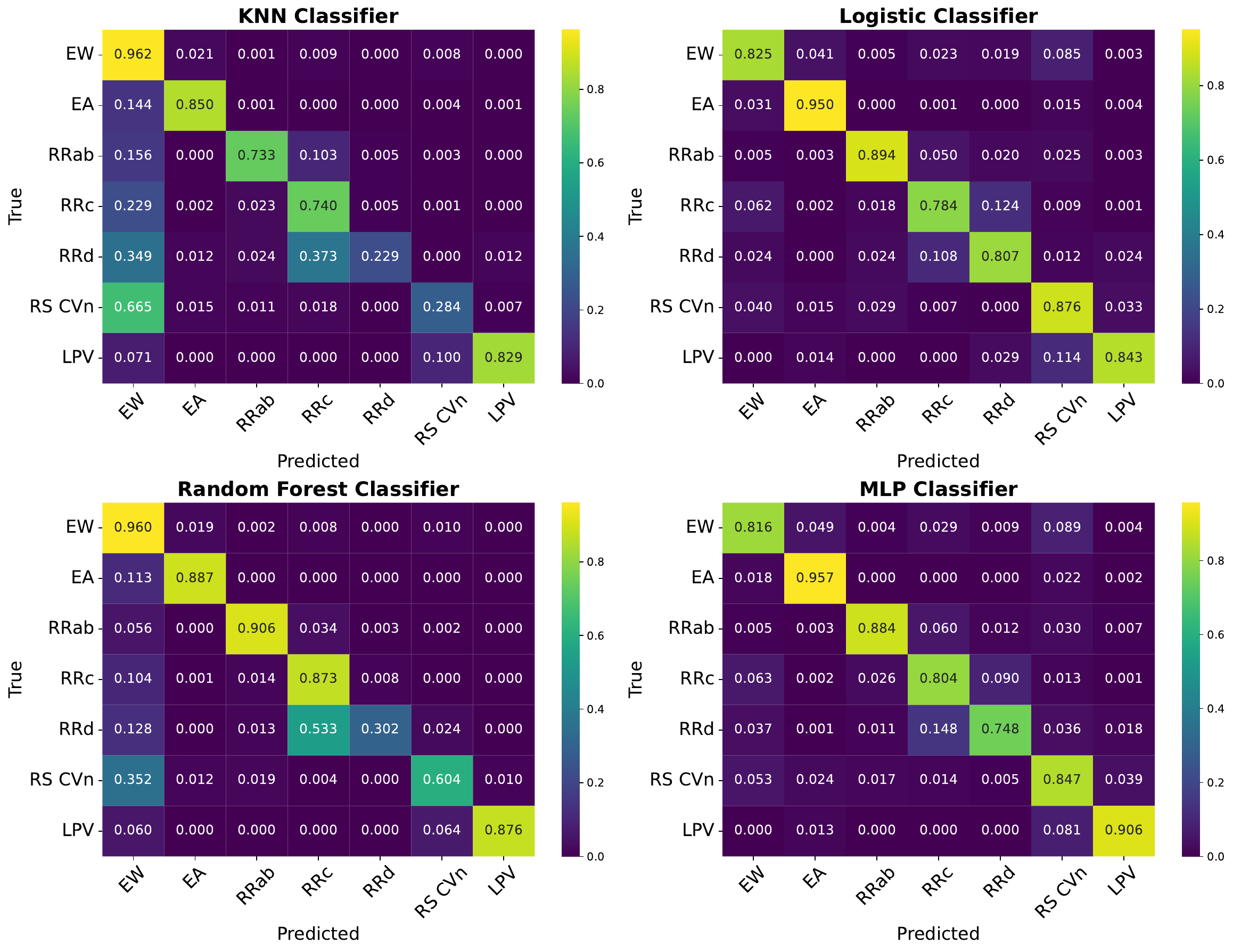}
  \caption{
  Confusion matrix of hand-crafted features on four classifiers.
  }
  \label{fig:confusion_hf}
\end{figure*}

\begin{figure*}
  \centering
  \vspace{-1em}
  \includegraphics[width=\linewidth]{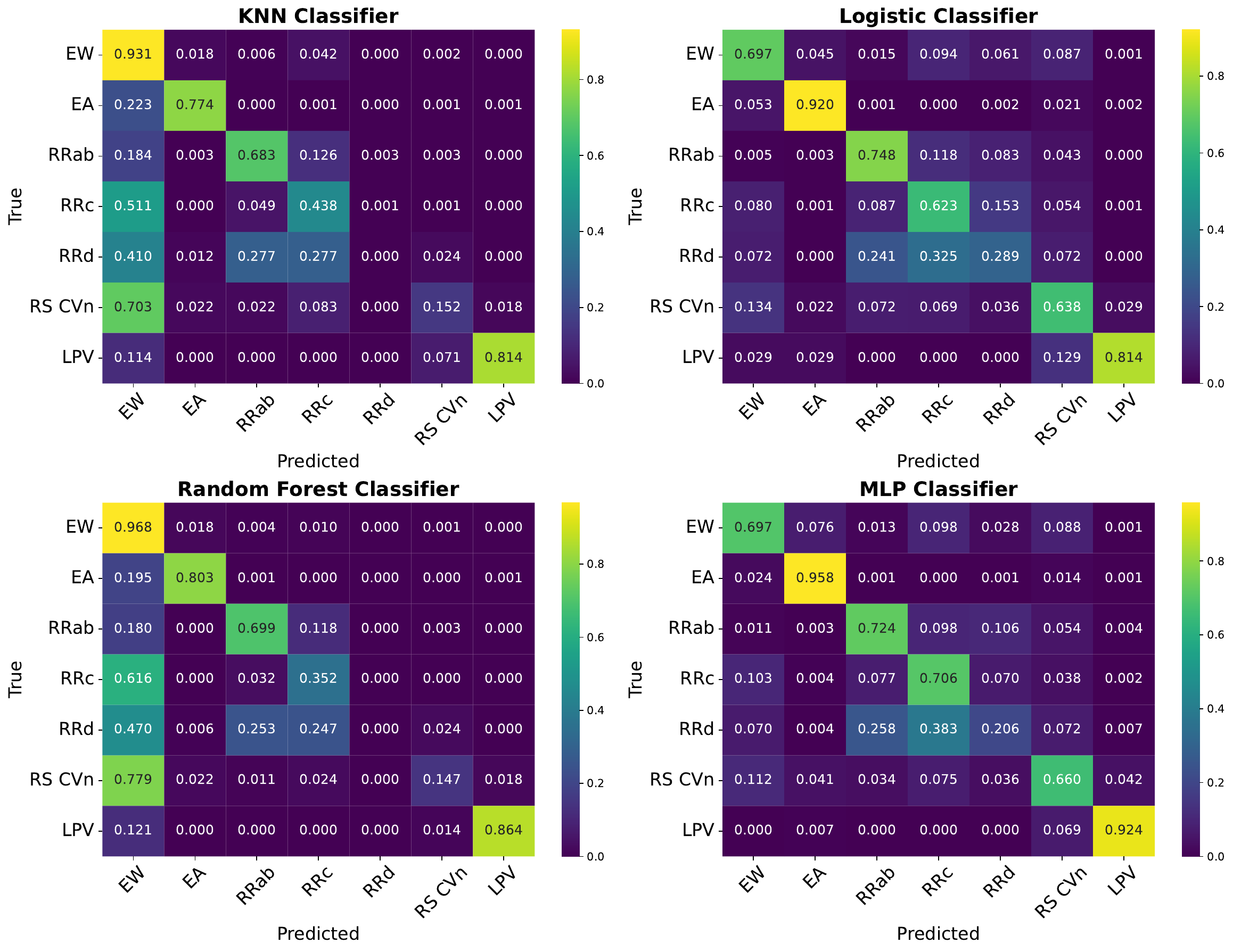}
  \caption{
  Confusion matrix of \texttt{Chronos-Bolt} on four classifiers.
  }
  \label{fig:confusion_bolt}
  \vspace{-0.5em}
\end{figure*}

\begin{figure*}
  \centering
  \includegraphics[width=\linewidth]{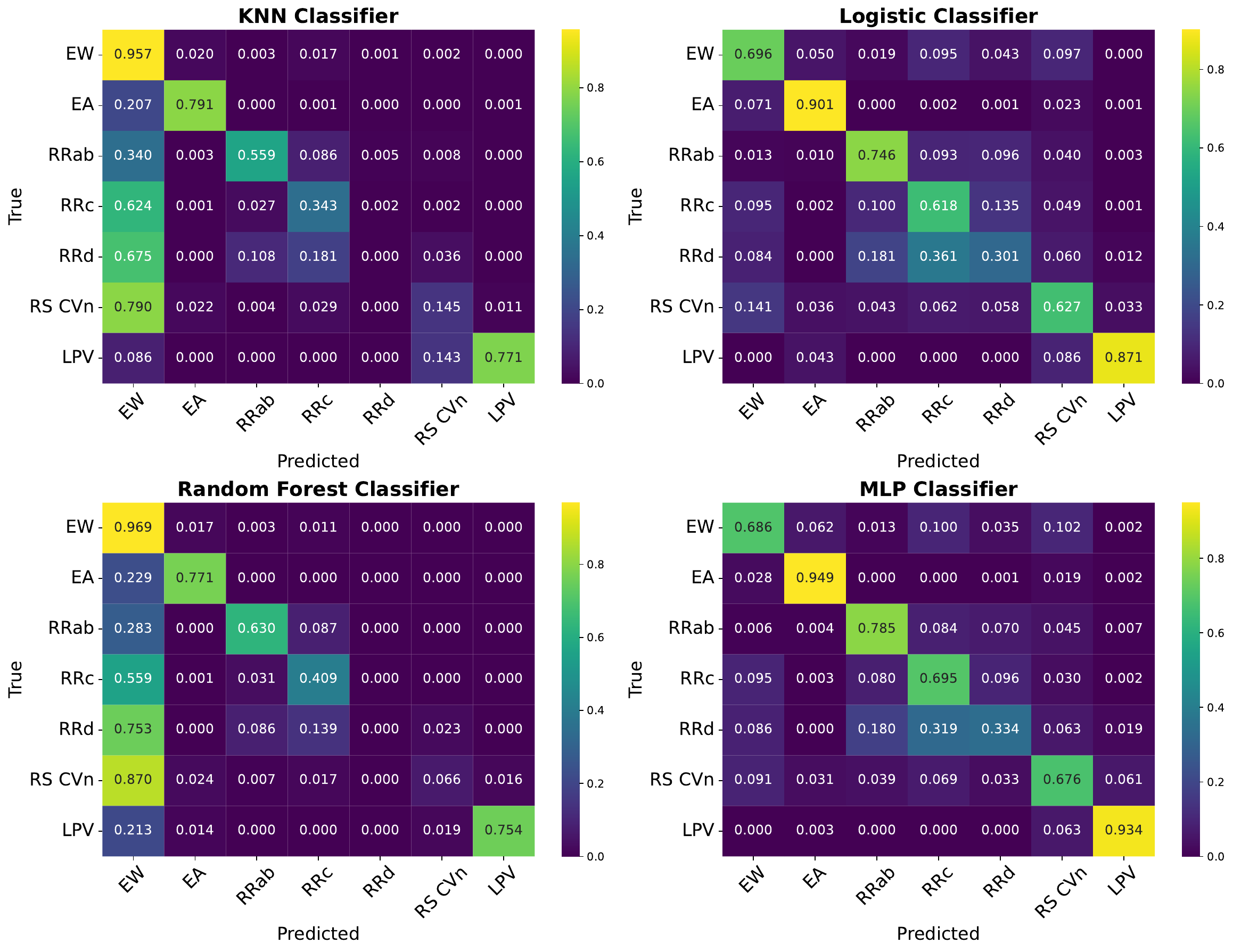}
  \caption{
  Confusion matrix of \texttt{Moirai} on four classifiers.
  }
  \label{fig:confusion_moirai}
  \vspace{-0.5em}
\end{figure*}

\begin{figure*}
  \centering
  \includegraphics[width=\linewidth]{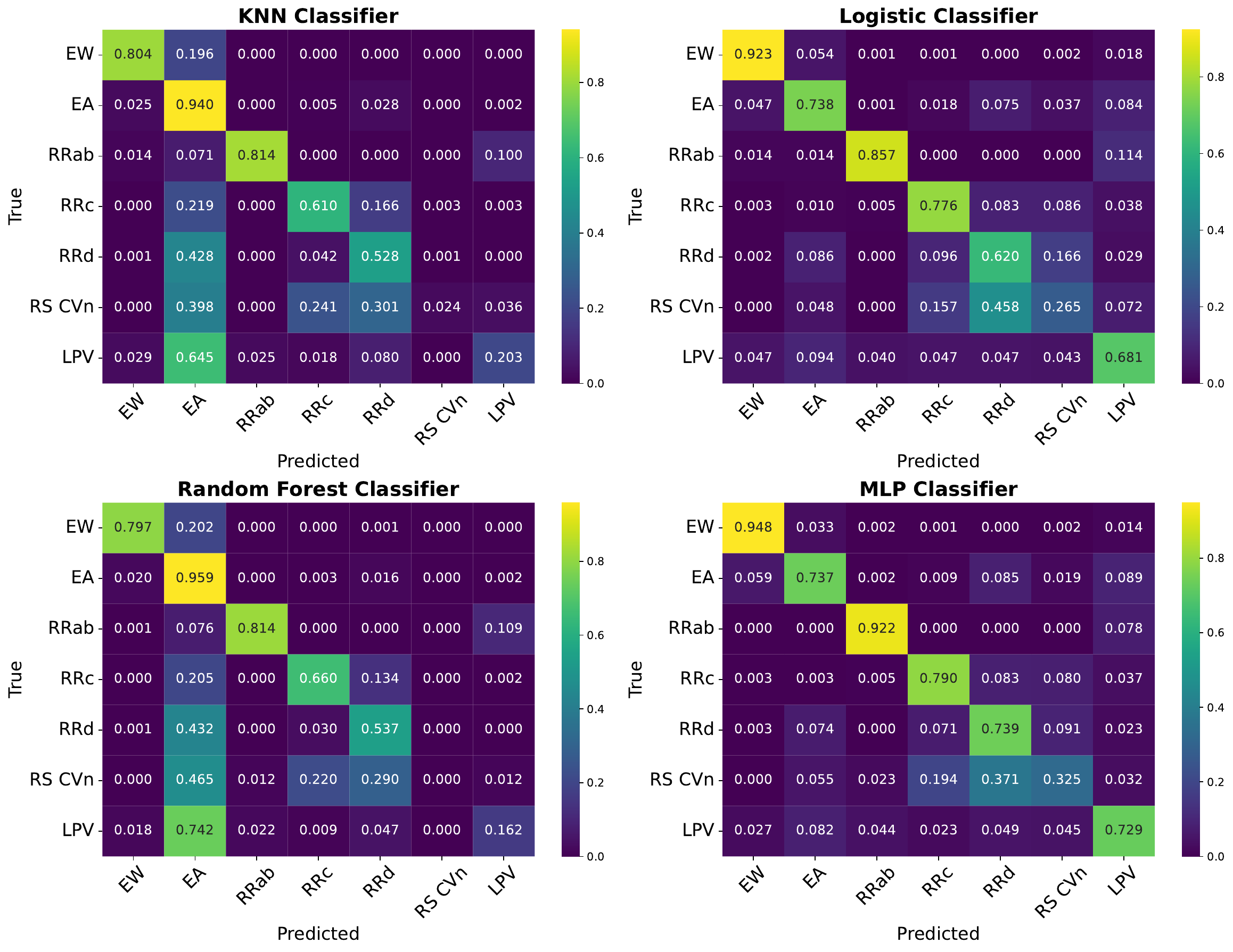}
  \caption{
  Confusion matrix of \texttt{Time-MoE} on four classifiers.
  }
  \label{fig:confusion_Time-MoE}
  \vspace{-0.5em}
\end{figure*}

\begin{figure*}
  \centering
  \includegraphics[width=\linewidth]{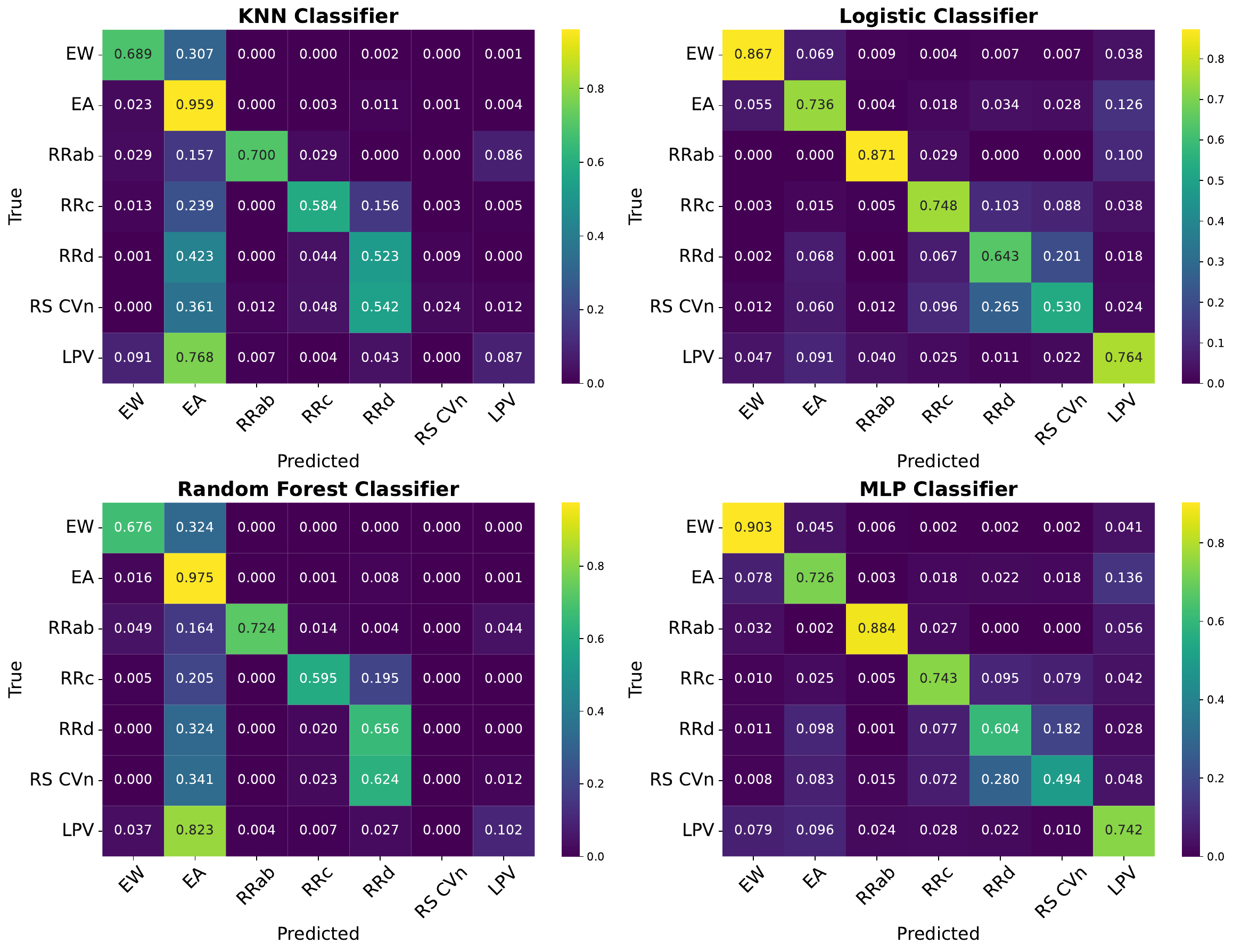}
  \caption{
  Confusion matrix of \texttt{Mantis+} on four classifiers.
  }
  \label{fig:confusion_Mantis+}
  \vspace{-0.5em}
\end{figure*}

\begin{figure*}
  \centering
  \includegraphics[width=\linewidth]{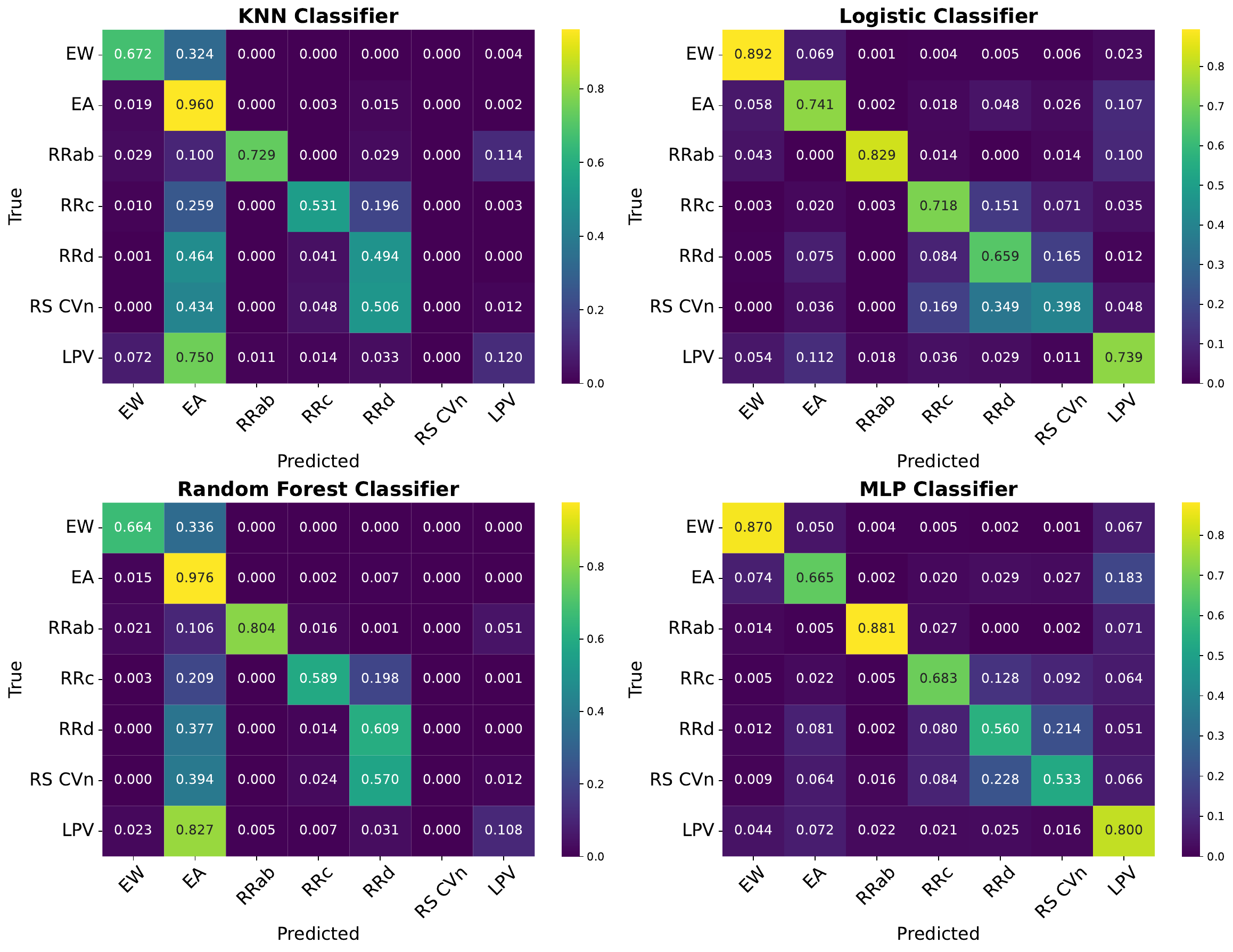}
  \caption{
  Confusion matrix of \texttt{MantisV2} on four classifiers.
  }
  \label{fig:confusion_MantisV2}
  \vspace{-0.5em}
\end{figure*}

\begin{figure*}
  \centering
  \includegraphics[width=\linewidth]{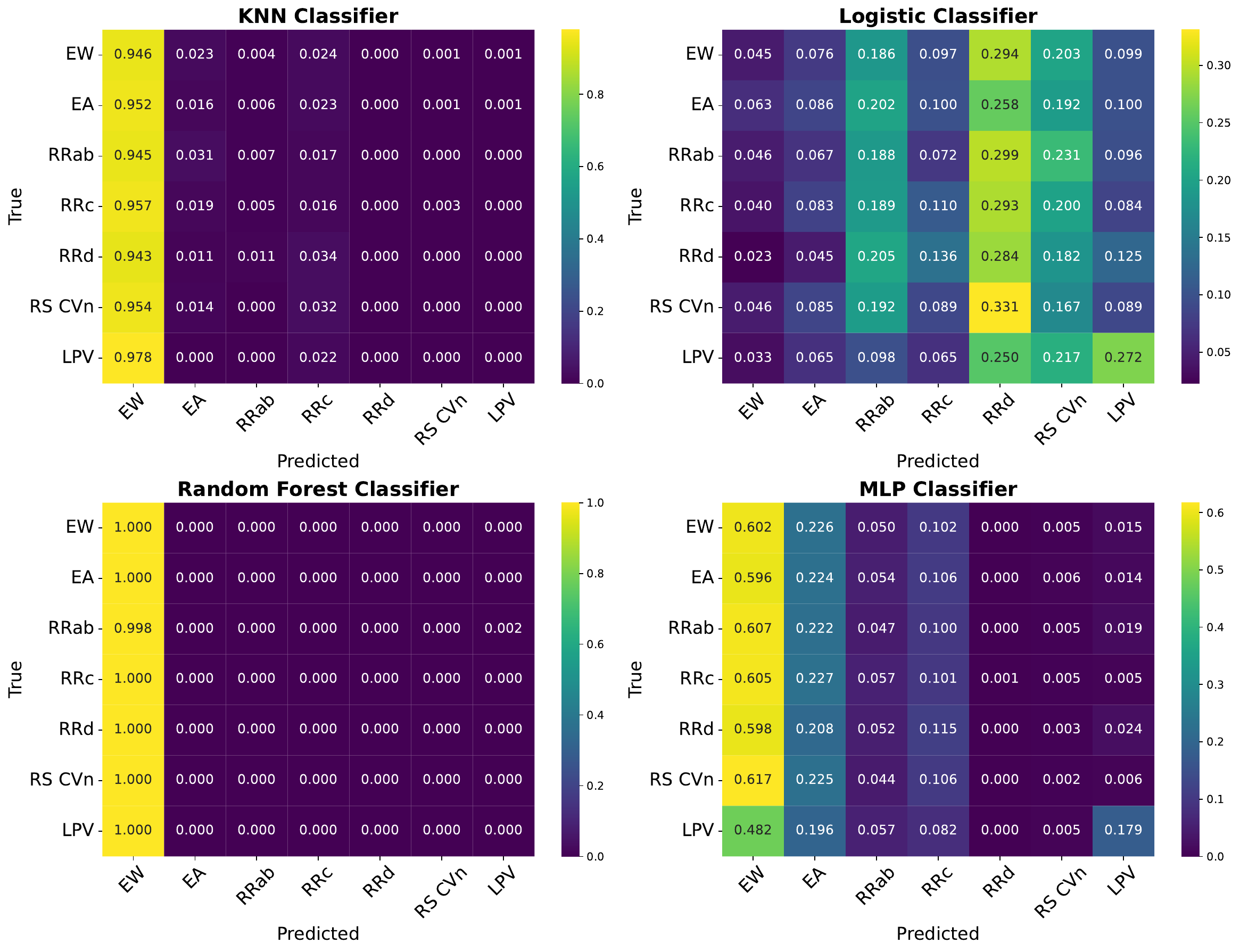}
  \caption{
  Confusion matrix of \texttt{Astromer-1} on four classifiers.
  }
  \label{fig:confusion_astromer_1}
  \vspace{-1em}
\end{figure*}

\begin{figure*}
  \centering
  \includegraphics[width=\linewidth]{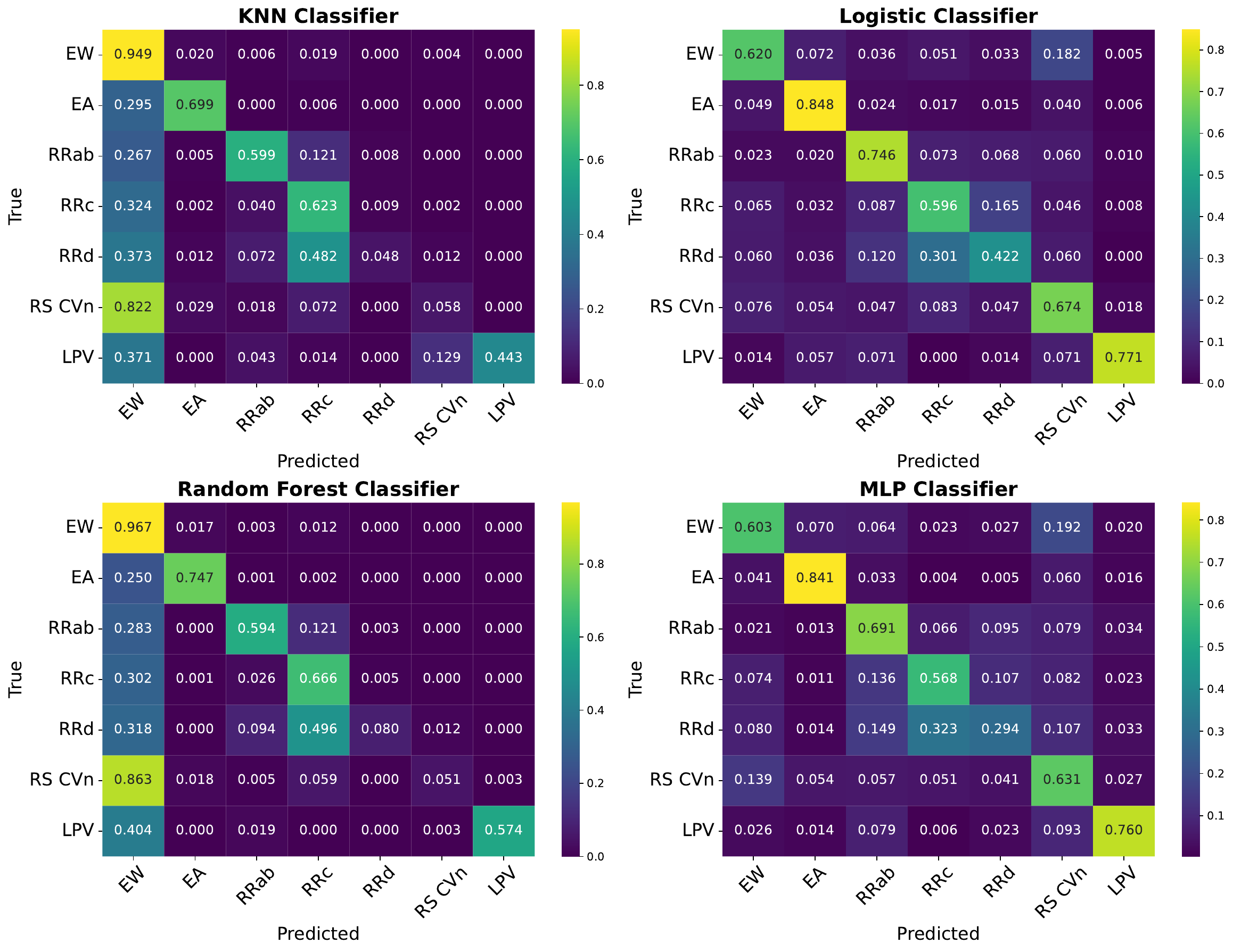}
  \caption{
  Confusion matrix of \texttt{Astromer-2} on four classifiers.
  }
  \label{fig:confusion_astromer_2}
  \vspace{-0.5em}
\end{figure*}

\blue{To provide detailed information about classification performance across different variable star classes, we report the confusion matrices for all embeddings and the hand-crafted features in this section.
For the random forest and MLP classifiers, the per-class accuracies are obtained by averaging over the results of $10$ runs.}

\end{document}